\documentclass[12pt]{article}
\usepackage{epsfig,amssymb}

\catcode`\@=11
\def \thesection {\arabic{section}.}

\def \be  {\begin{equation}}
\def \ee  {\end{equation}}
\def \ba  {\begin{eqnarray}}
\def \ea  {\end{eqnarray}}
\def \baa {\begin{eqnarray*}}
\def \eaa {\end{eqnarray*}}
\def \bb  {\begin {thebibliography} }
\def \eb  {\end{thebibliography}}
\def \lab #1 {\label{#1}}

\newcommand \bi [1] {\bibitem{#1}}
\newcommand\re[1]{(\ref{#1})}
\def \qqquad {\qquad\quad}
\def \qqqquad {\qquad\qquad}
\def \matrix #1 {\left(\begin{array}{cc} #1 \end{array}\right)}
\def \Tr {\mathop{\rm Tr}\nolimits}
\def \tr {\mathop{\rm tr}\nolimits}
\def \Im {\mathop{\rm Im}\nolimits}
\def \Re {\mathop{\rm Re}\nolimits}

\def \e  {\mathop{\rm e}\nolimits}
\newcommand\lr[1]{{\left({#1}\right)}}
\newcommand \widebar [1] {\overline{#1}}

\newcommand \vev [1] {\langle{#1}\rangle}

\newcommand \ket [1] {|{#1}\rangle}
\newcommand \bra [1] {\langle {#1}|}

\newcommand{\as}{\ifmmode\alpha_{\rm s}\else{$\alpha_{\rm s}$}\fi}
\def \CO {{\cal O}}

\font\cmss=cmss12 
\def\inbar{\,\vrule height1.5ex width.4pt depth0pt}
\def\IC{\relax\hbox{$\inbar\kern-.3em{\rm C}$}}
\def\IZ{\relax{\hbox{\cmss Z\kern-.4em Z}}}
\def\IR{{\hbox{{\rm I}\kern-.2em\hbox{\rm R}}}}

\def\IP{{\hbox{{\rm I}\kern-.2em\hbox{\rm P}}}}
\def\II{\hbox{{1}\kern-.25em\hbox{l}}}

\def\numberbysection{\@addtoreset{equation}{section}
                     \def\theequation{\thesection\arabic{equation}}}
\numberbysection

\newcommand \mybf[1] {\mbox{\boldmath$ {#1} $}}

\begin{document}

\begin{titlepage}
\begin{flushright}
\begin{tabular}{l}
LPT--Orsay--01--75\\
UB--ECM--PF--01/05\\
hep-th/0107193
\end{tabular}
\end{flushright}

\vskip3cm
\begin{center}
  {\large \bf
  Noncompact Heisenberg spin magnets from high-energy QCD  \\[2mm] I.~Baxter $Q-$operator and Separation of Variables}

\def\thefootnote{\fnsymbol{footnote}}%
\vspace{1cm} {\sc S.\'{E}. Derkachov}${}^1$, {\sc
G.P.~Korchemsky}${}^2$
          and {\sc A.N.~Manashov}${}^3$\footnote{
Permanent address:\ Department of Theoretical Physics,  Sankt-Petersburg State
University, St.-Petersburg, Russia}
\\[0.5cm]

\vspace*{0.1cm} ${}^1$ {\it
Department of Mathematics, St.-Petersburg Technology Institute,\\
St.-Petersburg, Russia
                       } \\[0.2cm]
\vspace*{0.1cm} ${}^2$ {\it
Laboratoire de Physique Th\'eorique%
\footnote{Unite Mixte de Recherche du CNRS (UMR 8627)},
Universit\'e de Paris XI, \\
91405 Orsay C\'edex, France
                       } \\[0.2cm]
\vspace*{0.1cm} ${}^3$
 {\it
Department d'ECM, Universitat de Barcelona,\\
08028 Barcelona, Spain} 

\vskip2cm
{\bf Abstract:\\[10pt]} \parbox[t]{\textwidth}{
We analyze a completely integrable two-dimensional quantum-mechanical model that
emerged in the recent studies of the compound gluonic states in multi-color QCD
at high energy. The model represents a generalization of the well-known
homogenous Heisenberg spin magnet to infinite-dimensional representations of the
$SL(2,\mathbb{C})$ group and can be reformulated within the quantum inverse
scattering method. Solving the Yang-Baxter equation, we obtain the $R-$matrix for
the $SL(2,\mathbb{C})$ representations of the principal series and discuss its
properties. We explicitly construct the Baxter $Q-$operator for this model and
show how it can be used to determine the energy spectrum. We apply Sklyanin's
method of the Separated Variables to obtain an integral representation for the
eigenfunctions of the Hamiltonian. We demonstrate that the language of Feynman
diagrams supplemented with the method of uniqueness provide a powerful technique
for analyzing the properties of the model.}
\vskip1cm

\end{center}

\end{titlepage}

\newpage

{\small \tableofcontents}

\newpage

\section{Introduction}

In this paper, we study a completely integrable quantum-mechanical model of $N$
interacting spinning particles in two-dimensional space. The model can be
thought of as a generalization of the well-known periodic one-dimensional ${\rm
XXX}$ Heinsenberg spin chain magnet \cite{QISM,XXX,ABA} -- each particle defines
the site of the spin chain and the interaction occurs between spins of two
neighboring particles. The spin operators associated with each particle are the
generators of an infinite-dimensional principal series representation of the
$SL(2,\mathbb{C})$ group and that is why we shall refer to the model as a
noncompact spin magnet.

The motivation for studying such models comes from two different and seemingly
unrelated physical problems. The first of them has to do with the Regge phenomena
in QCD. As was shown in \cite{L2,BKP}, the high-energy behavior of the scattering
amplitudes in the generalized leading-logarithmic approximation is governed by
the contribution of the color-singlet compound states, built from an arbitrary
number $(N=2,3,...)$ of gluons and defined as ground states of the effective QCD
Hamiltonian. The effective Hamiltonian acts on the two-dimensional transverse
coordinates of the gluons and exhibits remarkable properties of integrability
\cite{L1,FK}. Namely, in the limit of the multi-color QCD, the wave function of
the $N-$gluon compound state turns out to be identical to the ground state of
noncompact Heisenberg spin chain model with the number of sites equal to the
number of gluons, $N$. The quantum space in each site of this model is
parameterized by two-dimensional gluon transverse coordinates and corresponds to
the spin $s=0$ principal series representation of the $SL(2,\mathbb{C})$ group.
At $N=2$ the ground state is known as the BFKL Pomeron
\cite{L2}. At $N=3$ it gives rise to the Odderon \cite{JW} and for higher $N$ the
corresponding ground states define the unitarity corrections to the scattering
amplitudes. In spite of a lot of efforts, the solution of the Schr\"odinger
equation for $N\ge 3$ gluon compound states still represents an outstanding
theoretical problem in QCD.

Another motivation for studying the noncompact spin magnets comes from the
problem of solving the quantum completely integrable models defined on an
infinite-dimensional phase space. The best known example of such models is given
by the one-dimensional Toda chain model \cite{Toda}. One of their main features
is that they can not be solved using the conventional Algebraic Bethe Ansatz
(ABA) method \cite{QISM,XXX,ABA} and one has to rely on more advanced methods
\cite{Bax,SoV}. The noncompact two-dimensional $SL(2,\mathbb{C})$ spin magnets
belong to the same class of models and their solution represents a theoretical
challenge. As we will show in this paper, these two-dimensional models have many
features in common with the known one-dimensional integrable models. Namely, they
admit the same $R-$matrix representation as $\rm XXX$ Heisenberg compact spin
chain and part of their spectrum can be reconstructed using the ABA approach. At
the same time, similar to the Toda model, the quantum space of the system is
infinite-dimensional and, in general, the eigenstates do not admit the highest
weight representation. This means that the ABA approach can not provide the full
set of the eigenstates of the model and, therefore, it is not complete.

Our approach to solving the Schr\"odinger equation for the $SL(2,\mathbb{C})$
spin magnet relies on the methods of the Baxter $Q-$operator \cite{Bax} and the
Separation of Variables (SoV) developed by Sklyanin \cite{SoV}. The former allows
to determine the energy spectrum of the model, while the latter provides an
integral representation of the corresponding eigenfunctions. In recent years,
both methods have been developed and successfully applied to solving the Toda
chain model
\cite{SoV,PG,KL} and high spin generalizations of the homogenous ${\rm XXX}$ Heisenberg
$SL(2,\mathbb{R})$ spin magnet \cite{SD}. In this paper, we shall extend these
results to the $SL(2,\mathbb{C})$ Heisenberg spin magnets.

The central point of our analysis is the representation of different operators in
the model (the $R-$matrix, Hamiltonian, Baxter $Q-$operator {\it etc.}) by their
integral kernels defined on the two-dimensional plane. In this way, remarkable
integrability properties of the model that are usually expressed in the form of
operator identities, like Yang-Baxter equations, are translated into
(complicated) integral relations between the corresponding kernels. Adopting the
language of the two-dimensional Feynman integrals, one can associate the kernels
with particular Feynman diagrams and develop the diagrammatical representation
for the above identities. One of the main findings of this paper is that, using
the diagrammatical approach, one can establish integrability properties of the
$SL(2,\mathbb{C})$ spin magnet without doing any actual calculations by applying
two elementary identities between the Feynman diagrams known in QCD as the
``uniqueness relations'' \cite{uniq,uniq1}. Following this approach, we find the
solution of the Yang-Baxter equation for the $R-$matrix corresponding to the
$SL(2,\mathbb{C})$ principal series representation, construct the Baxter
$Q-$operator and the unitary transformation to the Separated Variables and,
finally, establish different relations between them that allow to solve the
model.

The paper is organized as follows. In Section 2 we introduce the Hamiltonian of
the model and show that it admits the $R-$matrix representation. We obtain the
expression for the $R-$matrix by solving the Yang-Baxter equation for the
principal series $SL(2,\mathbb{C})$ representation and discuss its general
properties. This allows to prove a complete integrability of the model and
establish its symmetry properties. Section 3 is devoted to the construction of
the Baxter $Q-$operator for the $SL(2,\mathbb{C})$ spin magnet. Our approach is
similar to the one used before for the Toda chain \cite{PG} and the homogenous
$SL(2,\mathbb{R})$ Heisenberg spin magnets \cite{SD}. Examining the properties of
the obtained expressions, we establish the relation between the transfer matrices
of the model and the product of the Baxter $Q-$operators. This relation allows to
express the Hamiltonian of the model in terms the $Q-$operator and reduce the
original Schr\"odinger equation to the problem of finding the solutions to the
Baxter equation on the eigenvalues of the $Q-$operator under the additional
conditions imposed by the analytical properties of the $Q-$operator and its
asymptotic behavior at infinity. In Section 5, we apply Sklyanin's method of
Separation of Variables to obtain an integral representation for the
eigenfunctions of the Hamiltonian. We construct the unitary transformation to the
separated variables and demonstrate its relation with the Baxter $Q-$operators.
We obtain the quantization conditions on the separated variables and calculate
the integration measure on the space of the eigenstates in the SoV
representation. Concluding remarks are presented in Section 6. Appendix A
provides a detailed description of the method of uniqueness which is intensively
used throughout the paper. In Appendix B, we describe the properties of the
$SL(2,\mathbb{C})$ transfer matrices including the fusion identities and their
relation with the Baxter $Q-$operators. The relation between the obtained
integral representation of the eigenstates and the Algebraic Bethe Ansatz method
is discussed in the Appendix C.

\section{The quantum noncompact spin chain model}

\subsection{Definition of the model}

The noncompact spin chain model is a quantum mechanical system of $N$ interacting
particles on a two-dimensional $(x,y)-$plane. 
In high-energy QCD, this plane corresponds to transverse gluonic degrees of
freedom. It becomes convenient to define the position of the particles on the
plane by introducing complex holomoprhic and antiholomorphic coordinates
($k=1,...,N$)
\be
z_k=x_k+iy_k\,,\qquad \bar z_k=x_k-iy_k\,.
\label{z}
\ee
We associate with each particle a pair of mutually commuting holomorphic
and antiholomorphic spin operators,
$S_\alpha^{(k)}$ and $\bar S_\alpha^{(k)}$, satisfying the standard commutation
relations $[S_\alpha^{(k)},S_\beta^{(n)}]=
i\varepsilon_{\alpha\beta\gamma}\delta^{kn}S_\gamma^{(k)}$ and similarly for
$\bar S_\alpha^{(k)}$. They act on the quantum space of the $k-$th
particle, $V^{(s_k,\bar s_k)}$, and can be represented as the following
differential operators
\ba
&& S_0^{(k)}=z_k\partial_{z_k}+s_k\,,\quad S_-=-\partial_{z_k}\,,\quad
S_+^{(k)}=z_k^2\partial_{z_k}+2s_kz_k\,,
\nonumber
\\
&&
\bar S_0^{(k)}=\bar z_k\partial_{\bar z_k}+\bar s_k\,,\quad
\bar S_-=-\partial_{\bar z_k}\,,\quad
\bar S_+^{(k)}=\bar z_k^2\partial_{\bar z_k}+2\bar s_k\bar z_k\,,
\label{SL2-spins}
\ea
so that $ (S^{(k)})^2=(S_0^{(k)})^2+(S_+^{(k)}S_-^{(k)}+S_-^{(k)}S_+^{(k)})/2 =
s_k(s_k-1) $ and similar for the antiholomorphic Casimir operator $(\bar
S^{(k)})^2$. The spin operators defined in this way are the generators of the
unitary principal series representation of the $SL(2,\mathbb{C})$ group
\be
\Psi(z_k,\bar z_k)\to\Psi'(z_k,\bar z_k)=(cz_k+d)^{-2s_k} (\bar c\bar z_k+\bar
d)^{-2\bar s_k}\Psi(z_k',{\bar z}_k')\,,
\label{tg}
\ee
where
\be
z_k'= \frac{az_k+b}{cz_k+d}\,,\qquad
\bar z_k'= \frac{\bar a\bar z_k+\bar b}{\bar c\bar z_k+\bar d}
\label{SL2-trans-z}
\ee
with $k=1,...,N$, $ad-bc=\bar a\bar d -\bar b\bar c=1$ and the complex parameters
$s_k$ and $\bar s_k$ specified below (see Eq.~\re{s-spin}). In what follows we
shall assume that the model is homogenous and the particles have the same spin,
$s_k=s$ and $\bar s_k=\bar s$ for $k=1,...,N$.

The Hamiltonian of the model, ${\cal H}_N$, describes the interaction between $N$
noncompact $SL(2,\mathbb{C})$ spins attached to the particles and it has the following
general form
\be
{\cal H}_N = H_N + \widebar H_N\,, \qquad [H_N,\widebar H_N]=0\,,
\label{H-full}
\ee
where $H_N$ and $\widebar H_N$ act on the holomorphic and antiholomorphic
coordinates of the particles, respectively, and therefore commute. The
(anti)holomorphic Hamiltonian is given by the sum of two-particle Hamiltonians
describing the nearest neighbor interaction between the corresponding
(anti)holomorphic spins with periodic boundary conditions
\be
H_N=\sum_{k=1}^N H(J_{k,k+1})\,,\qquad
\widebar H_N=\sum_{k=1}^N H(\bar J_{k,k+1})\,,\qquad
H(J)=\psi(1-J) + \psi(J) - 2\psi(1)
\label{H-2P}
\ee
with $\psi(x)=d\ln\Gamma(x)/dx$ being the Euler function and $J_{N,N+1}=J_{N,1}$.
Here, $J_{k,k+1}$ and $\bar J_{k,k+1}$ are defined through the Casimir operators
for the sum of the spins
\footnote{Although this equation defines the operators $J_{k,k+1}$ up to substitution
$J_{k,k+1}\to 1-J_{k,k+1}$, the two-particle Hamiltonian \re{H-2P} is invariant
under this transformation.\label{f3}}
\be
J_{k,k+1}(J_{k,k+1}-1)=(S^{(k)}+S^{(k+1)})^2
\label{J}
\ee
with $S^{(N+1)}_\alpha=S^{(1)}_\alpha$, and $\bar J_{k,k+1}$ is defined
similarly.

This paper is devoted to solving the Schr\"odinger equation
\be
{\cal H}_N \, \Psi(\vec z_1, \vec z_2, ..., \vec z_N) = E_N\, \Psi(\vec z_1, \vec
z_2, ..., \vec z_N)
\label{Sch-eq}
\ee
with the eigenstates $\Psi(\vec z_1,..., \vec z_N)$ being
single-valued functions on the plane $\vec z=(z,\bar z)$, normalizable
with respect to the $SL(2,\mathbb{C})$ invariant scalar product
\be
\|\Psi\|^2 = \int d^2 z_1 d^2 z_2 ... d^2 z_N  \,|\Psi(\vec z_1, \vec z_2, ..., \vec z_N)|^2
\label{SL2-norm}
\ee
with $d^2z=dx dy=dz d\bar z/2$. The equation \re{Sch-eq} has appeared for the
first time in the analysis of the high-energy asymptotics of the partial waves in
the multi-color QCD \cite{L1}. It was found \cite{L1,FK}, that it possesses the
set of mutually commuting conserved charges whose number is large enough for the
Schr\"odinger equation \re{Sch-eq} to be completely integrable.

The total spin of the $N-$particle system is one of the conserved charges.
Indeed, the Hamiltonian \re{H-2P} is a function of two-particle Casimir operators
and, therefore, it commutes with the operators of the total spin $S_\alpha=\sum_k
S_\alpha^{(k)}$ and $\bar S_\alpha=\sum_k \bar S_\alpha^{(k)}$, acting on the
quantum space of the system $V_N\equiv V^{(s_1,\bar s_1)}\otimes V^{(s_2,\bar
s_2)}\otimes...\otimes V^{(s_N,\bar s_N)}$. This implies that the eigenstates can
be classified according to the irreducible representations of the
$SL(2,\mathbb{C})$ group, $V^{(h,\bar h)}$, parameterized by the spins $(h,\bar
h)$. The eigenstates $\Psi(\vec z)$, belonging to $V^{(h,\bar h)}$ are labelled
by the center-of-mass coordinate $\vec z_0$ and they can be chosen to have the
following $SL(2,\mathbb{C})$ transformation properties
\be
\Psi(\vec z_k'-\vec z_0') = (cz_0+d)^{2h}(\bar c\bar z_0 +\bar d)^{2\bar h}
\prod_{k=1}^N(cz_k+d)^{2s_k}(\bar c\bar z_k +\bar d)^{2\bar s_k}
\Psi(\vec z_k-\vec z_0)
\label{SL2-trans}
\ee
with $z_0$ and $\bar z_0$ transformed in the same way as $z_k$ and $\bar z_k$,
Eq.~\re{SL2-trans-z}. As a consequence, they diagonalize the Casimir operators
corresponding to the total spin of the system, $S^2=S_0^2+(S_+S_-+S_-S_+)/2$,
\be
\left(S^2-h(h-1)\right)\Psi(\vec z_1, \vec z_2, ..., \vec z_N)
=\left(\bar S^2-\bar h(\bar h-1)\right)\Psi(\vec z_1, \vec z_2, ..., \vec z_N)=0\,.
\label{S2-h}
\ee

The complex parameters $(s_k,\bar s_k)$ and $(h,\bar h)$ entering \re{SL2-spins}
and \re{SL2-trans} parameterize the irreducible $SL(2,\mathbb{C})$
representations. For the principal series representation they satisfy the
conditions \cite{group}
\footnote{\label{foot-1}
Throughout the paper we indicate by bar symbol the variables belonging to the
antiholomorphic sector. Notice that, in general,
the corresponding variables in two sectors are not complex conjugated to each
other.}
\be
s_k-\bar s_k=n_{s_k}\,,\quad s_k+(\bar s_k)^*=1
\label{s-cond}
\ee
and have the following form
\be
s_k=\frac{1+n_{s_k}}2+i\nu_{s_k}\,,\qquad \bar s_k=\frac{1-n_{s_k}}2+i\nu_{s_k}
\label{s-spin}
\ee
with $\nu_{s_k}$ being real and $n_{s_k}$ being integer or half-integer.%
\footnote{Since
the unitary representations labeled by the spins $(s,\bar s)$ and $(1-s,1-\bar
s)$ are unitary equivalent and are related to each other through the intertwining
relation (see Eq.~\re{intertwin} below) one can choose $n_{s_k}$ in \re{s-spin}
to be nonnegative.} The spins $(h,\bar h)$ are given by similar expressions with
$n_{s_k}$ and $\nu_{s_k}$ replaced by $n_h$ and $\nu_h$, respectively. The
parameter $n_{s_k}$ has the meaning of the two-dimensional Lorentz spin of the
particle, whereas $\nu_{s_k}$ defines its scaling dimension. To see this, one
performs a $2\pi-$rotation of the particle on the plane, $z\to  z\exp(2\pi i)$
and $\bar z\to\bar z\exp(-2\pi i)$ and finds from \re{SL2-trans} that the wave
function acquires a phase $\Psi(z_k,\bar z_k)\to (-1)^{2n_{s_k}} \Psi(z_k,\bar
z_k)$. For half-integer $n_{s_k}$ it changes the sign and the corresponding
unitary representation is spinor. Similarly, to define the scaling dimension,
$s+\bar s=1+2i\nu_{s_k}$, one performs the transformation $z\to\lambda z$ and
$\bar z\to\lambda\bar z$. We recall that for the homogenous spin chain one takes
$s_k=s$ and $\bar s_k=\bar s$ for arbitrary $k$. In what follows, for the sake of
simplicity, we will not consider the spinor $SL(2,\mathbb{C})$ representations
and choose $n_{s_k}$ in \re{s-spin} to be nonnegative integer.

We notice that the holomorphic and antiholomorphic spin generators \re{SL2-spins}
as well as the Casimir operators \re{J} are conjugated to each other with respect
to the scalar product \re{SL2-norm} (see footnote to \re{J})
\be
\left[S_\alpha^{_{(k)}}\right]^\dagger = - \bar S_\alpha^{_{(k)}}\,,
\qquad
\left[J_{k,k+1}\right]^\dagger = 1-\bar J_{k,k+1}\,.
\label{S-dagger}
\ee
This implies that $H_N^\dagger =\bar H_N$ in Eq.~\re{H-full} and, as a
consequence, the Hamiltonian \re{H-full} is hermitian on the space of functions
endowed with the scalar product \re{SL2-norm}, ${\cal H}_N^\dagger = {\cal H}_N$.
Thus, the energy $E_N$ in \re{Sch-eq} is real and the corresponding eigenstates
are orthogonal to each other with respect to \re{SL2-norm}. Since the Euler
$\psi-$function has poles at negative integer arguments, the holomorphic and
antiholomorphic Hamiltonians, Eq.~\re{H-2P}, are unbounded operators. One can
verify however \cite{K1} that, thanks to the properties of the principal $SL(2,\mathbb{C})$
series, Eqs.~\re{s-cond} and \re{s-spin}, the poles are cancelled in their sum
\re{H-full} leading to the Hamiltonian ${\cal H}_N$ which is bounded from below.

\subsection{$R-$matrix}

Let us show that the model defined in the previous section can be described using
the $R-$matrix approach \cite{QISM}. For this we will need the expression for the
$R-$matrix acting on the tensor product of two $SL(2,\mathbb{C})$
representations, $V^{(s_1,\bar s_1)}\otimes V^{(s_2,\bar s_2)}$. The general
expression for the $R-$matrix is well known for the Heisenberg spin magnets in
the case of arbitrary higher spin $SL(2,\mathbb{R})$ representations
\cite{KRS,XXX} and to the best of our knowledge its $SL(2,\mathbb{C})$ analog has
not been discussed in the literature.

To find the $R-$matrix for the infinite-dimensional $SL(2,\mathbb{C})$
representations, we introduce the Lax operators in the holomorphic and
antiholomorphic sectors
\ba
L_s(u)&=&u+i(\sigma\cdot S)=\matrix{u+iS_0 & iS_- \\ iS_+ & u-iS_0}
\,,\qquad
\nonumber
\\
\bar L_{\bar s}(\bar u)&=&\bar u+i(\sigma\cdot \bar S)=
\matrix{\bar u+i\bar S_0 & i\bar S_- \\ i\bar S_+ & \bar u-i\bar S_0}
\label{Lax}
\ea
with $u$ and $\bar u$ being arbitary complex parameters and $\sigma_\alpha$ being
Pauli matrices. These operators act on the quantum space $V^{(s,\bar s)}$ and
coincide with similar expressions for the Lax operator in the Heisenberg spin
chain model \cite{QISM,XXX,ABA}. We define the $R-$matrix by requiring that the
Lax operators have to satisfy the commutation relations
\ba
&& L_{s_1}(v) L_{s_2}(v+u) R_{(s_1,\bar s_1),(s_2,\bar s_2)}(u,\bar u) =
R_{(s_1,\bar s_1),(s_2,\bar s_2)}(u,\bar u)L_{s_2}(v+u)L_{s_1}(v)\,,
\nonumber
\\
&&
\bar L_{\bar s_1}(\bar v) \bar L_{\bar s_2}(\bar v+\bar u)
R_{(s_1,\bar s_1),(s_2,\bar s_2)}(u,\bar u) = R_{(s_1,\bar s_1),(s_2,\bar
s_2)}(u,\bar u)
\bar L_{\bar s_2}(\bar v+\bar u)\bar L_{\bar s_1}(\bar v)\,.
\label{LLR}
\ea
Here, the operator $R_{(s_1,\bar s_1),(s_2,\bar s_2)}(u,\bar u)$ depends on two spectral
parameters, $u$ and $\bar u$, and it acts on the tensor product
$V^{(s_1,\bar s_1)}\otimes V^{(s_2,\bar s_2)}$ labeled by the $SL(2,\mathbb{C})$
spins $(s_1,\bar s_1)$ and $(s_2,\bar s_2)$.

\subsubsection{Yang-Baxter equation}

Solving the Yang-Baxter equation \re{LLR}, we use the method proposed in
\cite{KSS,DKK}. It is based on the representation of the $R-$operator by its
integral kernel
\footnote{Here, we assume that there exists the region of the spectral parameters, $u$ and $\bar
u$, in which the integral is convergent. For $u$ and $\bar u$ outside this region
the kernel of the $R-$operator is defined by the analytical continuation.}
\be
\left[R_{(s_1,\bar s_1),(s_2,\bar s_2)}(u,\bar u) \Psi\right] (\vec z_1,\vec z_2)
\equiv \int d^2 w_1 \int d^2 w_2 \, R_{u,\bar u}(\vec z_1,\vec z_2\,|\vec w_1,\vec w_2)
\,\Psi(\vec w_1,\vec w_2)\,,
\label{R-Psi}
\ee
where $\vec z_i=(z_i,\bar z_i)$ and $\vec w_i=(w_i,\bar w_i)$ are two-dimensional
vectors with the (anti)holomorphic coordinates defined in
\re{z} and $\Psi(\vec w_1,\vec w_2)$ is an arbitrary test function belonging to
$V^{(s_1,\bar s_1)}\otimes V^{(s_2,\bar s_2)}$. To find the
kernel, $R_{u,\bar u}(\vec z_1,\vec z_2\,|\vec
w_1,\vec w_2)$, we require that the $R-$matrix defined in this way has to satisfy
the Yang-Baxter equations \re{LLR} and, in addition, $R_{u,\bar u}(\vec z_1,\vec z_2\,|\vec
w_1,\vec w_2)$ has to be a single-valued function on the two-dimensional plane.
As we shall see in a moment, the latter condition imposes constraints
on the possible values of the spectral parameters $u$ and $\bar u$.

\begin{figure}[t]
\centerline{{\epsfxsize12.0cm\epsfbox{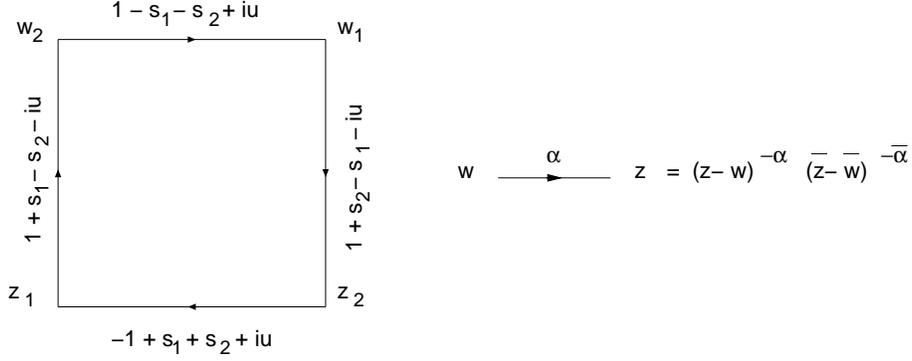}}}
\caption[]{Feynman diagram representation of the kernel of the $R-$matrix,
Eq.~\re{R-kernel}. Each side of the square represents a two-dimensional
propagator  and the arrow indicates in which order the difference of coordinates
is taken.}
\label{R-figure}
\end{figure}

The dependence of the kernel of the $R-$matrix on the holomorphic coordinates is
fixed by the first relation in \re{LLR}. Applying its both sides to the same test
function $\Psi(\vec z_1,\vec z_2)$, one substitutes the Lax operators and the
$R-$operator by their expressions, Eqs.~\re{Lax} and
\re{R-Psi}, and integrates by parts in the r.h.s.\ using the identity
\ba
\left[R(u,\bar u) S^{(1)}_\alpha \Psi\right]
(\vec z_1,\vec z_2) &\equiv& \int d^2 w_1 \int d^2 w_2 \, R_{u,\bar u} (\vec
z_1,\vec z_2\,|\vec w_1,\vec w_2)
\,\lr{S^{(s_1)}_\alpha(w_1)\Psi(\vec w_1,\vec w_2)}
\nonumber
\\
&=& -\int d^2 w_1 \int d^2 w_2 \, \lr{S^{(1-s_1)}_\alpha(w_1)R_{u,\bar u}
(\vec z_1,\vec z_2\,|\vec w_1,\vec w_2)}
\,\Psi(\vec w_1,\vec w_2)\,.
\label{by-parts}
\ea
Here, $S_\alpha^{(s_1)}(w_1)$ and $S_\alpha^{(1-s_1)}(w_1)$ denote the
differential operators acting on the holomorphic coordinates $w_1$. They and given
by Eqs.~\re{SL2-spins} with the spin $s_k$ replaced by $s_1$ and $1-s_1$,
respectively. In this way, the first relation in \re{LLR} can be replaced
by the following matrix equation
\ba
&&\lr{v+i\sigma\cdot S^{(s_1)}(z_1)}\lr{v+u+i\sigma\cdot S^{(s_2)}(z_2)}
R_{u,\bar u}(\vec z_1,\vec z_2\,|\vec w_1,\vec w_2)
\nonumber
\\
&=&\lr{u+v-i\sigma\cdot S^{(1-s_2)}(w_2)}\lr{v-i\sigma\cdot S^{(1-s_1)}(w_1)}
R_{u,\bar u}(\vec z_1,\vec z_2\,|\vec w_1,\vec w_2)\,,
\label{LLR-matrix}
\ea
which leads to an overcompleted system of the differential equations on the
kernel $R_{u,\bar u}$. It turns out that the system has a unique solution. The
simplest way to find it is \cite{DKK} to project the both sides of
\re{LLR-matrix} by the vector $(-w_2,1)$ from the left and by the vector
$(1,z_2)$ from the right and, then, match the coefficients in front of powers of
$u$ and $v$. One obtains two first-order differential equations on $R_{u,\bar
u}$, which fix the dependence of the kernel on the holomorphic coordinates $z_1$
and $w_1$ up to arbitrary prefactor depending on the specral parameters.
Repeating similar analysis for the second equation in \re{LLR}, we restore the
dependence of the kernel on the antiholomorphic coordinates. The resulting
expression for the solution to
\re{LLR} is given by the product of the holomorphic and antiholomorphic kernels
\ba
&&\hspace*{-10mm} R_{u,\bar u}(\vec z_1,\vec z_2\,|\vec w_1,\vec w_2) =
\rho_{_R}(u,\bar u)
\nonumber\\[2mm]
&&
\times
(w_2-z_1)^{iu-s_1+s_2-1}(z_1-z_2)^{-iu-s_1-s_2+1}(w_1-w_2)^{-iu+s_1+s_2-1}
(z_2-w_1)^{iu+s_1-s_2-1}
\label{R-kernel}
\\
&&
\times
(\bar w_2-\bar z_1)^{i\bar u-\bar s_1+\bar s_2-1} (\bar z_1-\bar z_2)^{-i\bar
u-\bar s_1-\bar s_2+1} (\bar w_1-\bar w_2)^{-i\bar u+\bar s_1+\bar s_2-1}
(z_2-w_1)^{iu+s_1-s_2-1}
\nonumber
\ea
with the normalization factor $\rho_{_R}(u,\bar u)$ being an arbitrary function
of the spectral parameters. The obtained expression for the kernel of the
$R-$matrix can be translated into the language of two-dimensional Feynman
diagrams as shown in Fig.~\ref{R-figure}. As we will demonstrate below, this
diagrammatical representation becomes extremely useful in solving the model.

Examining the expression for the kernel, Eq.~\re{R-kernel}, we notice that
$R_{u,\bar u}(\vec z_1,\vec z_2\,|\vec w_1,\vec w_2)$ acquires a nonzero
monodromy as $\vec z_1$ encircles the points $\vec z_2$ and $\vec w_2$ on the
plane and similarly for other arguments of the $R-$matrix. For
the kernel to be a single-valued function, the corresponding monodromies should
cancel in the r.h.s.\ of \re{R-kernel}. In general, for the function of the form
$(w-z_1)^\alpha (\bar w-\bar z_1)^{\bar \alpha}$ this condition amounts to the
Lorentz spin $\alpha-\bar \alpha$ to be integer. Applying the same condition to
\re{R-kernel} and taking into account the expressions for the spins $s_1$ and
$s_2$, Eqs.~\re{s-spin} with $n_{s_1}$ and $n_{s_2}$ integer, we find that the
spectral parameters $u$ and $\bar u$ have to satisfy the additional condition
\be
i(u-\bar u) = n
\label{u-bar u}
\ee
with $n$ being an integer.

\begin{figure}[t]
\centerline{{\epsfxsize12.0cm\epsfbox{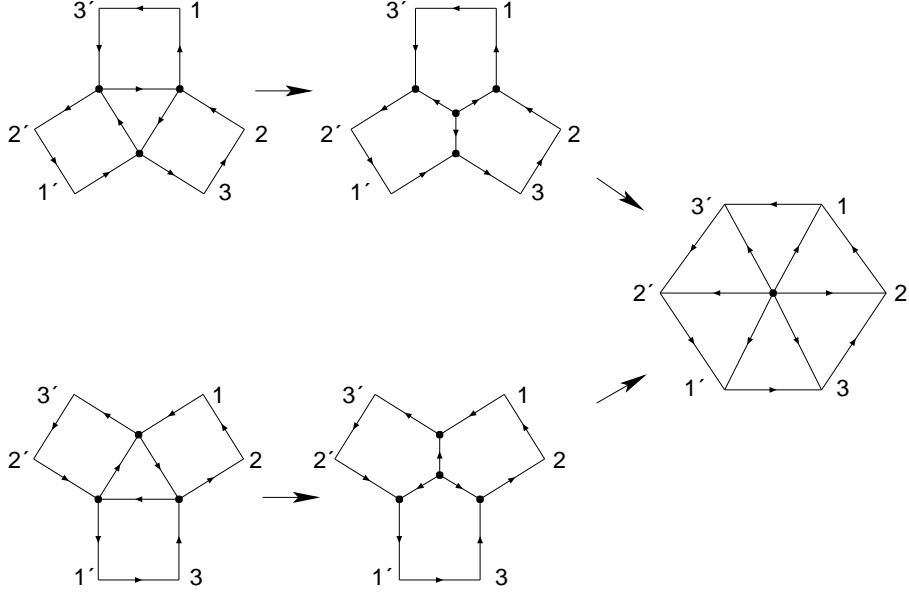}}}
\caption[]{Diagrammatical proof of the Yang-Baxter relation based on the
``uniqueness'' relations. Integration over the position of internal vertices (fat
points) is tacitly assumed.}
\label{YB-figure}
\end{figure}

Using the expression for the kernel \re{R-kernel} it becomes straightforward to
show that the $R-$matrix defined by Eqs.~\re{R-Psi} and \re{R-kernel} satisfies
the Yang-Baxter equation
\be
R_{12}(u,\bar u) R_{13}(v,\bar v) R_{23}(v-u,\bar v- \bar u) = R_{23}(v-u,\bar v-
\bar u) R_{13}(v,\bar v) R_{12}(u,\bar u)\,,
\label{YB}
\ee
where we used a shorthand notation for $R_{kn}(u,\bar u)\equiv R_{(s_k,\bar
s_k),(s_n,\bar s_n)}(u,\bar u)$. Going over to the integral representation
\re{R-Psi}, the product of the $R-$operators in the l.h.s.\ of \re{YB}
has the following kernel
\ba
\lefteqn{\Big[R_{12}(u,\bar u) R_{13}(v,\bar v)R_{23}(v-u,\bar v-\bar u)
\Big](\vec z_1,\vec z_2,\vec z_3|\vec w_1,\vec w_2,\vec w_3)=}
\nonumber
\\
& &\qquad
\int d^2y_1\,d^2y_2\,d^2 y_3\, R_{u,\bar u}(\vec z_1,\vec z_2|\vec y_1,\vec y_2)
R_{v,\bar v}(\vec y_1,\vec z_3|\vec w_1,\vec y_3) R_{v-u,\bar v-\bar u}(\vec
y_2,\vec y_3|\vec w_2,\vec w_3)
\,.
\ea
One finds similar expression for the r.h.s.\ of \re{YB}. The proof of \re{YB} can
be carried out diagrammatically as shown in Fig.~\ref{YB-figure}, without doing
any calculations. Replacing each $R-$matrix by a square (see Fig.~\ref{R-figure})
one represents the l.h.s.\ of
\re{YB} by the left Feynman diagram in the lower line in Fig.~\ref{YB-figure}.
This diagram consists of three squares and each pair of squares shares a common
vertex. The central triangle formed by the sides of three squares turns out to be
``unique'' (see Appendix A for the definitions) and it can be transformed into
the ``star'' diagram, using the uniqueness relation
\re{uniq}. Then, three new stars with the centers located in the vertices of the
central triangle are also unique and can be transformed back into triangles
giving rise to a hexagon diagram. Performing the same set of transformations on
the r.h.s.\ of \re{YB} (the left diagram in the upper line in
Fig.~\ref{YB-figure}), one arrives at the same hexagon diagram, thus proving the
Yang-Baxter equation.

\begin{figure}[t]
\begin{center}
{\epsfxsize17.0cm\epsfbox{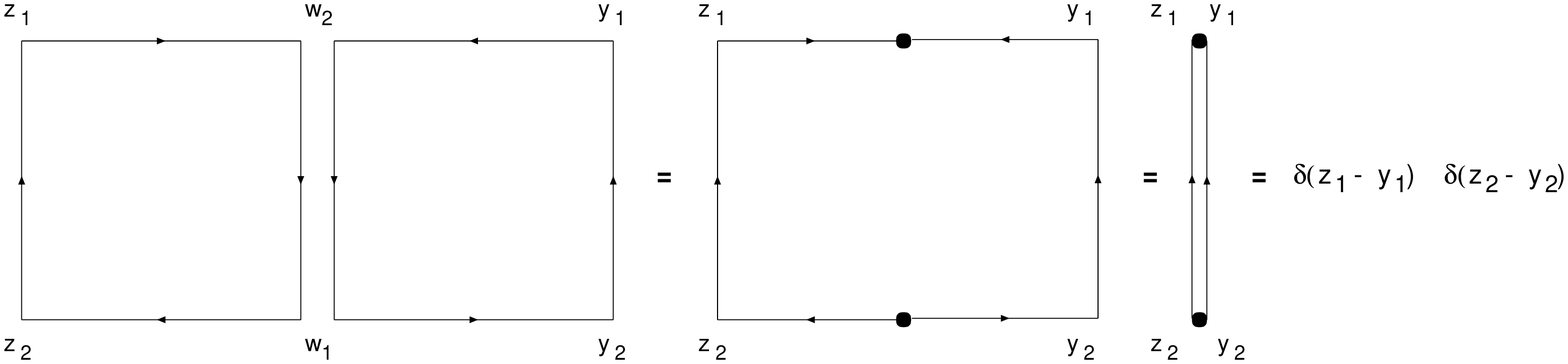}}
\end{center}
\caption[]{Diagrammatical proof of Eq.~\re{Ru-u}. Two lines connecting $\vec w_1$ and
$\vec w_2$ cancel each other and the integration over the position of these
points is performed using the chain relation \re{chain-rule-delta}.}
\label{Fig-unit}
\end{figure}

Following similar procedure one can show that the $R-$operator \re{R-kernel}
satisfies the $T-$inversion relation (see Fig.~\ref{Fig-unit})
\ba
\label{Ru-u}
&&R_{(s_1,\bar s_1),(s_2,\bar s_2)}(u,\bar u) R_{(s_1,\bar s_1),(s_2,\bar
s_2)}(-u,-\bar u)
=\\
&&\ \ \ \ \ \ \ \ \ \ \ \  \ \ \ \ \  \ \ \ \  \ \ \ \ \
\frac{\pi^4\,\rho_{_R}(u,\bar u)\,\rho_{_R}(-u,-\bar u)}
{a(s_1-s_2+iu)\,a(\bar s_2 -\bar s_1+i\bar u )\,a(s_1-s_2-iu)\, a(\bar s_2 -\bar
s_1-i\bar u )}\times\II\,,
\nonumber
\ea
where the function $a(x)$ is defined in \re{a-a} and $\II$ stands for the
identity operator on $V^{(s_1,\bar s_1)}\otimes V^{(s_2,\bar s_2)}$. Defining the
normalization factor $\rho_{_R}(u,\bar u)$ we fix, for later convenience, the
coefficient in front of the identity operator in Eq.\re{Ru-u} to be equal to
unity
\be
\label{rho-u}
\rho_{_R}(u,\bar u)=\frac{1}{\pi^2}\,a( s_1-s_2+iu)\,a(\bar s_2-\bar s_1-i\bar u)\,.
\ee

\subsubsection{Eigenvalues of the $R-$matrix}

It follows from the Yang-Baxter equation \re{LLR} that the $R-$matrix commutes
with the sum the $SL(2,\mathbb{C})$ spins and, therefore, it is invariant
under the $SL(2,\mathbb{C})$ transformations \re{tg} and \re{SL2-trans-z}%
\footnote{To see this, one takes the limit in \re{LLR} as
$v\to\infty$ for $u={\rm fixed}$  and similarly for the antiholomorphic sector.}
\be
[R_{12}(u,\bar u),S_\alpha^{_{(1)}}+S_\alpha^{_{(2)}}]= [R_{12}(u,\bar u),\bar
S_\alpha^{_{(1)}}+\bar S_\alpha^{_{(2)}}]=0\,.
\ee
This implies that the $R-$operator is a function of the Casimir operators
$J_{12}$ and $\bar J_{12}$ defined in \re{J}. In order to find the explicit
form of this function or, equivalently, to calculate the eigenvalues of the
$R-$matrix, one has to decompose the tensor product $V^{(s_1,\bar s_1)}\otimes
V^{(s_2,\bar s_2)}$ over the irreducible components $V^{(h,\bar h)}$ and
take into account that the operators, $J_{12}$, $\bar J_{12}$ and $R_{12}(u,\bar u)$
are diagonal on $V^{(h,\bar h)}$ simultaneously.

The projector onto the principal series spin$-(h,\bar h)$ representation,
$\Pi^{(h,\bar h)}$, is defined as
\be
\Psi^{(h,\bar h)}(\vec z)=\int d^2 w_1\, d^2 w_2 \,
\Pi^{(h,\bar h)}(\vec w_1-\vec z,\vec w_2-\vec z)\Psi^{(s_1,\bar s_1),(s_2,\bar s_2)}
(\vec w_1,\vec w_2)\,,
\ee
where $\Psi^{(h,\bar h)}(\vec z)$ belongs to $V^{(h,\bar h)}$ and the kernel
$\Pi^{(h,\bar h)}(\vec w_1-\vec z,\vec w_2-\vec z)$ is given by \cite{group}
\ba
 \Pi^{(h,\bar h)}(\vec w_1-\vec z,\vec w_2-\vec z)&=&
(w_1-w_2)^{h+s_1+s_2-2}\,(w_2-z)^{s_2-h-s_1}\,(z-w_1)^{s_1-h-s_2}
\nonumber
\\
&\times& (\bar w_1-\bar w_2)^{\bar h+\bar s_1+\bar s_2-2}\, (\bar w_2-\bar
z)^{\bar s_2-\bar h-\bar s_1}\, (\bar z-\bar w_1)^{\bar s_1-\bar h-\bar s_2}\,.
\label{projector}
\ea
Requiring this kernel to be a well-defined function on the plane, one finds that
the possible values of the spins $(h,\bar h)$ are given by general
$SL(2,\mathbb{C})$ expressions
\re{s-spin}
\be
h=\frac{1+n_h}2+i\nu_h\,,\qquad
\bar h = \frac{1-n_h}2 + i\nu_h
\label{h}
\ee
with $\nu_h$ arbitrary real and $n_h$ integer.
One can verify that the projectors $\Pi^{(h,\bar h)}(\vec z_1-\vec z_0,\vec z_2-\vec z_0)$ are orthogonal
to each other with respect to the scalar product \re{SL2-norm} for different $\vec z_0$
and the $SL(2,\mathbb{C})$ spins $(h,\bar h)$ such that $n_h\ge 0$.

\begin{figure}[t]
\centerline{{\epsfysize3.5cm\epsfbox{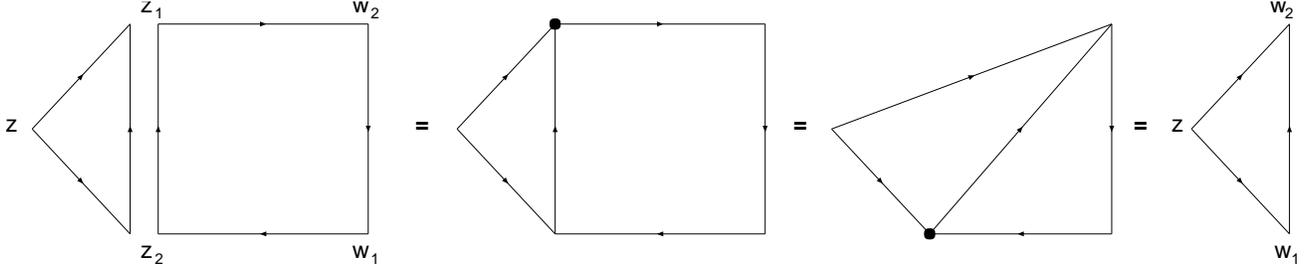}}}
\caption[]{Diagrammatical calculation of the eigenvalues of the $R-$matrix.}
\label{Fig-R-eig}
\end{figure}

By the definition, the projector diagonalizes the Casimir operator
$(S_1+S_2)^2=h(h-1)$ and $(\bar S_1+\bar S_2)^2=\bar h(\bar h-1)$ leading to
\be
\Pi^{(h,\bar h)}\,J_{12}=h\,\Pi^{(h,\bar h)}\,,\qquad
\Pi^{(h,\bar h)}\,{\bar J}_{12}={\bar h}\,\Pi^{(h,\bar h)}
\,.
\label{J=s}
\ee
To find the eigenvalue of the $R-$matrix we calculate the product of the
operators $\Pi^{(h,\bar h)}R_{12}(u,\bar u)$. Replacing the operators by their
kernels, Eqs.~\re{projector} and \re{R-kernel}, one gets
\be
\int d^2 z_1 d^2 z_2 \,\Pi^{(h,\bar h)}(\vec z_1-\vec z,\vec z_2-\vec z) R_{u,\bar u}
(\vec z_1,\vec z_2|\vec w_1,\vec w_2) =R_{h,\bar h}(u,\bar u)\,\Pi^{(h,\bar
h)}(\vec w_1-\vec z,\vec w_2-\vec z)\,,
\ee
where the r.h.s.\ is fixed up to factor $R_{h,\bar h}(u,\bar u)$ by the
$SL(2,\mathbb{C})$ transformation properties \re{SL2-trans-z}. The calculation of
the integral in the l.h.s.\ can be perform diagrammatically as shown in
Fig.~\ref{Fig-R-eig}. The corresponding diagram is obtained by gluing together
the triangle ($\Pi^{(h,\bar h)}$) and the square ($R_{u,\bar u}$) along the line
connecting the points $\vec z_1$ and $\vec z_2$. In the resulting diagram, these
points are the centers of two stars, which turn out to be unique. Subsequently
applying the uniqueness ``star-triangle'' relations, one obtains the triangle
diagram, which is equal to the projector multiplied by a $c-$valued factor
$R_{h,\bar h}(u,\bar u)$ depending on the $SL(2,\mathbb{C})$ spins $h$ and $\bar
h$. Replacing them by the corresponding operators, Eq.~\re{J=s}, one obtains the
operator form of the $R-$matrix
\ba
R_{h,\bar h}(u,\bar u) &=&
\frac{\Gamma(\bar s_2-\bar s_1+i\bar u)
\Gamma(1+\bar s_1-\bar s_2+i\bar u)}{
\Gamma( s_2- s_1-i u)\Gamma(1+ s_1- s_2-i u)}\,\times\,
\,\frac{\Gamma(1-\bar h-i\bar u )\,\Gamma(\bar h-i \bar u)}{
\Gamma(1-h +i u)\,\Gamma( h +i u)}\,,
\nonumber
\\
R_{12}(u,\bar u) &=& R_{h,\bar h}(u,\bar u)\bigg|_{h=J_{12},\,\bar h=\bar
J_{12}}\,.
\label{R-operator}
\ea
Remarkably enough, the obtained expression for the $R-$matrix is factorized into
the product of the holomorphic and antiholomorphic operators. Each of them
formally coincides with the well-known expression for the $R-$matrix for high
spin representations of the $SL(2,\mathbb{R})$ group \cite{KRS}. However, the
important difference with the latter case is that the spectral parameters in two
sectors, $u$ and $\bar u$, are not arbitrary anymore and have to satisfy the
additional condition \re{u-bar u}.

\subsubsection{Unitary $R-$matrix}

For the homogenous spin chain, $s_1=s_2=s$ and $\bar s_1=\bar s_2=\bar
s$, the general expression for the $R-$matrix, Eq.~\re{R-operator}, simplifies to
\be
R_{12}(u,\bar u)=
\frac{\Gamma(i\bar u)
\Gamma(1+i\bar u)}{
\Gamma(-i u)\Gamma(1-i u)}\,\times\,
\,\frac{\Gamma(1-\bar J_{12}-i\bar u )\,\Gamma(\bar J_{12}-i \bar u)}{
\Gamma(1-J_{12} +i u)\,\Gamma( J_{12} +i u)}\,.
\label{R-homo}
\ee
This operator acts on the tensor product $V\otimes V$ with $V\equiv V^{(s,\bar
s)}$ and has the following properties.

At $u=\bar u=0$ the $R-$matrix \re{R-homo} coincides with the permutation operator $P_{12}$%
\footnote{Since the spectral parameters have to satisfy the condition \re{u-bar u},
the limit of \re{R-homo} at  $u=\bar u=0$ has to calculated by putting $u=\bar u$
and sending $u\to 0$ afterwards.}
\be
R_{12}(0,0) = (-1)^{1+J_{12}-\bar J_{12}} = 
-P_{12}\,,
\label{R(0)}
\ee
which is defined on $V\otimes V$ as
\be
P_{12}\,\Psi(\vec z_1,\vec z_2)=\Psi(\vec z_2,\vec z_1)\,.
\label{P12}
\ee
To verify \re{R(0)}, one projects its both sides onto $\Pi^{(h,\bar h)}$ and
takes into account \re{h}.

Taking into account \re{S-dagger}, one finds that the $R-$operator satisfies the
following relations
\be
\left[R_{12}(u,\bar u)\right]^\dagger = R_{12}(-\bar u^*,-u^*)\,,\qquad
R_{12}(u,\bar u)R_{12}(-u,-\bar u)=\II\,.
\label{R-dagger}
\ee
According to \re{R-dagger}, the $R-$matrix is not a unitary operator on $V\otimes
V$ for arbitrary $u$ and $\bar u$ satisfying \re{u-bar u}. However, there exists
a region of the spectral parameters
\be
\bar u= u^*\,,
\label{u-const}
\ee
in which the unitarity holds
\be
\left[R_{12}(u,\bar u)\right]^\dagger R_{12}(u,\bar u)=\II\,.
\label{R-unit}
\ee
Combining \re{u-const} together with \re{u-bar u}, we find that the
$R-$matrix is a unitary operator on  $V\otimes V$ for the spectral parameters
of the general form
\be
u=\nu -\frac{in}{2}\,,\qquad
\bar u = \nu + \frac{in}{2}
\label{u-unit}
\ee
with $\nu$ real and $n$ integer.

\subsection{Complete integrability}

\begin{figure}[t]
\centerline{{\epsfxsize12cm\epsfbox{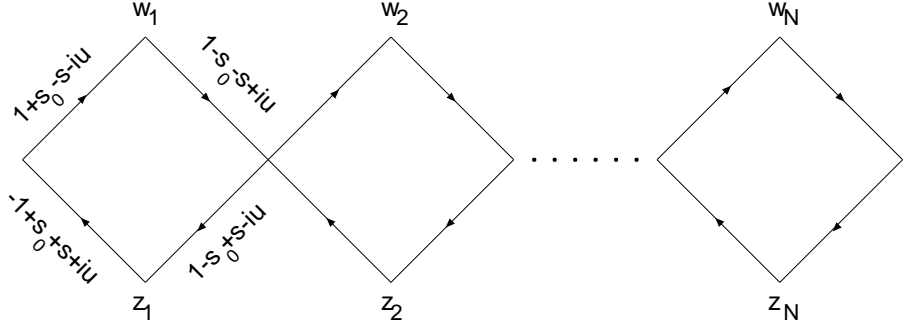}}}
\caption[]{Diagrammatical representation of the transfer matrix
 $\mathbb{T}_N^{(s_0,\bar s_0)}(u,\bar u)$ defined in \re{t-N}. The left- and rightmost vertices
are located at the same point and integration over its position as well as over
the position of the remaining internal vertices is implied.}
\label{Fig-T}
\end{figure}

Let us show that the $R-$matrix \re{R-homo} defines a completely integrable
quantum-mechanical system with the Hamiltonian given by
Eqs.~\re{H-full}--\re{H-2P}. Applying the $R-$matrix approach and following the
standard procedure \cite{QISM,XXX,ABA}, we define the family of the transfer
matrices $\mathbb{T}_N^{(s_0,\bar s_0)}(u,\bar u)$ parameterized by the spins
$(s_0,\bar s_0)$ and acting on the quantum space of the system $V^N$
\be
\mathbb{T}_N^{(s_0,\bar s_0)}(u,\bar u) = \Tr_{(s_0,\bar s_0)}
\left[R_{(s_0,\bar s_0),(s_1,\bar s_1)}(u,\bar u)
R_{(s_0,\bar s_0),(s_2,\bar s_2)}(u,\bar u) ... R_{(s_0,\bar s_0),(s_N,\bar
s_N)}(u,\bar u)\right]\,.
\label{t-N}
\ee
Here, the trace is taken over the auxiliary $SL(2,\mathbb{C})$ representation
space $V^{(s_0,\bar s_0)}$. We recall that the spin chain is homogenous, so that
$s_1= ... = s_N=s$ and $\bar s_1=...=\bar s_N=\bar s$. The spins $(s_0,\bar s_0)$
have the form \re{s-spin} and, in general, they are different from the spins
$(s,\bar s)$. Substituting the $R-$matrices in \re{t-N} by their integral
representation, \re{R-Psi} and \re{R-kernel}, one can evaluate the kernel of the
transfer matrix $\mathbb{T}_N^{(s_0,\bar s_0)} (u,\bar u)$ as $N-$fold
convolution of the $R-$kernel with periodic boundary conditions. The
diagrammatical representation of the transfer matrix $\mathbb{T}_N^{(s_0,\bar
s_0)}(u,\bar u)$ is shown in Fig.~\ref{Fig-T}.

Invoking the standard arguments \cite{QISM,XXX,ABA}, one finds from the
Yang-Baxter equation, Eq.~\re{YB}, that the transfer matrices form the family of
mutually commuting $SL(2,\mathbb{C})$ invariant operators
\be
 [\mathbb{T}_N^{(s_0,\bar s_0)}(u,\bar u),\mathbb{T}_N^{(s_0',\bar s_0')}(v,\bar v)]=
 [S_\alpha,\mathbb{T}_N^{(s_0,\bar s_0)}(u,\bar u)]=
 [\bar S_\alpha,\mathbb{T}_N^{(s_0,\bar s_0)}(u,\bar u)]=0
\label{T-T-com}
\ee
and, therefore, serve as the generating functions of the integrals of motion and the
Hamiltonian of the model. The latter is obtained from the fundamental transfer
matrix $\mathbb{T}^{(s,\bar s)}(u,\bar u)$, for which the auxiliary space
$V^{(s,\bar s)}$ coincides with the quantum space of a single particle, $V$,
and the $R-$matrices entering \re{t-N} are given by \re{R-homo}.

Examining the expansion of the fundamental transfer matrix
$\mathbb{T}^{(s,\bar s)}(u,\bar u)$ around the origin, $u=\bar u=0$, and
using the properties of the $R-$operators, Eqs.~\re{R-homo} and \re{R(0)}, one finds
\be
{\cal H}_N = i \left[\frac{d~~}{du}\ln \mathbb{T}_N^{(s,\bar s)}(u,u)
\right]\bigg|_{u=0}= H_{12} + ... + H_{N-1,N}+ H_{N,1}\,,
\label{H-from-R}
\ee
where the two-particle Hamiltonian $H_{k,k+1}$ is given by
\ba
H_{12}&=&-i P_{12} \frac{d}{du}R_{(s_1,\bar s_1),(s_2,\bar s_2)}(u,u)\bigg|_{u=0}
\nonumber
\\
&=&\psi(J_{12})+\psi(1-J_{12})-2\psi(1)+\psi(\bar J_{12})+\psi(1-\bar
J_{12})-2\psi(1)\,,
\label{H12}
\ea
with the Casimir operators $J_{12}$ and $\bar J_{12}$ defined in \re{J}. Here, we
substituted the $R-$matrix by its operator expression \re{R-homo} and used
\re{R(0)}. Comparing \re{H-from-R} with \re{H-full} we conclude that two
Hamiltonians are identical.

To identify the total set of the integrals of motion of the model, one constructs
the auxiliary holomorphic monodromy matrix \cite{QISM,XXX,ABA}
\be
T_N(u)= L_{1}(u) L_2(u) ... L_{N}(u)
      =\matrix{A_N(u) & B_N(u) \\C_N(u) & D_N(u)}
\label{monodromy}
\ee
and similarly for antiholomorphic monodromy operator $\bar T_N(u)$. Replacing the
Lax operators by their expressions $L_k\equiv L_{s_k}(u)$, Eq.~\re{Lax}, one
obtains the operators $A_N$, $...$, $D_N$ in the form of polynomials in the
spectral parameter. Their asymptotics at large $u$ is given by
\ba
&& A_N(u) = u^N + iS_0 u^{N-1} + \CO(u^{N-2})\,,\qquad B_N(u) = i S_- u^{N-1} +
\CO(u^{N-2})\,,
\nonumber
\\[3mm]
&& D_N(u) = u^N - iS_0 u^{N-1} + \CO(u^{N-2})\,,\qquad C_N(u) = i S_+ u^{N-1} +
\CO(u^{N-2})
\,,
\label{B-asym}
\ea
where $S_\alpha=\sum_{k=1}^N S_\alpha^{_{(k)}}$ is the total $SL(2,\mathbb{C})$
spin of $N$ particles. These operators act on the quantum space of the system and
they are related to their antiholomorphic counterparts as
\be
[A_N(u)]^\dagger = \bar A_N(u^*)\,,\qquad [B_N(u)]^\dagger = \bar B_N(u^*)
\label{A-dagger}
\ee
and similarly for the remaining operators $C_N$ and $D_N$.

Taking the trace of the monodromy matrix \re{monodromy} we define the auxiliary
transfer matrix
\be
t_N(u)= A_N(u) + D_N(u) = 2u^N + q_2 u^{N-2} + ... + q_N
\label{t-aux}
\ee
and similarly for $\bar t_N(\bar u)$. Combining together \re{A-dagger} and
\re{t-aux} one finds that
\be
[t_N(u)]^\dagger = \bar t_N(u^*)\,,\qquad q_k^\dagger = \bar q_k
\ee
for $k=2,...,N$. Here, the operators $q_k$ ($\bar q_k$) are given by certain
linear combinations of the product of $k$ spin operators. They can be rewritten
using \re{SL2-spins} as $k-$th order differential operators acting on (anti)holomorphic
coordinates. For instance,
\be
q_2 =-\sum_{k>n} 2 (S^{(k)}\, S^{(n)}) =-S^2+Ns(s-1)=
\sum_{k>n}^N (z_k-z_n)^{2(1-s)}\partial_{z_k}\partial_{z_n}
(z_k-z_n)^{2s}+2Ns(s-1)\,,
\ee
where $S^2=h(h-1)$ is the Casimir operator corresponding to the total spin of $N$
particles, Eq.~\re{S2-h}.

It follows from the Yang-Baxter equations \re{LLR} and \re{YB} that the auxiliary
transfer matrices $t_N(u)$ and $\bar t_N(\bar u)$ commute with the transfer
matrices $\mathbb{T}^{(s_0,\bar s_0)}(v,\bar v)$ and, in addition, satisfy the
same relations \re{T-T-com}. Together with Eqs.~\re{t-aux} and \re{H-from-R}
these relations imply that the operators $(q_k,\bar q_k)$ for $k=2,...,N$ form
the set of $2N-2$ mutually commuting $SL(2,\mathbb{C})$ invariant integrals of
motion
\be
[{\cal H}_N,q_k] =[{\cal H}_N,\bar q_k]=[{\cal H}_N,S_\alpha]=[{\cal H}_N,\bar
S_\alpha]=0
\label{H-inv}
\ee
and
\be
[q_k, q_n] =  [\bar q_k, \bar q_n] =[q_k,S_\alpha]=[\bar q_k,\bar S_\alpha]=0\,.
\ee
Two additional operators can be added to this set due to the $SL(2,\mathbb{C})$
invariance of the Hamiltonian
\re{H-inv}. It is convenient to choose them as particular projections of the total spins $S_\alpha$ and
$\bar S_\alpha$
\be
p=iS_-=-i\sum_{k=1}^N \partial_{z_k}\,,\qquad
\bar p=i\bar S_- = - i\sum_{k=1}^N \partial_{\bar z_k}\,,
\ee
which have the meaning of the total momenta of $N$ particles. Then, the
eigenstates of the Hamiltonian with a definite value of the momenta $\vec
p=(p,\bar p)$ are given by
\be
\Psi_{\vec p,\{q,\bar q\}}(\vec z_1,\vec z_2,...,\vec z_N) =
\int d^2 z_0 \e^{iz_0p+i\bar z_0 \bar p}\, \Psi(\vec z_1-\vec z_0, \vec z_2 -
\vec z_0, ..., \vec z_N-\vec z_0)
\ee
with $\vec z_0$ being the center-of-mass of the system.

Thus, the Schr\"odinger equation \re{Sch-eq} possesses the set of $2N$ mutually
commuting conserved charges, $\vec p$ and $\{q_k,\bar q_k\}$ $(2\le k \le N)$,
and, therefore, is completely integrable. This implies that, firstly, the
Hamiltonian of the model can be expressed as a function of the integrals of
motion
\be
{\cal H}_N = {\cal H}_N(q_2, \bar q_2; ...; q_N, \bar q_N)
\label{H-q}
\ee
and, secondly, the wave function $\Psi_{\vec p,\{q,\bar q\}}(\vec z_1,\vec
z_2,...,\vec z_N)$ can be defined as a simultaneous eigenstate of the integrals
of motion. The corresponding energy levels $E_{\{q,\bar q\}}$ can be obtained
from \re{H-q} by replacing the operators $\{q_k,\bar q_k\}$ by their
corresponding eigenvalues. Notice that the Hamiltonian can not depend on the
momentum operator $\vec p$ due to the $SL(2,\mathbb{C})$ invariance \re{H-inv}.
The explicit form of the dependence \re{H-q} will be established in the
Section~3.5 using the method of the Baxter $Q-$operator.

\subsubsection{Special case: $N=2$}

At $N=2$ the Schr\"odinger equation \re{Sch-eq} can be solved exactly. In this
case, the Hamiltonian of the model is given by the two-particle kernel \re{H-2P},
${\cal H}_2= 2 H(J_{12})+ 2H(\bar J_{12})$ and its spectrum can be found by
diagonalizing simultaneously the Casimir operators $J_{12}$ and $\bar J_{12}$, as
well as the momentum operators $p$ and $\bar p$. Taking into account
Eqs.~\re{projector} and \re{J=s} and putting $s_1=s_2=s$ one finds the
eigenstates as
\ba
&&\Psi_{\vec p,(h,\bar h)}(\vec z_1,\vec z_2)=\int d^2 z_0 \e^{iz_0 \bar p +
i\bar z_0 p}\,
\Psi_{h,\bar h}(\vec z_1-\vec z_0,\vec z_2-\vec z_0)\,,
\nonumber
\\
&&\Psi_{h,\bar h}(\vec z_1-\vec z_0,\vec z_2-\vec z_0)
=[z_1-z_2]^{h-2s} [z_1-z_0]^{-h} [z_2-z_0]^{-h}\,,
\label{WF-N=2}
\ea
where the notation was introduced for
$[z_j-z_k]^\alpha\equiv (z_j-z_k)^\alpha (\bar z_j-\bar z_k)^{\bar \alpha}$.
Here, the spins $h=(1+n_h)/2+i\nu_h$ and $\bar h=(1-n_h)/2+i\nu_h$ are the
eigenvalues of the $SL(2,\mathbb{C})$ Casimir operators $J_{12}$ and $\bar J_{12}$,
Eq.~\re{S2-h}. One verifies that \re{WF-N=2} satisfies \re{SL2-trans}.
The corresponding energy is equal to
\be
E_{N=2}(h,\bar h) = 4\,{\rm Re}\left[\psi(1-h)+\psi(h)-2 \psi(1)\right] =8\,{\rm
Re}\left[\psi\lr{\frac{1+|n_h|}2+i\nu_h}- \psi(1)\right]\,.
\label{E-N=2}
\ee
Minimizing this expression with respect to integer $n_h$ and real $\nu_h$ one
finds that the ground state corresponds to $h=\bar h=1/2$, or equivalently
$n_h=\nu_h=0$
\be
E^{(0)}_{N=2}=-16 \ln 2\,.
\label{E0-N=2}
\ee
For $s=0$ and $\bar s=1$ the expressions \re{WF-N=2} and \re{E-N=2} are well
known in QCD as defining the spectrum of the two-gluon color singlet compound
states. The ground state energy \re{E0-N=2} determines the intercept of the BFKL
Pomeron \cite{L2}.

\subsection{Discrete symmetries}

The Hamiltonian \re{H-from-R} is invariant under the cyclic and mirror permutations of the
particles. The generators of these transformations, $\mathbb{P}$ and $\mathbb{M}$, respectively,
are defined as follows
\be
 [\mathbb{P}\,\Psi](\vec z_1,...,\vec z_{N-1},\vec z_N) = \Psi(\vec z_2,...,\vec
z_N,\vec z_1)
\,, \qquad[\mathbb{M}\,\Psi](\vec z_1,...,\vec z_{N-1},\vec z_N) =
\Psi(\vec z_N,\vec z_{N-1},...,\vec z_1)
\label{P-cyclic}
\ee
so that $[{\cal H}_N,\mathbb{P}]=[{\cal H}_N,\mathbb{M}]=0$. Obviously, the operators
$\mathbb{P}$ and $\mathbb{M}$ are identical at $N=2$. As we will show below, this
symmetry allows to establish some general properties of the spectrum of the model.

By the definition \re{P-cyclic}, the operators $\mathbb{P}$ and
$\mathbb{M}$ do not commute and satisfy the following relations
\be
\mathbb{P}^N=\mathbb{M}^2=\II\,,\qquad
\mathbb{P}^\dagger=\mathbb{P}^{-1}=\mathbb{P}^{N-1}\,,\qquad\mathbb{M}^\dagger=\mathbb{M} \,,\qquad
\mathbb{P}\,\mathbb{M}=\mathbb{M}\,\mathbb{P}^{-1}=\mathbb{M}\,\mathbb{P}^{N-1}\,,
\label{MP}
\ee
or equivalently $(\mathbb{M}\mathbb{P})^2=\II$. As a consequence, the operators
$\mathbb{P}$ and $\mathbb{M}$ can not be diagonalized simultaneously and, therefore,
the eigenvalues of the Hamiltonian could possess, in general, definite quantum numbers only
with respect to one of them.

Let us examine the action of the permutations on the integrals of motion, $q_k$ and
$\bar q_k$, or equivalently on the auxiliary transfer matrices $t_N(u)$ and $\bar t_N(\bar u)$.
Using the definition \re{monodromy} and \re{t-aux}, we find that the transfer matrices are
invariant under the cyclic permutations, $\mathbb{P}^\dagger t_N(u)\mathbb{P}
=t_N(u)$, while under the mirror permutation they are transformed  as
\ba
\mathbb{M} \,t_N(u)\, \mathbb{M} &=& \tr \lr{L_N(u) ... L_2(u) L_1(u)}
\nonumber
\\
&=& (-1)^N \tr\lr{L_1(-u)L_2(-u) ... L_N(-u)} = (-1)^N t_N(-u)\,,
\ea
where the second relation follows from the property of the transposed Lax
operator, Eq.~\re{Lax}, $L^t(u) = -\sigma_2 L(-u)\sigma_2$. Replacing the
auxiliary transfer matrix by its expression \re{t-aux}, we find
\be
\mathbb{M}\, q_k = (-1)^k q_k \,\mathbb{M}\,,\qquad \mathbb{P}\, q_k = q_k \,\mathbb{P}
\label{q-M}
\ee
and similar relations hold for the antiholomorphic charges. Since the Hamiltonian
is invariant under the mirror permutations, it has to satisfy the following
relation as a function of the conserved charges, Eq.~\re{H-q},
\be
{\cal H}(q_k,\bar q_k) = \mathbb{M}{\cal H}(q_k,\bar q_k)\mathbb{M}={\cal
H}(\mathbb{M}q_k\mathbb{M},\mathbb{M}\bar q_k\mathbb{M})={\cal H}\lr{(-1)^{k}
q_k,(-1)^{k} \bar q_k}\,.
\ee
This implies that, firstly, the eigenstates of the Hamiltonian corresponding to two
different sets of the quantum numbers, $\{q_k,\bar q_k\}$ and $
\{(-1)^kq_k,(-1)^k\bar q_k\}$, have the same energy
\be
E_N\lr{q_k,\bar q_k} = E_N\lr{(-1)^k q_k, (-1)^k\bar q_k}
\ee
and, secondly, all energy levels of the Hamiltonian except those with $q_{2k+1}=\bar
q_{2k+1}=0$ $(k=1,2,...)$ are (at least) double degenerate and the corresponding wave
functions are related as
\be
\Psi_{\vec p,\{(-1)^k q_k,(-1)^k\bar q_k\}}(\vec z_1,\vec z_2,...,\vec z_N)
=\left[\mathbb{M}\,\Psi_{\vec p,\{q_k,\bar q_k\}}\right](\vec z_1,\vec
z_2,...,\vec z_N) =\Psi_{\vec p,\{q_k,\bar q_k\}}\,(\vec z_N,...,\vec z_2,\vec
z_1)\,.
\label{M-Psi}
\ee

One concludes from \re{q-M}, that among two operators, $\mathbb{P}$ and
$\mathbb{M}$, only the first one commutes simultaneously with the Hamiltonian and
the conserved charges, and therefore, it is diagonalized by the eigenstates
$\Psi_{\vec{p}\{q,\bar q\}}$. The corresponding eigenvalues define the
quasimomentum $\theta$ of the state
\be
\label{cyclic}
[\mathbb{P}\,\Psi_{\vec{p}\{q,\bar q\}}](z_1,...,z_N) = \e^{i\theta(q,\bar q)}
\,\Psi_{\vec{p}\{q,\bar q\}}(z_1,...,z_N)\,.
\ee
The complete integrability of the model implies that the quasimomentum is a
function of the total set of the integrals of motion $\{q,\bar q\}$. Since $\mathbb{P}^N=\II$,
its eigenvalues are quantized as
\be
\theta(q,\bar q)=2\pi \frac{k}{N}\,,\qquad  \mbox{for $k=0,1,...,N-1$}\,.
\label{theta}
\ee
Applying the operator of mirror permutations, $\mathbb{M}$, to the both sides of
\re{cyclic} and taking into account \re{M-Psi} and \re{MP}, we find that the
quasimomentum of the eigenstate $\Psi_{\vec p,\{(-1)^k q_k,(-1)^k\bar q_k\}}$ is
equal to $-\theta(q,\bar q)$. This leads to the following relation
\be
\theta\lr{(-1)^kq_k,(-1)^k\bar q_k} = -~\theta\lr{q_k,\bar q_k}\,.
\label{theta-sym}
\ee
As a consequence, the eigenstate of the Hamiltonian with the quantum
numbers $q_{2k+1}=\bar q_{2k+1}=0$ $(k=1,2,...)$ has a vanishing quasimomentum and,
therefore, is symmetric under the cyclic permutations of $N$ particles.

Using the solutions to \re{cyclic}, one can construct the eigenstates of the
operator of mirror permutations $\mathbb{M}$
\be
\Psi_{\vec{p}\{q,\bar q\}}^{(\pm)}=\frac{1\pm \mathbb{M}}2~\Psi_{\vec{p}\{q,\bar q\}}
\,,\qquad
\mathbb{M}\,\Psi_{\vec{p}\{q,\bar q\}}^{(\pm)}=\pm~\Psi_{\vec{p}\{q,\bar
q\}}^{(\pm)}\,.
\ee
Although these states do not diagonalize the integrals of motion, $\{q,\bar q\}$,
they are the eigenstates of the Hamiltonian with the same energy
$E_N(q,\bar q)$. We find from  \re{MP} and \re{cyclic} that the operator of cyclic
permutations $\mathbb{P}$ acts on them as
\be
\mathbb{P}\,\Psi_{\vec{p}\{q,\bar q\}}^{(\pm)}
= \frac12\lr{\e^{i\theta}\pm\e^{-i\theta}\mathbb{M}}\Psi_{\vec{p}\{q,\bar q\}}\,.
\ee
Notice that at $\theta=0$ and $\theta=\pi$ the states $\Psi_{\vec{p}\{q,\bar
q\}}^{(\pm)}$ diagonalize the operator $\mathbb{P}$ and, therefore, have a
definite parity with respect to the cyclic and mirror permutations
simultaneously.

Defining the eigenstates of the model, one has to choose between two (equivalent)
sets of the states, $\Psi_{\vec{p}\{q,\bar q\}}$ and $\Psi_{\vec{p}\{q,\bar
q\}}^{(\pm)}$. The former set is consistent with the integrability properties of
the model, while the latter is more suitable for high-energy QCD as it reveals
the Bose properties of the corresponding $N-$gluon states.

\section{Baxter $Q-$operator}

Solving the Schr\"odinger equation \re{Sch-eq}, we shall apply the powerful
method of the $Q-$operator \cite{Bax}. This method plays an important role in the
theory of integrable models as it provides an alternative to the conventional
Algebraic Bethe Ansatz. It is based on the existence of the operator $Q$ that
acts on the quantum space of the model and satisfies a finite-difference Baxter
operator relations. In contrast with the ABA, the method of the $Q-$operator does
not assume the existence of highest weight representation for the eigenstates
and, as a consequence, it has a wider range of applicability. Both methods become
equivalent if the eigenvalues of the $Q-$operator are restricted to be
polynomials in a spectral parameter. In this case, the Baxter equations are
reduced to the Bethe equations on the roots of these polynomials. It is not
obvious, however, whether polynomial $Q-$operators furnish all relevant physical
solutions to the Baxter equation, or equivalently the ABA method is complete.
This turns out to be the case for the $SL(2,\mathbb{R})$ Heisenberg spin magnets
\cite{K1,SD}, while for the noncompact, $SL(2,\mathbb{C})$ spin chain one has to
go beyond the class of polynomial solutions.

In this Section we shall construct a general (nonpolynomial) $Q-$operator for the
homogeneous $SL(2,\mathbb{C})$ spin chain that we shall denote as
$\mathbb{Q}(u,\bar u)$. This operator acts on the quantum space of the model
$V\otimes ... \otimes V$, with $V\equiv V^{(s,\bar s)}$, and depends on two
spectral parameters $u$ and $\bar u$. As we will see in a moment, for
$\mathbb{Q}(u,\bar u)$ to be a well-defined operator, these parameters have to
satisfy the same condition \re{u-bar u} as those for the $R-$matrix. Following
\cite{Bax}, we require that the operator $\mathbb{Q}(u,\bar u)$ has to satisfy
the relations
\begin{itemize}
\item{Commutativity:}
\be
\label{com-Q}
[\,\mathbb{Q}(u,\bar u),\,\mathbb{Q}(v,\bar v)\,]~=~0\,.
\ee
\item{$Q-t$ relations:}
\be
\label{com-t-Q}
[\,{t}_N(u),~\mathbb{Q}(u,\bar u)\,]=[\,\bar t_N(\bar u),~\mathbb{Q}(u,\bar
u)\,]~=~0\,.
\ee
\item{Baxter equations:}
\ba
\label{def-Qh}
t_N(u)\,\mathbb{Q}(u,\bar u)&=&(u+is)^N\, \mathbb{Q}(u+i,\bar u)~+~(u-is)^N\,
\mathbb{Q}(u-i,\bar u)\,,\\
\label{def-Qa}
\bar t_N(\bar u)\,\mathbb{Q}(u,\bar u)&=&(\bar u+i\bar s)^N\,
 \mathbb{Q}(u,\bar u+i)~+~
(\bar u-i\bar s)^N\, \mathbb{Q}(u,\bar u-i)\,,
\ea
\end{itemize}
where $t_N(u)$ and $\bar t_N(\bar u)$ are the auxiliary transfer matrices defined
in \re{t-aux}.

According to \re{com-Q}, the Baxter $\mathbb{Q}$-operator and the
auxiliary transfer matrices, $t_N(u)$ and $\bar t_N(\bar u)$, share the common set
of the eigenfunctions
\be
\mathbb{Q}(u,\bar u)\,\Psi_{\vec p,\{q,\bar q\}}(\vec z_1,\vec z_2,...,\vec z_N)
= Q_{\{q,\bar q\}}(u,\bar u) \,\Psi_{\vec p,\{q,\bar q\}}(\vec z_1,\vec
z_2,...,\vec z_N)\,,
\label{Q-values}
\ee
which, at the same time, are the solutions to the Schr\"odinger equation
\re{Sch-eq} The eigenvalues of the $Q-$operator, $Q(u,\bar u)$, satisfy the same
Baxter equations, Eqs.~\re{def-Qh} and \re{def-Qa}, with the auxiliary transfer
matrices \re{t-aux} replaced by their corresponding eigenvalues.

Our approach to constructing the $Q-$operator for the $SL(2,\mathbb{C})$ spin
chain is inspired by previous works~\cite{PG,SD}, in which the Baxter
$Q-$operator was constructed for the periodic Toda chain and the homogenous
$SL(2,\mathbb{R})$ Heisenberg spin magnet. Namely, we shall represent the
$Q-$operator by its integral kernel
\be
[\mathbb{Q}(u,\bar u)\,\Psi](\vec z_1,\vec z_2,...,\vec z_N) =\int d^2 w_1 ...
\int d^2 w_N \, Q_{u,\bar u}(\vec z_1,\vec z_2,...,\vec z_N|
\vec w_1,\vec w_2,...,\vec w_N)\Psi(\vec w_1,\vec w_2,...,\vec w_N)
\label{Q-kernel}
\ee
and find the explicit form of the kernel by solving Eqs.~\re{com-Q}--\re{def-Qa}.

\subsection{Baxter equations}

According to the definition \re{t-aux}, the auxiliary transfer matrix $t_N(u)$ is
a polynomial of degree $N$ in the $SL(2,\mathbb{C})$ spin operators
$S_\alpha^{_{(k)}}$. Using \re{t-aux}, $t_N(u)$ can be expressed as the $N-$order
differential operator acting on the holomorphic coordinates $z_k$. Its
substitution into the l.h.s.\ of \re{def-Qh} leads to a complicated holomorphic
differential equation on the kernel $Q_{u,\bar u}$. One finds similar
antiholomorphic equation from \re{def-Qa}.

Instead of trying to solve the resulting differential equations we follow the
approach developed in~\cite{PG,SD}. It allows to find the exact solution to
\re{def-Qh} and \re{def-Qa} by using the invariance of the auxiliary transfer
matrix $t_N(u)=\tr T_N(u)$ under local (gauge) rotations of the Lax operators
\be
L_k(u)\to \widetilde L_{k}(u) = M_k^{-1}\,L_k(u)\,M_{k+1}\,,\qquad T_N(u) \to
\widetilde T_N(u) = M_1^{-1}\, T_N(u)\, M_1\,,
\label{rot}
\ee
where $M_k$ are arbitrary $2\times2$ matrices, such that $M_{N+1}=M_{1}$ and
$\det M_k\neq 0$. Let us choose the matrices $M_k$ as
\be
\label{S-trick}
M_k=\left(\begin{array}{cc}1&0\\y_k&1\end{array}\right)\,,\qquad
M_k^{-1}=\matrix{1&0 \\ -y_k&1}
\ee
with $y_1,...,\,y_N$ being arbitrary gauge parameters. The matrix elements of the
rotated Lax operator~\re{rot} are given by
\ba
\label{L-rot}
&&[\widetilde L_k]_{_{11}}=u+is+i(z_k-y_{k+1})\partial_{z_k}\,, \ \ \ \ \ \ \
[\widetilde L_k]_{_{22}}=u-is-i(z_k-y_{k})\partial_{z_k}\,, \nonumber \\[3mm]
&& [\widetilde L_k]_{_{21}}=i(z_k-y_k)(z_k-y_{k+1})\partial_{z_k}+
(u+is)(z_k-y_k)+(-u+is)(z_k-y_{k+1})\,,
\ea
while $[\widetilde L_k]_{_{12}}$ remains unchanged.

Let us now define the following
function
\be
\label{Q-l}
Y_{u,\bar u}^{(s,\bar s)}(\vec z,\vec y)=
\prod_{k=1}^N\,[z_k-y_k]^{-s-iu}\,[z_k-y_{k+1}]^{-s+iu}\,,
\ee
where $[z-y]^{-s\pm iu}\equiv (z-y)^{-s\pm iu} (\bar z-\bar y)^{-\bar s\pm i\bar
u}$, $\vec y_{N+1}=\vec y_1$ and $\vec z\equiv (\vec z_1,...,\vec z_N)$. It
depends on $N$ auxiliary vectors $\vec y\equiv (\vec y_1,...,\vec y_N)$, as well
as the spins $(s,\bar s)$ and the spectral parameters $(u,\bar u)$. The unique
feature of the function $Y_{u,\bar u}^{(s,\bar s)}(\vec z,\vec y)$ is that it is
annihilated by the off-diagonal matrix element of the gauge transformed Lax
operator,
\be
[\tilde L_k]_{_{21}}\,Y_{u,\bar u}^{(s,\bar s)}(\vec z,\vec y)=0
\ee
for $k=1,...,N$. Therefore, applying $Y_{u,\bar u}^{(s,\bar s)}(\vec z,\vec y)$
to the matrix elements of the Lax operator $\tilde L_k$, one finds that the
latter takes the form of the upper triangular matrix. The product of such
matrices can be easily calculated leading to the monodromy operator of the form
\ba
\widetilde T_N(u) \,Y_{u,\bar u}^{(s,\bar s)}(\vec z,\vec y)
&=& \matrix{\prod_{k=1}^N [\widetilde L_k(u)]_{_{11}} & \ast \\ 0 & \prod_{k=1}^N
[\widetilde L_k(u)]_{_{22}}}
\,Y_{u,\bar u}^{(s,\bar s)}
\nonumber
\\
&=&
\matrix{(u-is)^N\,Y_{u-i,\bar u}^{(s,\bar s)} & \ast \\
0 &  (u+is)^N\,Y_{u+i,\bar u}^{(s,\bar s)}}
\,,\label{T-tilde-Y}
\ea
where in the last relation we used the explicit expression for the diagonal
elements of the Lax operator~\re{L-rot}. Taking the trace in the both sides of
this relation, we obtain the following relation for the transfer matrix $t_N=\tr
\widetilde T_N(u)$
\be
\label{t-Y}
t_N(u; S^{(s)}(z))\,Y_{u,\bar u}^{(s,\bar s)}(\vec z,\vec y) =
(u-is)^N\,Y_{u-i,\bar u}^{(s,\bar s)}(\vec z,\vec y)~+~ (u+is)^N\,Y_{u+i,\bar
u}^{(s,\bar s)}(\vec z,\vec y)\,.
\ee
Here, we indicated explicitly that the transfer matrix $t_N(u)$ is expressed in
terms of the differential operators $S^{(s)}(z)$ acting on the $z-$coordinates of
the particles and defined in \re{SL2-spins}. Remarkably enough, the relation
\re{t-Y} takes the form of the Baxter equation \re{def-Qh} and, as we shall see
later in this Section, it allows to construct the $Q-$operator.

Repeating similar analysis for the transfer matrix in the antiholomorphic sector,
$\bar t_N(\bar u)$, one can show that the same function \re{Q-l} satisfies the
antiholomorphic Baxter relation
\be
\label{t-bar-Y}
\bar t_N(\bar u;\bar S^{(\bar s)}(\bar z))\,Y_{u,\bar u}^{(s,\bar s)}(\vec z,\vec y)=
(\bar u-i\bar s)^N\,Y_{u,\bar u-i}^{(s,\bar s)}(\vec z,\vec y)~+~ (\bar u+i\bar
s)^N\,Y_{u,\bar u+i}^{(s,\bar s)}(\vec z,\vec y)\,.
\ee
Notice that in order for the function $Y_{u,\bar u}^{(s,\bar s)}(\vec z,\vec y)$
to be well-defined on the plane, the spectral parameters $u$ and $\bar u$ have to
satisfy the condition \re{u-bar u}.

\subsubsection{Properties of the $Y-$function}

Before we proceed with constructing the Baxter $\mathbb{Q}-$operator, let us
discuss some properties of the function $Y_{u,\bar u}(\vec z,\vec y)$, which will
be used below.

Since the monodromy matrix $\widetilde T_N(u)$ is related to the operator $T_N(u)$
by the gauge transformation \re{rot}, its matrix
elements can be expressed, using \re{monodromy}, in terms of the operators $A_N(u)$, $...$, $D_N(u)$ and
the gauge parameter $y_1$. Substituting the resulting expressions into
\re{T-tilde-Y}, we find that the function $Y_{u,\bar u}^{(s,\bar s)}(\vec z,\vec
y)$ satisfies the following relations for arbitrary $y_k$
\be
\left[y_1^2 B_N(u)+y_1 (A_N(u) -D_N(u)) - C_N(u)\right] Y_{u,\bar u}^{(s,\bar s)}(\vec z,\vec y)=0
\label{Y-1}
\ee
and
\ba
&&
\left[A_N(u)+y_1 B_N(u)\right] Y_{u,\bar u}^{(s,\bar s)}(\vec z,\vec y)
=(u-is)^N\,Y_{u-i,\bar u}^{(s,\bar s)}(\vec z,\vec y)\,,
\nonumber
\\
&&
\left[D_N(u)-y_1 B_N(u)\right] Y_{u,\bar u}^{(s,\bar s)}(\vec z,\vec y)
=(u+is)^N\,Y_{u+i,\bar u}^{(s,\bar s)}(\vec z,\vec y)\,.
\label{Y-2}
\ea
We recall that the operators $A_N(u)$,
$...$, $D_N(u)$ act on the holomorphic $z-$coordinates of $N-$particles and do not
depend on the gauge parameters $y_k$.

Let us examine the relations \re{Y-1} and \re{Y-2} in the limit $y_1\to \infty$.
At large $y_1$, we use the definition \re{Q-l} and expand the $Y-$function as
\be
Y_{u,\bar u}^{(s,\bar s)}(\vec z, \vec y) = [-y_1]^{2s}
\left\{\Lambda_{u,\bar u}^{(s,\bar s)}(\vec z_1, ..., \vec z_N|\vec y_2, ...,\vec y_N) +
{\cal O}(1/y_1,1/\bar y_1)\right\}\,,
\ee
with the leading term given by
\be
\Lambda_{u,\bar u}^{(s,\bar s)}(\vec z_1, ..., \vec z_N|\vec y_2, ...,\vec y_N)~=~[z_1-y_2]^{-s+iu}\,
\left(\prod_{k=2}^{N-1}\,[z_k-y_k]^{-s-iu}\,[z_k-y_{k+1}]^{-s+iu}\right)\,
[z_N-y_N]^{-s-iu}\,.
\label{Psi}
\ee
Then, one finds from \re{Y-1} and \re{Y-2} that the function $\Lambda_{u,\bar
u}^{(s,\bar s)}$, defined in this way, satisfies the following relations
\be
B_N(u) \Lambda_{u,\bar u}^{(s,\bar s)}(\vec z\,|\,\vec y)=0\,,\qquad A_N(u)
\Lambda_{u,\bar u}^{(s,\bar s)}= (u+is)^N
\Lambda_{u+i,\bar u}^{(s,\bar s)}\,,\qquad D_N(u)
\Lambda_{u,\bar u}^{(s,\bar s)}= (u-is)^N \Lambda_{u-i,\bar u}^{(s,\bar s)}\,.
\label{Psi-prop}
\ee
Obviously, the same relations hold in the antiholomorphic sector. As we will show
in Sect.~4.2, the function $\Lambda_{u,\bar u}^{(s,\bar s)}(\vec z\,|\,\vec y)$
becomes a building block in the construction of the unitary transformation to the
Separated Variables.

\subsubsection{The kernel of the $\mathbb{Q}-$operator}
\label{Q-sect}

Taking into account the properties of the function $Y_{u,\bar u}^{(s,\bar s)}$,
Eqs.~\re{t-Y} and \re{t-bar-Y}, we look for the kernel of the Baxter
$\mathbb{Q}-$operator in the form
\be
\label{Q-kern}
Q_{u,\bar u}(\vec z\,|\,\vec w)=\int d^2y \, Y_{u,\bar u}^{(s,\bar s)}(\vec
z,\vec y)\, Z_R(\vec y,\vec w)\,,
\ee
where $d^2 y=d^2y_1 ...d^2 y_N$ and $Z_R$ is assumed to be a well-defined
function on the plane, independent on the spectral parameters $u$ and $\bar u$.
Substituting this ansatz into \re{Q-kernel} one finds that, thanks to
Eqs.~\re{t-Y} and \re{t-bar-Y}, the Baxter equations \re{def-Qh} and \re{def-Qa}
are satisfied for arbitrary function $Z_R(\vec y,\vec w)$.

To fix the form of the function $Z_R(\vec y,\vec w)$ in \re{Q-kern} we require
that the $\mathbb{Q}-$operator has to commute with the auxiliary transfer
matrices, Eq.~\re{com-t-Q}. It is convenient to represent the same condition in
the form of the Baxter equations, Eqs.~\re{def-Qh} and \re{def-Qa}, with the
transfer matrices and the $\mathbb{Q}-$operator interchanged in the l.h.s.\ of
these equations. Applying the operator $\mathbb{Q}(u,\bar u)\, t_N(u)$ to an
arbitrary test function and replacing the $\mathbb{Q}-$operator by its integral
representation, one integrates by parts to arrive at the relation analogous to
\re{by-parts}
\be
\left[\,\mathbb{Q}(u,\bar u)\, t_N(u)\Psi\right](\vec z) = \int d^2 w\ \lr{ t_N(u; -S^{(1-s)}(w))
Q_{u,\bar u}(\vec z\,|\,\vec w)}\Psi(\vec w)\,,
\ee
where $d^2 w\equiv d^2 w_1...\,d^2 w_N$ and the auxiliary transfer matrix in the
r.h.s.\ is obtained from \re{t-aux} and \re{monodromy} by replacing the
holomorphic $SL(2,\mathbb{C})$ generators in the expression for the Lax operators
as $S_\alpha^{(s)}\to -S_\alpha^{(1-s)}$. In this way, the condition \re{com-t-Q}
on the $\mathbb{Q}-$operator can be formulated as
\be
(-1)^N\, t_N(-u; S^{(1-s)}(w))\,Q_{u,\bar u}(\vec z\,|\,\vec w)=
(u+is)^N\,Q_{u+i,\bar u}(\vec z\,|\,\vec w)+(u-is)^N\,Q_{u-i,\bar u}(\vec
z\,|\,\vec w)\,,
\label{conj-B}
\ee
where we took into account that the auxiliary transfer matrix is a polynomial of
degree $N$ in the spectral parameter with the coefficient in front of $u^k$
proportional to the $(N-k)$th power of the spin operators. Similar relation holds
for the antiholomorphic auxiliary transfer matrix $\bar t_N(-\bar u; \bar
S^{(1-\bar s)}(\bar w))$.

Let us look for the solution to \re{conj-B} in the form
\be
Q_{u,\bar u}(\vec z\,|\,\vec w)=\left[a(s+iu,\bar s-i\bar u)\right]^N\,
\widetilde Q_{-u,-\bar u}(\vec w\,|\,\vec z)
\label{Q-tilde}
\ee
with the functions $a(x)$ defined in \re{a-a}. Substituting this ansatz into
\re{conj-B} and changing the spectral parameter as $u\to -u$, one finds that the
function $\widetilde Q_{u,\bar u}$ satisfies the Baxter equation
\be
t_N(u,S^{(1-s)}(w))\,\widetilde Q_{u,\bar u}(\vec w\,|\,\vec z)~=~(u+i(1-s))^N\,
\widetilde Q_{u+i,\bar u}(\vec w\,|\,\vec z)~+~(u-i(1-s))^N\,
\widetilde Q_{u-i,\bar u}(\vec w\,|\,\vec z)\,,
\ee
which coincides with \re{t-Y} once one changes the spin as $s\to 1-s$. Therefore,
the general solution \re{Q-tilde} to the Baxter equation \re{conj-B} and its
antiholomorphic counterpart can be written as
\be
\label{Q-2}
Q_{u,\bar u}(\vec z \,|\,\vec w)=\left[a(s+iu,\bar s-i\bar u)\right]^N
\int d^2 y\, Z_L(\vec y,\vec z)\,Y_{-u,-\bar u}^{(1-s,1-\bar s)}(\vec w,\vec y)
\,,
\ee
with $Z_L(\vec y,\vec z)$ being a well-defined function on the plane,
independent on the spectral parameters $u$ and $\bar u$.

\begin{figure}[t]
\centerline{{\epsfxsize16.0cm\epsfbox{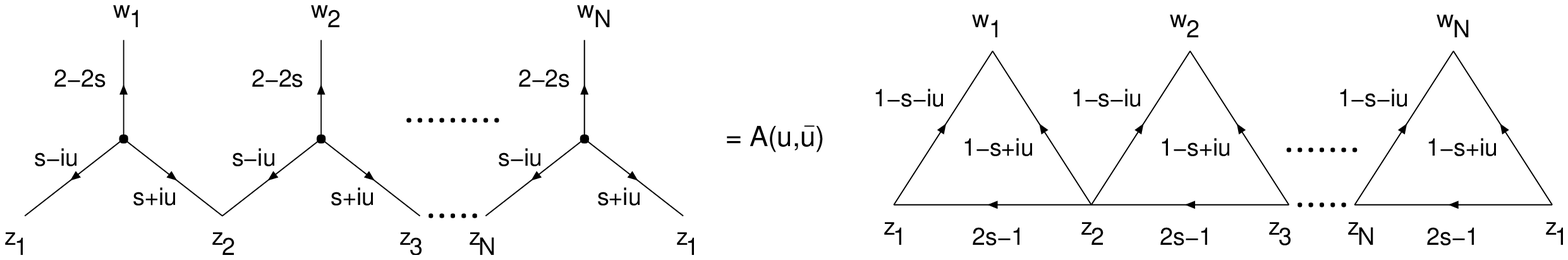}}}
\caption[]{Diagrammatical representation for the kernel of the
operator $\mathbb{Q}_+(u,\bar u)$ defined in Eq.~\re{kern+}. $A(u,\bar
u)=\left[\pi\,a(2-2s,s+iu,\bar s-i\bar u)\right]^N.$}
\label{Q1-f}
\vskip 1cm
\centerline{{\epsfxsize16.0cm\epsfbox{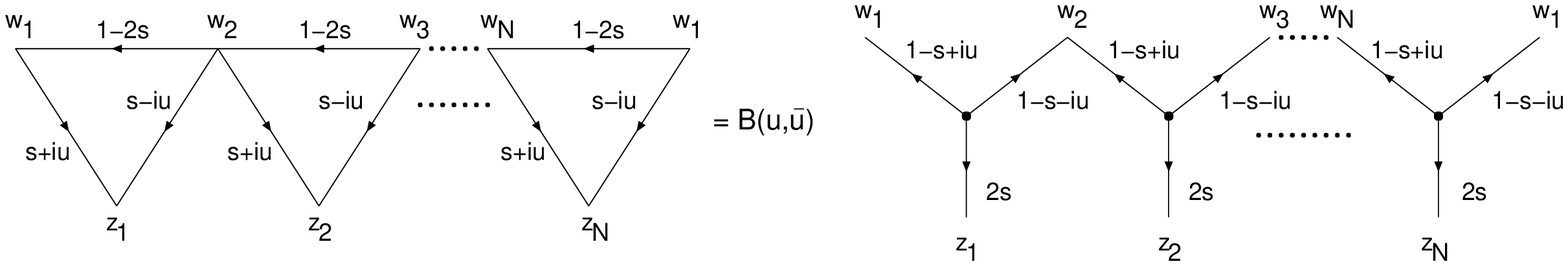}}}
\caption[]{Diagrammatical representations for the kernel of the
operator $\mathbb{Q}_-(u,\bar u)$ defined in  Eq.~\re{kern-}. $B(u,\bar
u)=\left[a(1-2s,s+iu,\bar s-i\bar u)/\pi\right]^N$.}
\label{Q2-f}
\end{figure}

Comparing two different representations for the kernel of the
$\mathbb{Q}-$operator, Eqs.~\re{Q-kern} and \re{Q-2}, we notice that \re{Q-kern}
fixes the dependence of $Q_{u,\bar u}(\vec z \,|\,\vec w)$ on the
$z-$coordinates, whereas \re{Q-2} fixes its $w-$dependence. We find that
\re{Q-kern} and \re{Q-2} become compatible provided that $Z_R$ is given by the
following expressions
\be
\label{Z}
Z_R^+(\vec y,\vec w)=\prod_{k=1}^{N}  [w_{k-1}-y_{k}]^{2s-2}\,,\qquad Z_R^-(\vec
y,\vec w)=\prod_{k=1}^{N} \delta^2(w_k-y_k) [w_k-w_{k+1}]^{2s-1}\,,
\ee
with $w_0=w_N$ and $w_{N+1}=w_1$. Substitution of \re{Z} into
\re{Q-kern} yields two different expressions for the kernel of the $\mathbb{Q}-$operator
\ba
\label{kern+}
Q_{u,\bar u}^{_{(+)}}(\vec z\,|\,\vec w)&=&\int d^2y\,
\prod_{k=1}^N
[z_k-y_k]^{-s-iu}\,[z_{k-1}-y_{k}]^{-s+iu}\,[w_{k-1}-y_{k}]^{2s-2}
\\
&=&
\left[a(2-2s,s+iu,\bar s-i\bar u)\pi\right]^N
\prod_{k=1}^N [w_k-z_k]^{s-1+iu}\,[w_k-z_{k+1}]^{s-1-iu}\,
[z_{k}-z_{k+1}]^{1-2s}\,,
\nonumber
\ea
and
\ba
\label{kern-}
Q_{u,\bar u}^{_{(-)}}(\vec z\,|\,\vec w)&=&
\prod_{k=1}^N\,[z_k-w_k]^{-s-iu}\,[z_k-w_{k+1}]^{-s+iu}\,[w_k-w_{k+1}]^{2s-1}
\\
&=&\left[a(1-2s,s+iu,\bar s-i\bar u)/\pi\right]^N
\int d^2y\, \prod_{k=1}^N [w_{k}-y_k]^{s-1+iu}\,[w_{k-1}-y_{k}]^{s-1-iu}\,[z_{k-1}-y_{k}]^{-2s}\,.
\nonumber
\ea
Here, the second relation in both equations is obtained from the star-triangle
identity \re{uniq}. The diagrammatical representation of the kernels $Q_{u,\bar
u}^{_{(+)}}$ and $Q_{u,\bar u}^{_{(-)}}$ is shown in Figs.~\ref{Q1-f} and
\ref{Q2-f}, respectively. Using this representation, one checks that the
expressions \re{kern+} and \re{kern-} match into Eqs.~\re{Q-kern} and \re{Q-2}
simultaneously.

The obtained expressions for the kernels, Eqs.~\re{kern+} and \re{kern-}, define
two different operators, $\mathbb{Q}_+(u,\bar u)$ and $\mathbb{Q}_-(u,\bar u)$.
By the construction, they commute with the auxiliary transfer matrices,
Eq.~\re{com-t-Q}, and satisfy the Baxter equation \re{def-Qh} and \re{def-Qa}.
For the kernels of these operators, Eqs.~\re{kern+} and \re{kern-}, to be
well-defined on the plane one has to require that, similar to the $R-$matrix, the
spectral parameters $u$ and $\bar u$ have to satisfy the condition \re{u-bar u}.
Finally, in order to identify $\mathbb{Q}_+(u,\bar u)$ and $\mathbb{Q}_-(u,\bar
u)$ as the Baxter ${Q}-$operators we have to show that they fulfil the
commutativity condition \re{com-Q}.

\subsection{Commutativity condition}

\begin{figure}[t]
\centerline{{\epsfxsize12.0cm\epsfbox{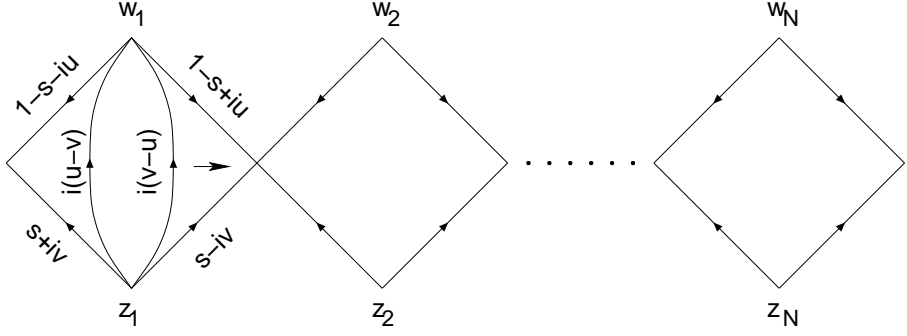}}}
\caption[]{Diagrammatical representation of the kernel
$X_{u,\bar u\,;\,v,\bar v}(\vec z,\vec w)$ defined in Eq.~\re{X-def} as a
periodic chain of rhombuses. The rightmost vertex is identified with the leftmost
one. To prove the commutativity relation \re{perm-sym}, one inserts two lines
with opposite indices into the left rhombus between the points $\vec z_1$ and
$\vec w_1$.}
\label{QQ}
\end{figure}

Let us show that the operators $\mathbb{Q}_+(u,\bar u)$ and $\mathbb{Q}_-(u,\bar
u)$ with the kernels defined in \re{kern+} and \re{kern-} commute with each other
for different values of the spectral parameters

\be
 [\, \mathbb{Q}_+(u,\bar u), \mathbb{Q}_+(v,\bar v)\, ]
=[\, \mathbb{Q}_-(u,\bar u), \mathbb{Q}_-(v,\bar v)\, ] =[\, \mathbb{Q}_+(u,\bar
u), \mathbb{Q}_-(v,\bar v)\, ] =0
\label{com-Q-Q}
\ee
and satisfy the following exchange relations
\ba
&&
\mathbb{Q}_+(u,\bar u)\,\mathbb{Q}_-(v,\bar v)
=\mathbb{Q}_+(v,\bar v)\,\mathbb{Q}_-(u,\bar u)\,,
\label{prod-Q-Q}
\\
&&\mathbb{Q}_-(u,\bar u)\,\mathbb{Q}_+(v,\bar v) =\mathbb{Q}_-(v,\bar
v)\,\mathbb{Q}_+(u,\bar u)\,.
\nonumber
\ea
We start with the product of two operators
\be
 [\,\mathbb{Q}_-(v,\bar v)\,\mathbb{Q}_+(u,\bar u)\,](\vec z\,|\,\vec w) =
\int d^2 y\,Q_{v,\bar v}^{_{(-)}}(\vec z\,|\,\vec y)\, Q_{u,\bar u}^{_{(+)}}(\vec y\,|\,\vec w)
\ee
and substitute them by the corresponding Feynman diagrams. It becomes convenient
to use the left diagram in Fig.~\ref{Q2-f} for the operator $\mathbb{Q}_-(v,\bar
v)$ and the right diagram in Fig.~\ref{Q1-f} for the operator
$\mathbb{Q}_+(u,\bar u)$. Gluing together two sets of the triangles we notice
that their common lines with the indices $(1-2s)$ and $(2s-1)$ annihilate each
other and the resulting Feynman diagram takes the form shown in Fig.~\ref{QQ}.
The corresponding Feynman integral can be written in a simple form by introducing
notation for the following function
\be
X_{v,\bar v;u,\bar u}(\vec z\,|\,\vec w)=
\left[a(s+iu,\bar s-i\bar u)\right]^N\int d^2 y\,
Y_{v,\bar v}^{(s,\bar s)}(\vec y,\vec z)\, Y_{-u,-\bar u}^{(1-s,1-\bar s)}(\vec
y,\vec w)\,,
\label{X-def}
\ee
with the $Y-$functions defined in \re{Q-l}. In this way, one arrives at
\be
[\mathbb{Q}_-(v,\bar v)\,\mathbb{Q}_+(u,\bar u)]\,(\vec z\,|\,\vec w) =\left[\pi
a(2-2s)\right]^N \, X_{v,\bar v;u,\bar u}(\vec z\,|\,\vec w)\,.
\label{Q-+}
\ee
Following the same steps, we now calculate the product of the same operators but
in an opposite order, $[\,\mathbb{Q}_+(u,\bar u)\,\mathbb{Q}_-(v,\bar v)\,](\vec
z\,|\,\vec w)$. This time we replace two operators by the left diagram in
Fig.~\ref{Q1-f} and the right diagram in Fig.~\ref{Q2-f}, respectively, and
obtain two sets of the star-diagrams glued together through common vertices.
Integration over the position of these vertices can be easily performed using
\re{chain-rule-delta} and it gives rise to the $\delta-$function connecting the
centers of the star-diagrams. As a result, one arrives at the diagram, which is
similar to the one shown in Fig.~\ref{QQ}, leading to
\be
 [\,\mathbb{Q}_+(u,\bar u)\,\mathbb{Q}_-(v,\bar v)\,](\vec z\,|\,\vec w) =
\left[\pi a(2-2s)\right]^N \, X_{u,\bar u;v,\bar v}(\vec
z\,|\,\vec w)\,. \label{Q+-}
\ee
Comparing \re{Q-+} and \re{Q+-} we notice that
their r.h.s.\ differ from each other by interchanging the spectral parameters.
Then, the commutativity of the $\mathbb{Q}_+(u,\bar u)$ and $\mathbb{Q}_-(v,\bar
v)$ as well as the relations \re{prod-Q-Q} follow from the following symmetry
property of the $X-$function
\be
X_{u,\bar u\,;\,v,\bar v}(\vec z|\vec w) = X_{v,\bar v\,;\,u,\bar u}(\vec z|\vec
w)\,.
\label{perm-sym}
\ee

\begin{figure}[t]
\centerline{{\epsfxsize16.0cm\epsfbox{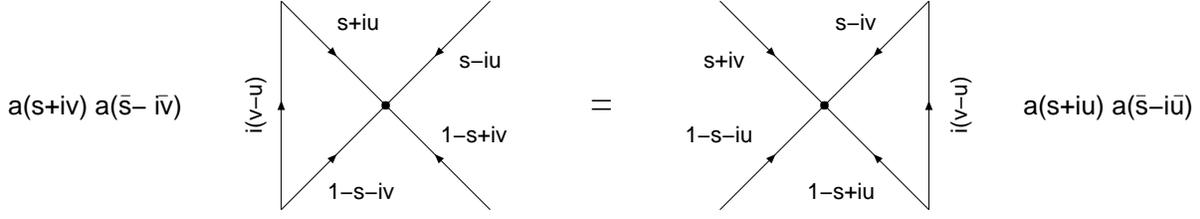}}}
\caption[]{The permutation identity.}
\label{perm}
\end{figure}

The proof of \re{perm-sym} is based on the uniqueness relations, Eq.~\re{uniq},
and it can be carried out using the diagrammatical representation of the function
$X_{u,\bar u\,;\,v,\bar v}$. It also relies on the permutation identity shown in
Fig.~\ref{perm}. To verify this identity it is sufficient to turn the ``unique''
triangles in the both sides of the relation into unique stars and check that the
resulting diagrams coincide. Turning to \re{perm-sym}, we use the diagrammatical
representation of the function $X_{u,\bar u\,;\,v,\bar v}$ and insert two
additional propagators with the indices $i(u-v)$ and $-i(v-u)$, respectively,
into one of the rhombuses as shown in Fig.~\ref{QQ}. Since the sum of the indices
vanishes, this transformation does not change the function. Subsequently applying
the permutation identity (see Fig.~\ref{perm}), we move the line with the index
$i(v-u)$ to the right of the diagram until it returns to its initial position and
annihilates the line with an opposite index, thanks to periodic boundary
conditions along the chain of rhombuses in Fig.~\ref{QQ}. In this way, one
arrives at the initial diagram, in which the spectral parameters $u$ and $v$ are
interchanged, thus proving the relation
\re{perm-sym}.

Let us now turn to the first two relations in \re{com-Q-Q} and examine the
product $\mathbb{Q}_+(u,\bar u)\,\mathbb{Q}_+(v,\bar v)$. Representing the first
operator by the right diagram in Fig.~\ref{Q1-f} and the second one by the left
diagram, one obtains the following Feynman integral
\be
[\mathbb{Q}_+(u,\bar u)\,\mathbb{Q}_+(v,\bar v)](\vec z\,|\,\vec w)=
\left[\pi a(2-2s)\right]^N
\int d^2 y\, X_{v,\bar v;u,\bar u}(\vec y\,|\,\vec z)
\prod_{k=1}^N [z_k-z_{k+1}]^{1-2s}[w_{k}-y_{k+1}]^{2(1-s)}\,.
\ee
Due to the symmetry property \re{perm-sym}, the r.h.s.\ of this relation is
symmetric under permutation of the spectral paramaters and, as a consequence, the
operators $\mathbb{Q}_+(u,\bar u)$ and $\mathbb{Q}_+(v,\bar v)$ commute. The
proof of the commutativity of the operators $\mathbb{Q}_-(u,\bar u)$ and
$\mathbb{Q}_-(v,\bar v)$ goes along the same lines.

Finally, $\mathbb{Q}_+(u,\bar u)$ and $\mathbb{Q}_-(u,\bar u)$ are the
$SL(2,\mathbb{C})$ invariant operators
\be
 [\,\mathbb{Q}_+(u,\bar u), S_\alpha\,]=[\,\mathbb{Q}_-(u,\bar u), S_\alpha\,]=0
\label{Q-inv}
\ee
with $S_\alpha=\sum_{k=1}^N S_\alpha^{_{(k)}}$ being the total $SL(2,\mathbb{C})$
spin of the system. The same property implies that the kernels of the
$\mathbb{Q}-$operators have to transform under the $SL(2,\mathbb{C})$
transformations \re{SL2-trans-z} as
\be
Q_{u,\bar u}^{(\pm)}(\vec z\,'\,|\,\vec w\,')=
\lr{\prod_{k=1}^N [c w_k + d]^{2-2s} [c z_k + d]^{2s}}\,
Q_{u,\bar u}^{(\pm)}(\vec z\,|\,\vec w)\,.
\ee
This relation can be verified using the explicit form of the kernels,
Eqs.~\re{kern+} and \re{kern-}.

\subsection{Properties of the $\mathbb{Q}-$operator}

In the previous Section we have constructed two different Baxter $Q-$operators,
$\mathbb{Q}_+(u,\bar u)$ and $\mathbb{Q}_-(u,\bar u)$. Each of them satisfies the
defining relations, Eqs.~\re{com-Q} -- \re{def-Qa}, as well as the additional
relations \re{prod-Q-Q} and \re{Q-inv}. Let us show that these two operators are
conjugated to each other with respect to the scalar product \re{SL2-norm}.

\subsubsection{Conjugated $\mathbb{Q}-$operator}

The kernel of the operator $\mathbb{Q}^\dagger(u,\bar u)$ conjugated to the
Baxter $\mathbb{Q}-$operator is defined as follows
\be
\left[[\mathbb{Q}(u,\bar u)]^\dagger \,\Psi\right](\vec z_1,...,\vec z_N)=
\int d^2 w\, \lr{Q^{}_{u,\bar u}(\vec w_1,...,\vec w_N\,|\,\vec z_1,...,\vec z_N)}^*\,
\Psi(\vec w_1,...,\vec w_N)\,.
\label{Q-dag-def}
\ee
Calculating $[\mathbb{Q}_+(u,\bar u)]^\dagger$, one substitutes its kernel,
Eq.~\re{kern+}, into the r.h.s.\ of \re{Q-dag-def} and notices, using \re{s-cond},
that $\lr{Q^{_{(+)}}_{u,\bar u}(\vec w\,|\,\vec z)}^*$ turns out to be
proportional to the kernel $Q^{_{(-)}}_{\bar u^*,u^*}(\vec z\,|\,\vec w)$ defined
in \re{kern-}. This leads to the following relation between the corresponding
$\mathbb{Q}-$operators
\be
\left[\mathbb{Q}_+(u,\bar u)\right]^\dagger= \left(a(1-2s,s+i\bar u^*,\bar s-iu^*)/\pi\right)^{-N}
\mathbb{Q}_-(\bar u^*,u^*)\,.
\label{Q+dagger}
\ee
The inverse relation looks like
\be
\left[\mathbb{Q}_-(u,\bar u)\right]^\dagger= \left[(a(2-2s,s+i\bar u^*,\bar s-iu^*)\,\pi\right)^{-N}
\mathbb{Q}_+(\bar u^*,u^*)\,.
\label{Q-dagger}
\ee
Since the operators $\mathbb{Q}_\pm^\dagger(u,\bar u)$ and $\mathbb{Q}_\mp(\bar
u^*,u^*)$ differ from each other by c-valued coefficient function, they commute
with each other for different values of the spectral parameters due to
\re{com-Q-Q}
\be
\left[\,[\mathbb{Q}_\pm(u,\bar u)]^\dagger,\mathbb{Q}_\mp(v,\bar
v)\,\right]=\left[\,[\mathbb{Q}_\pm(u,\bar u)]^\dagger,\mathbb{Q}_\pm(v,\bar
v)\,\right] =0\,.
\label{com-Q-Q-dag}
\ee
Additional properties of the operators $\mathbb{Q}_\pm^\dagger(u,\bar u)$ follow
from conjugation of the corresponding relations for the operators
$\mathbb{Q}_\pm(u,\bar u)$. In particular, the conjugated Baxter equations
\re{def-Qh} and \re{def-Qa} look as
\ba
t_N(u)\,[\mathbb{Q}_\pm(\bar u^*,u^*)]^\dagger &=& (u-i(1-s))^N
[\mathbb{Q}_\pm(\bar u^*-i,u^*)]^\dagger+(u+i(1-s))^N [\mathbb{Q}_\pm(\bar
u^*+i,u^*)]^\dagger\,,
\nonumber
\\[3mm]
\bar t_N(\bar u)\,[\mathbb{Q}_\pm(\bar u^*,u^*)]^\dagger &=& (\bar u-i(1-\bar
s))^N [\mathbb{Q}_\pm(\bar u^*,u^*-i)]^\dagger+(\bar u+i(1-\bar s))^N
[\mathbb{Q}_\pm(\bar u^*,u^*+i)]^\dagger\,.
\nonumber
\\[1mm]
{}\label{Bax-dagger}
\ea
Here, the spectral parameters $u$ and $\bar u$ are arbitrary complex numbers
satisfying the condition \re{u-bar u}.

The relations \re{Q+dagger} -- \re{Bax-dagger} can be further simplified if one
imposes the additional condition on the spectral parameters, $\bar u=u^*$,
leading to \re{u-unit}. As we will show in Sect.~4.4, it is for these values of
the spectral parameters that the Baxter operator $\mathbb{Q}(u,u^*)$ enters into
the expression for the eigenstates of the Hamiltonian in the representation of
the Separated Variables. Notice, however, that in the r.h.s.\ of the Baxter
equations \re{Bax-dagger} the holomorphic and antiholomorphic arguments of the
$Q-$operators are shifted by $\pm i$ independently and, therefore, one can not
obtain from \re{Bax-dagger} a closed system of equations on the operator
$\mathbb{Q}(u,u^*)$.

\subsubsection{Quasimomentum}

The Baxter $\mathbb{Q}-$operator commutes with the transfer matrices of the model
and, therefore, it could serve as a generating function of the Hamiltonian and
the integrals of motion. It is well-known from the theory of the compact spin
magnets that the main observables (like Hamiltonian, quasimomentum, {\it etc})
are determined by the behaviour of the Baxter $\mathbb{Q}-$function at the
special values of the spectral parameter $u=\pm is$ \cite{DKM}. As we will show
below, the same formulae hold for the noncompact spin magnet.

Let us examine the operator $\mathbb{Q}_+(u,\bar u)$ with the spectral parameters
given by $u=\pm is$ and $\bar u=\pm i\bar s$. Using the diagrammatical
representation of this operator, Fig.~\ref{Q1-f}, we find that the lines with the
indices $s\pm iu$ disappear and the integrals over the centers of the star
diagrams take the form of \re{chain-rule-delta} and give rise to the
$\delta-$function
\be
Q^{(+)}_{is,i\bar s}(\vec z\,|\,\vec w) =
\rho_s^{-N}
\prod_{k=1}^N \delta^{(2)}(z_k-w_k)\,,
\qquad
Q^{(+)}_{-is,-i\bar s}(\vec z\,|\,\vec w) =
\rho_s^{-N}
\prod_{k=1}^N \delta^{(2)}(z_{k+1}-w_k)\,,
\ee
where $\rho_s=1/(\pi^2 a(2\bar s,2-2s))=(n_s^2+4\nu_s^2)/\pi^2$. Going over to
the operators one gets
\be
\mathbb{Q}_+(is,i\bar s)=\rho_s^{-N}\,\times\II\,,
\qqquad
\mathbb{Q}_+(-is,-i\bar s)=\rho_s^{-N}\,\times \mathbb{P}\,,
\label{Q-is}
\ee
where $\mathbb{P}$ is the operator of the cyclic permutations defined in
\re{P-cyclic}. Putting $u=\pm is$ and $\bar u=\pm i\bar s$ in \re{Q+dagger} and
taking into account \re{Q-is}, one finds that the $a-$factor in the r.h.s.\ of
\re{Q+dagger} vanishes and, as a consequence, the operator $\mathbb{Q}_-(\pm i
\bar s^*,\pm i s^*)$ is divergent. A careful analysis shows that
\ba
&&\mathbb{Q}_-(-i\bar
s^*+\epsilon,-is^*+\epsilon)=\II\times\frac{(i\pi)^N}{\epsilon^N}
[1+\CO(\epsilon)]
\,,
\nonumber
\\
&&\mathbb{Q}_-(i\bar s^*-\epsilon,is^*-\epsilon)=
\mathbb{P}^{-1}\times\frac{(i\pi)^N}{\epsilon^N}
[1+\CO(\epsilon)]
\label{Q-i(1-s)}
\ea
as $\epsilon\to 0$, so that the $\mathbb{Q}_--$operator  exhibits the $N-$th
order pole. The detailed analysis of the analytical properties of the operators
$\mathbb{Q}_\pm(u,\bar u)$ will be performed in Sect.~3.5.

Using \re{Q-is} and \re{Q-i(1-s)}, one obtains the following relations for the
operator of quasimomentum $\theta(q,\bar q)$ defined in Eq.~\re{cyclic}
\be
\label{c-Q}
\theta=-i\ln \mathbb{P} = i\ln\frac{\mathbb{Q}_+(is,i\bar s)}{\mathbb{Q}_+(-is,-i\bar
s)} = i\ln\frac{\mathbb{Q}_-(i\bar
s^*-\epsilon,is^*-\epsilon)}{\mathbb{Q}_-(-i\bar
s^*+\epsilon,-is^*+\epsilon)}\bigg|_{\epsilon\to 0}
\,.
\ee
We recall that the eigenvalues of this operator have the general form \re{theta}
and satisfy the relation \re{theta-sym}.

\subsubsection{Properties of the eigenvalues}

As we have seen in Sect.~2.4, the Hamiltonian of the model is invariant under the
cyclic and mirror permutations generated by the operators $\mathbb{P}$ and
$\mathbb{M}$, respectively. Let us now examine the action of these operators on
the $\mathbb{Q}-$operators. Since the $\mathbb{Q}-$operators commute with each
other for different values of the spectral parameters, it follows from \re{Q-is}
that
\be
[\mathbb{Q}_\pm(u,\bar u),\mathbb{P}]=0\,.
\ee
The same property can be easily understood from Figs.~\ref{Q1-f} and \ref{Q2-f}
-- the diagrams representing the $\mathbb{Q}-$operators remain unchanged if one
performs simultaneously the cyclic permutations of the $\vec z-$ and $\vec
w-$vertices.

The tranformation properties of the $\mathbb{Q}-$operators under mirror
permutations become more complicated and can be expressed as
\be
\mathbb{Q}_+(-u,-\bar u)=\mathbb{M}\,\mathbb{P}^{-1}~\mathbb{Q}_+(u,\bar
u)~\mathbb{M}\,,\qquad
\mathbb{Q}_-(-u,-\bar u)=\mathbb{M}\,\mathbb{P}~\mathbb{Q}_-(u,\bar
u)~\mathbb{M}\,.
\label{Q-mirror}
\ee
To see this, one applies the operators $\mathbb{M}$ and $\mathbb{P}$ to the
diagrams shown in Figs.~\ref{Q1-f} and \ref{Q2-f} and finds that the diagrams
describing the r.h.s.\ of the relations \re{Q-mirror} look like mirror images of
those shown in Figs.~\ref{Q1-f} and \ref{Q2-f}. To bring them back to the
original form, one flips the whole diagram along the vertical axis and changes
simultaneously the sign of the spectral parameters as indicated in the l.h.s.\ of
\re{Q-mirror}. One can verify \re{Q-mirror} by putting $u=-is$ and $u=i\bar s^*$
in the first and the second relations, respectively, and substituting the
$\mathbb{Q}-$operators by their expressions
\re{Q-is} and \re{Q-i(1-s)}.

Applying the both sides of \re{Q-mirror} to the eigenstate $\Psi_{\vec p,\{q,\bar
q\}}$ one finds the relation between the corresponding eigenvalues of the
$\mathbb{Q}-$operators defined in \re{Q-values}
\ba
\mathbb{Q}_+(-u,-\bar u)\ket{\Psi_{\vec p,\{q,\bar q\}}}
&=&\mathbb{M}\,\mathbb{P}^{-1}~\mathbb{Q}_+(u,\bar u)~\ket{\Psi_{\vec
p,\{-q,-\bar q\}}}
\nonumber
\\
&=&\e^{-i\theta(-q,-\bar q)} Q^{(+)}_{\{-q,-\bar q\}} (u,\bar u)\ket{\Psi_{\vec
p,\{q,\bar q\}}}\,,
\ea
where $(-q,-\bar q) \equiv ((-1)^kq_k,(-1)^k\bar q_k)$. Here, in the second
relation we used Eq.~\re{M-Psi} and in third one Eqs.~\re{cyclic} and
\re{theta-sym}. Then, taking into account \re{cyclic} we find
\be
Q^{(+)}_{\{q,\bar q\}} (-u,- \bar u) = \e^{i\theta(q,\bar q)} Q^{(+)}_{\{-q,-\bar
q\}} (u,\bar u)\,.
\label{Q-eig-prop}
\ee
In a similar manner, we find from the second relation in \re{Q-mirror}
\be
Q^{(-)}_{\{q,\bar q\}} (-u,- \bar u) = \e^{-i\theta(q,\bar q)}
Q^{(-)}_{\{-q,-\bar q\}} (u,\bar u)\,.
\ee

Let us now consider the ratio of the operators
\be
\frac{\mathbb{Q}_+(u,\bar u)}{\mathbb{Q}_-(u,\bar u)}
\equiv \rho_s^{-N/2}\times\mathbb{W}
\label{W-def}
\ee
with $\rho_s=(n_s^2+4\nu_s^2)/\pi^2$. According to \re{prod-Q-Q}, the operator
$\mathbb{W}$ defined in this way does not depends on the spectral parameters.
The normalization factor is chosen in \re{W-def} in such a way that the operator
$\mathbb{W}$ is unitary
\be
\mathbb{W}^\dagger \, \mathbb{W} = \rho_s^{N}
\left[\frac{\mathbb{Q}_+(\bar u^*,u^*)}{\mathbb{Q}_-(\bar
u^*,u^*)}\right]^\dagger
\frac{\mathbb{Q}_+(u,\bar u)}{\mathbb{Q}_-(u,\bar u)}=\II\,,
\qquad \mathbb{W}\,{\Psi_{\vec p,\{q,\bar q\}}}=\e^{iw_{q,\bar
q}}{\Psi_{\vec p,\{q,\bar q\}}}\,.
\ee
Here, we used the fact that the operator $\mathbb{W}$ does not depend on the
spectral parameters and applied the identity \re{Q+dagger}. Since the eigenstates
${\Psi_{\vec p,\{q,\bar q\}}}(\vec z)$ diagonalize the operators
$\mathbb{Q}_\pm$, they also diagonalize $\mathbb{W}$. Then, we obtain from
\re{W-def} and \re{Q+dagger}
\be
\frac{Q^{(+)}_{\{q,\bar q\}}(u,\bar u)}{\left(Q^{(+)}_{\{q,\bar q\}} (\bar u^*,u^*)\right)^*}
=\e^{iw_{q,\bar q}} \, [a(s+iu,\bar s - i\bar u)\e^{i\varphi_s}]^N\,,
\label{Q-phase}
\ee
with $\e^{i\varphi_s}=[a(1-2s,2-2s)]^{1/2}$, or
$\varphi_s=\arg[\Gamma(1-n_s+2i\nu_s)\Gamma(-n_s+2i\nu_s)]$.

We shall use the relation \re{Q-phase} below to fix the phase of
$Q^{_{(+)}}_{\{q,\bar q\}}(u,u^*)$ at $\bar u=u^*$ and to calculate the residues
of $Q^{_{(+)}}_{\{q,\bar q\}}(u,\bar u)$ at its poles. For instance, putting
$u=\pm (i\bar s^*-\epsilon)$ and $\bar u=\pm (is^*-\epsilon)$ in
\re{Q-phase} and applying \re{Q-is}, we calculate the residue of
$Q^{(+)}_{\{q,\bar q\}}$ at the $N-$th order pole in $\epsilon$ as
\ba
&& Q^{(+)}_{\{q,\bar q\}}(-i\bar s^*+\epsilon,-i s^*+\epsilon)=\e^{iw_{q,\bar
q}}\frac{C_s^N}{\epsilon^N}\left[1+\CO(\epsilon)\right]\,,\qquad
\nonumber
\\[3mm]
&& Q^{(+)}_{\{q,\bar q\}}(i\bar s^*-\epsilon,i s^*-\epsilon)=\e^{-i\theta_{q,\bar
q}} Q^{(+)}_{\{q,\bar q\}}(-i\bar s^*+\epsilon,-i
s^*+\epsilon)\left[1+\CO(\epsilon)\right]
\label{Q-pole}
\ea
with $C_s=i\pi\rho_s^{-1/2}=i[(n_s^2+4\nu_s^2)/\pi^4]^{-1/2}$ and $\bar s=1-s^*$.

\subsubsection{Reduction formulae}

The operator $\mathbb{Q}_+(u,\bar u)$ acts on the quantum space of the system
$V_1\otimes...\otimes V_N$, with $V_k\equiv V^{(s,\bar s)}$, and depends explicitly
on the number of particles $N$. There exist the relations connecting the Baxter
operators defined for the different number of particles.

To obtain the reduction relations, one examines the kernel of the operator,
$Q_{u,\bar u}^{_{(+)}}(\vec z|\vec w)$, in the limit when two particles with
coordinates, say, $\vec z_N$ and $\vec z_1$ approach each other on the plane,
$|\vec z_N-\vec z_1|\to 0$. Using the diagrammatical representation of the kernel
-- the left diagram in Fig.~\ref{Q1-f}, one finds that in this limit two lines
connecting the center of the left unique star diagram with the points $\vec z_N$
and $\vec z_1$ merge and produce a single line with the index $2s$. Then, the
resulting integral over the center of the star takes the form
\re{uniq-deg} and gives rise to $\delta(\vec z_N-\vec w_N)$
\ba
&&Q_{u,\bar u}^{_{(+)}}(\vec z_1, ...,\vec z_{N-1},\vec z_N|\vec w_1, ...,\vec
w_{N-1},\vec w_N)\bigg|_{\vec z_N=\vec z_1}
\label{Q-red}
\\
&& \hspace*{30mm}=
\pi^2 a(2s,2(1-\bar s))\, \delta(\vec z_N-\vec w_N)\,
Q_{u,\bar u}^{_{(+)}}(\vec z_1, ...,\vec z_{N-1}|\vec w_1, ...,\vec w_{N-1})\,.
\nonumber
\ea
Here, the $\mathbb{Q}_+-$operator in the r.h.s.\ acts on the quantum space of
$N-1$ particles, $V_1\otimes ...\otimes V_{N-1}$. One can continue the reduction
procedure by putting any pair of the nearest neighbors atop of each other, $\vec
z_k=\vec z_{k+1}$. Each time the number of particles is reduced by 1 until one
reaches the limit $\vec z_1=\vec z_2=...=\vec z_N$ when the $N-$particle Baxter
$\mathbb{Q}-$operator is degenerated into the identity operator $\sim
\prod_k
\delta(\vec z_1-\vec w_k)$. Similar reduction formula holds for the kernel
of the $\mathbb{Q}_--$operator, Eq.~\re{kern-}, but in this case one merges the
right arguments of the kernel
\ba
&&Q_{u,\bar u}^{_{(-)}}(\vec z_1, ...,\vec z_{N-1},\vec z_N|\vec w_1, ...,\vec
w_{N-1},\vec w_N)\bigg|_{\vec w_N=\vec w_1}
\label{Q-red-minus}
\\
&& \hspace*{30mm}=
\pi a(2(1-s),s+iu,\bar s-i\bar u)\, \delta(\vec z_N-\vec w_N)\,
Q_{u,\bar u}^{_{(-)}}(\vec z_1, ...,\vec z_{N-1}|\vec w_1, ...,\vec w_{N-1})\,.
\nonumber
\ea
Here, the additional factor comes from the $B-$factor in the r.h.s.\ of
Fig.~\ref{Q2-f}.

Let us now put $\vec z_1=\vec z_N$ in the both sides of the eigenproblem
\re{Q-values} and apply the relation \re{Q-red}
\ba
&&Q_{q,\bar q}^{(N)}(u,\bar u) \,\Psi_{\vec p,\{q,\bar q\}}(\vec z_1, ...,\vec
z_{N-1},\vec z_1)=
\frac{\pi^2}{n_s^2+4\nu_s^2}
\nonumber
\\
&&\hspace*{30mm}
\times \int d^2 w\,Q_{u,\bar u}^{_{(+)}}(\vec z_1, ...,\vec
z_{N-1}|\vec w_1, ...,\vec w_{N-1})
\Psi_{\vec p,\{q,\bar q\}}(\vec w_1,...,\vec w_{N-1},\vec z_1)\,,
\label{Q-reduc}
\ea
where the superscript $(N)$ in the l.h.s.\ indicates that $Q_{q,\bar
q}^{(N)}(u,\bar u)$ is the eigenvalue of the $N-$particle Baxter operator. Using
the property of completeness of the solutions to the eigenproblem
\re{Q-values}, we may expand $\Psi_{\vec p,\{q,\bar q\}}(\vec w_1,...,\vec
w_{N-1},\vec z_1)$ over the eigenstates of the $(N-1)-$particle Baxter operator,
$\Psi^{_{(N-1)}}_{\vec p',\{q',\bar q'\}}(\vec w_1,...,\vec w_{N-1})$ with the
expansion coefficients depending on $\vec z_1$. After its substitution into
\re{Q-reduc}, the integral in the r.h.s.\ is replaced by the sum over the eigensvalues
$Q^{_{(N-1)}}_{q',\bar q'}$. Then, comparing the $\vec z-$dependence of the both
sides of the resulting relation one arrives at overcompleted system of equations
on the expansion coefficients. Its consistency requirements provide the
quantization conditions on the integrals of motion $q_k$ and $\bar q_k$
\cite{KW}. Solving the system one finds the expansion coefficients and,
as a byproduct, reconstructs the expansion of the eigenfunctions and the
eigenvalues of the $N-$particle Baxter equation, $Q_{q,\bar q}^{_{(N)}}(u,\bar
u)$, over those corresponding to the system of $(N-1)-$particles
\cite{DKM-s}.

\subsubsection{Eigenvalues at $N=2$}

\begin{figure}[t]
\centerline{{\epsfysize5.8cm
\epsfbox{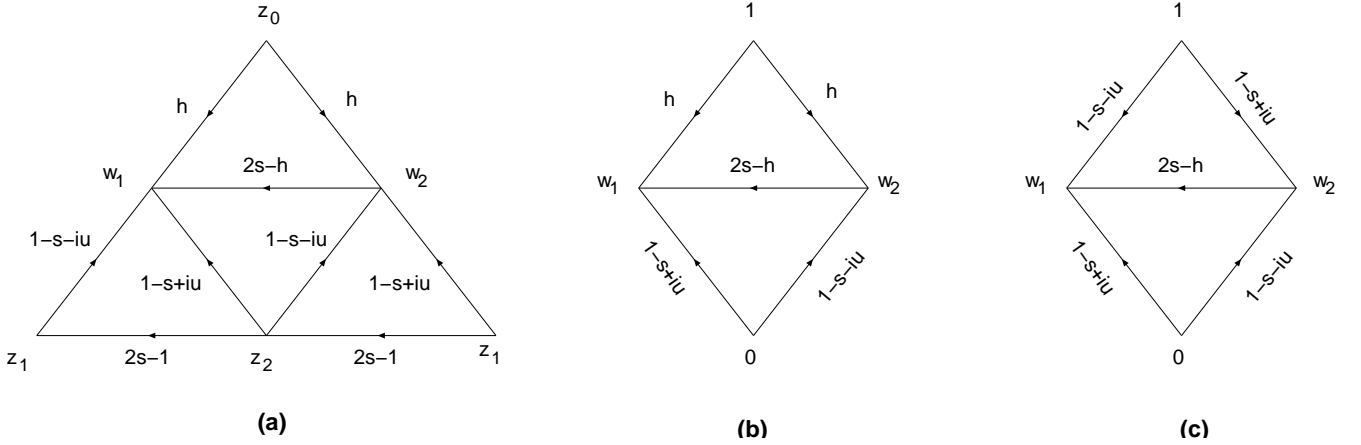}}}
\caption[]{The eigenvalue of the Baxter $\mathbb{Q}-$operator at $N=2$:
(a) general diagrammatical representation; (b) reduction at $\vec z_0=1$, $\vec
z_1\to\infty$ and $\vec z_2=0$; (b) reduction at $\vec z_0=\infty$, $\vec z_1=1$
and $\vec z_2=0$.}
\label{Fig-N=2}
\end{figure}

The eigenproblem for the Baxter operator $\mathbb{Q}_+(u,\bar u)$ can be solved
exactly for the system of $N=2$ particles. In this case, the eigenfunctions of
$\mathbb{Q}_+(u,\bar u)$ are given by \re{WF-N=2}. They are parameterized by the
pair of the $SL(2,\mathbb{C})$ spins $(h,\bar h)$ and by the total momentum $\vec
p$, or equivalently the center-of-mass coordinate $\vec z_0$. Their substitution
into \re{Q-values} yields
\be
\int d^2 w_1 d^2 w_2\, Q_{u,\bar u}(\vec z_1,\vec z_2| \vec w_1, \vec w_2)\,
\Psi_{h,\bar h}(\vec w_1-\vec z_0,\vec w_2-\vec z_0)
=Q_{h,\bar h}(u,\bar u)\,\Psi_{h,\bar h}(\vec z_1-\vec z_0,\vec z_2-\vec z_0)\,,
\label{Q-N=2}
\ee
where the function $\Psi_{h,\bar h}(\vec z_1,\vec z_2)$ was defined in
\re{WF-N=2}. Following the diagrammatical approach, we represent the integral
entering the l.h.s.\ of this relation as the Feynman diagram shown in
Fig.~\ref{Fig-N=2}a. To arrive at this diagram we replaced the kernel of the
$\mathbb{Q}-$operator by the right diagram in Fig.~\ref{Q1-f} at $N=2$ and
represented the eigenfunction $\Psi_{h,\bar h}(\vec w_1-\vec z_0,\vec w_2-\vec
z_0)$ as a (nonunique) triangle with the vertices at the points $\vec w_1$, $\vec
w_2$ and $\vec z_0$.

The both sides of the relation \re{Q-N=2} depend on two vectors $\vec z_1-\vec
z_0$ and $\vec z_2-\vec z_0$ that we may choose to our convenience. One possible
choice could be $\vec z_1\to\infty$, $\vec z_0=0$ and $z_2=\bar z_2=1$. In this
limit, four lines connecting the point $\vec z_1$ with the rest of the diagram
can be effectively removed and the resulting diagram takes the form shown in
Fig.~\ref{Fig-N=2}b. The corresponding Feynman integral defines the eigenvalues
of the Baxter operator $\mathbb{Q}_+(u,\bar u)$ at $N=2$
\be
Q_{h,\bar h}^{(N=2)}(u,\bar u)=
\int d^2 w_1\, d^2 w_2 \,
\frac{\lr{\pi\,a(2-2s,s+iu,\bar s-i\bar u)}^2}{[w_1]^{1-s+iu} [w_2]^{1-s-iu} [(w_1-1)]^{h} [(w_2-1)]^{h}
[(w_1-w_2)]^{2s-h}}\,.
\label{Q-eig-a}
\ee

One can find another (equivalent) representation for the same eigenvalue by
choosing in \re{Q-N=2} $\vec z_0\to\infty$, $\vec z_2=0$ and $z_1=\bar z_1=1$. In
this case, one can remove two lines attached to the point $\vec z_0$. The
resulting diagram takes the form shown in Fig.~\ref{Fig-N=2}c.
By the construction, $Q_{h,\bar h}^{(N=2)}(u,\bar u)$ satisfies the Baxter
equations, Eqs.~\re{def-Qh} and \re{def-Qa} for $N=2$, with the auxiliary
transfer matrices given by \re{t-aux}
\be
t_2(u)=2u^2-h(h-1)+2s(s-1)
\ee
and $\bar t_2(\bar u)$ is given by a similar expression in the antiholomorphic
sector.

The two-dimensional integral in \re{Q-eig-a} has a striking resemblance to
expressions for the correlation functions in two-dimensional conformal field
theories and, therefore, can be evaluated using the powerful technique developed
in \cite{DF}. The resulting expression for \re{Q-eig-a} is given by the sum over
the product of holomorphic and antiholomorphic ``conformal blocks''
\ba
Q_{h,\bar h}^{(N=2)}(u,\bar u)&=& c(s,h)\, \frac{\Gamma(1-\bar s-i\bar
u)\Gamma(1-\bar s+i\bar u)}{\Gamma(\bar s-i\bar u)\Gamma(\bar s+i\bar u)}
\nonumber
\\
&&\qqqquad\times \bigg\{ Q_0(u)\, \left(Q_0(-\bar u^*)\right)^* +(-1)^{n_h}\,
Q_0(-u)\, \left(Q_0(\bar u^*)\right)^*
\bigg\}\,,
\label{Q-2-2}
\ea
where $h=(1+n_h)/2+i\nu_h$, $\bar h=1-h^*$ and $Q_0(u)$ is given by
one-dimensional contour integral that can be expressed in terms of hypergeometric
series as
\be
Q_0(u) = {}_3F{}_2\lr{{{2s-h,\,2s-1+h,\,s+iu}\atop {2s,\,2s}}\bigg|1}\,.
\label{3F2}
\ee
The constant $c(s,h)$ in \re{Q-2-2} does not depend on the spectral parameter and it is fixed
by the normalization condition \re{Q-is}, $Q_{h,\bar h}(is,i\bar
s)=\pi^4/(n_s^2+4\nu_s^2)^2$.

As a matter of fact, the expression \re{3F2} has already appeared in the analysis
of the Baxter equation for the noncompact spin chains. It was introduced for the
first time in \cite{FK,K1} as the solution to the holomorphic Baxter equation for
the $SL(2,\mathbb{C})$ magnet of spin $s=1$ and $\bar s=0$. Later, the same
expression was identified \cite{SD} as the solution to Baxter equation for the
Heisenberg $SL(2,\mathbb{R})$ magnet of arbitrary spin $s$ and the total spin of
the system, $h-2s$, being nonnegative integer. In the latter case, $Q_0(u)$ is
reduced to a polynomial of degree $h-2s$ in $u$, which turns out to be identical
to the Hahn orthogonal polynomial.

The detailed analysis of the properties of the obtained expressions,
Eqs.~\re{Q-2-2} and \re{3F2}, and their generalization to higher $N$ will be
presented elsewhere \cite{DKM-s}.

\subsection{Relation to the Hamiltonian}

There exists the relation between the Hamiltonian of the model, \re{H-from-R},
and the Baxter $\mathbb{Q}-$operator. It allows to replace the original
Schr\"odinger equation \re{Sch-eq} by the spectral problem \re{Q-values} and
express the energy $E_N$ in terms of the eigenvalues of the operator
$\mathbb{Q}_+(u,\bar u)$.

The above relation follows, in its turn, from a more general relation between the
$SL(2,\mathbb{C})$ transfer matrices of arbitrary spin, ${\mathbb{T}}^{(s_0,\bar
s_0)}(u,\bar u)$, defined in \re{t-N}, and the Baxter $\mathbb{Q}-$operator (see
Appendix B for details). The simplest way to establish this relation is to
compare the Feynman diagrams describing the transfer matrix and the product of
$\mathbb{Q}-$operators, Figs.~\ref{Fig-T} and \ref{QQ}, respectively. It is easy
to see that, up to redefinition of the spins and the spectral parameters, two
diagrams are identical. In this way, we obtain the following relation
\be
\mathbb{T}_N^{(s_0,\bar s_0)}(u,\bar u) \sim
\mathbb{X}(u+i(1-s_0),\bar u+i(1-\bar s_0); u+ i s_0, \bar u+ i\bar s_0)\,,
\ee
which is valid for arbitrary $SL(2,\mathbb{C})$ spins $(s_0,\bar s_0)$ and the
spectral parameters $u$ and $\bar u$. Here, the notation was introduced for the
operator $\mathbb{X}(u,\bar u; v,\bar v)$ with the kernel $X_{u,\bar u\,;\,v,\bar
v}(\vec z,\vec w)$ defined in \re{X-def}. Taking into account the relation
between the $\mathbb{X}-$operator and the Baxter $\mathbb{Q}-$operators,
Eqs.~\re{Q-+} and \re{Q+-}, we obtain the expression for the transfer matrix in
terms of the operators $\mathbb{Q}_\pm$
\ba
\mathbb{T}_N^{(s_0,\bar s_0)}(u,\bar u) &=&
\rho_{_T}^{(s_0,\bar s_0)}(u,\bar u)\times
\mathbb{Q}_-\lr{u+i(1-s_0),\bar u+i(1-\bar s_0)}\,\mathbb{Q}_+\lr{u+ i s_0, \bar u+ i\bar
s_0}\,,
\nonumber
\\
&=&
\rho_{_T}^{(s_0,\bar s_0)}(u,\bar u)\times
\mathbb{Q}_-\lr{u+ i s_0, \bar u+ i\bar s_0}\,\mathbb{Q}_+\lr{u+i(1-s_0),\bar u+i(1-\bar s_0)}\,,
\label{T-Q+Q-}
\ea
where the spectral parameters satisfy the relation \re{u-bar u} and the
normalization factor $\rho_{_T}^{(s_0,\bar s_0)}(u,\bar u)$ can be calculated
from Eqs.~\re{rho-u} and \re{X-def} as
\be
\rho_{_T}^{(s_0,\bar s_0)}(u,\bar u)=\left[
\frac1
{\pi^3 a(2-2s)}
\frac{a(s_0-s+iu,\bar s-\bar s_0 -i \bar u)}{a(s-s_0+iu,\bar s_0+\bar s -i \bar u)}
\right]^N\,,
\label{rho-T}
\ee
The following
comments are in order.

According to \re{H-from-R}, the Hamiltonian of the model is given by the
logarithmic derivative of the fundamental transfer matrix $\mathbb{T}^{(s,\bar
s)}(u,\bar u)$ at $u=\bar u=0$. Since the spectral parameters have to satisfy the
condition \re{u-bar u}, to calculate the Hamiltonian it becomes sufficient to
analyze the transfer matrix for real values of the spectral parameters $u=\bar
u=u^*$. In this case, one puts $s_0=s$ in \re{T-Q+Q-} and substitutes the
$\mathbb{Q}_--$operator by its expression in terms of
$(\mathbb{Q}_+)^\dagger-$operator, \re{Q+dagger}, to find (see also \re{T-QQ})
\ba
\mathbb{T}_N^{(s,\bar s)}(u,u) &=&
\rho_{_Q}(u)\,
\left[\,\mathbb{Q}_+\lr{u-is,u-i\bar s}\right]^\dagger
\mathbb{Q}_+\lr{u+ i s,  u+ i\bar s}
\label{T-fund}
\\
&=&\widetilde\rho_{_Q}u^{2N}\,\left[\,\mathbb{Q}_+\lr{u-i\bar s^*,u-i
s^*}\right]^\dagger
\mathbb{Q}_+\lr{u+ i\bar s^*,  u+ is^*}
\nonumber
\ea
with $\bar s^*=1-s$ and the normalization factors
$$
\rho_{_Q}(u)=\left(
\frac{a(1-2\bar s+iu,1-i
u)}{a(2-2s-iu,1+iu)}\right)^N\widetilde\rho_{_Q}\,,\qquad
\widetilde\rho_{_Q}=\left(
\frac{n_s^2+4\nu_s^2}{\pi^4}
\right)^N\,.
$$
One verifies, using \re{Q-is}, that $\mathbb{T}^{(s,\bar
s)}(0,0)=(-1)^N\mathbb{P}^{-1}$. Together with \re{H-from-R} this leads to
$\mathbb{T}^{(s,\bar s)}_N(u,u)=(-1)^N\exp\lr{-i\theta_N - iu {\cal H}_N+
\CO(u^2)}.$ We find the expression for the Hamiltonian ${\cal H}_N$ by substituting
\re{T-fund} into this relation
\be
{\cal H}_N = \varepsilon_N+i\frac{d}{du} \ln\mathbb{Q}_+(u+is,u+i\bar
s)\bigg|_{u=0} -\left[i\frac{d}{du} \ln\mathbb{Q}_+(u-is,u-i\bar
s)\bigg|_{u=0}\right]^\dagger\,,
\label{H-from-Q}
\ee
where the notation was introduced for the additive normalization constant
\be
\varepsilon_N=i\frac{d}{du} \ln\rho_{_Q}(u) \bigg|_{u=0}=2N\Re \left[\psi(2s)+\psi(2-2s)-2\psi(1)\right]\,.
\ee
Eq.~\re{H-from-Q} establishes the relation between the Hamiltonian and the Baxter
$\mathbb{Q}-$operator. We recall that the latter satisfies the Baxter equations,
Eqs.~\re{def-Qh} and \re{def-Qa}, and depends on the integrals of motion
$\{q,\bar q\}$. Then, Eq.~\re{H-from-Q} provides the explicit form of the
dependence \re{H-q}.

Applying the eigenstate $\Psi_{\vec{p}\{q,\bar q\}}$ to the both sides of
\re{H-from-Q} we calculate the energy as
\be
E_N(q,\bar q)=\varepsilon_N + i\lr{\ln Q_{q,\bar q}^{}(is,i\bar s)}'+i
\lr{\ln\, [Q_{q,\bar q}^{}(-is,-i\bar s)]^*}'\,,
\ee
where prime denotes derivative with respect to the spectral parameter and
$Q_{q,\bar q}(u,\bar u)\equiv Q^{(+)}_{\{q,\bar q\}}(u,\bar u)$ stands for the eigenvalue of the operator
$\mathbb{Q}_+(u,\bar u)$. We can further simplify this expression by using the
properties of the eigenvalues \re{Q-eig-prop}
\be
E_N(q,\bar q)=\varepsilon_N + i\lr{\ln Q_{q,\bar q}^{}(is,i\bar s)}' - i
\lr{\ln\, [Q_{-q,-\bar q}^{}(is,i\bar s)]^*}'\,,
\ee
where $(-q,-\bar q)\equiv ((-1)^kq_k,(-1)^k \bar q_k)$. Taking into account that
$E_N(q,\bar q)$ is invariant under the replacement $(q,\bar q)\to (-q,-\bar q)$
(see Sect.~2.4) one gets from the last relation the expression for the energy
\be
E_N(q,\bar q)=\varepsilon_N - \Im \frac{d}{du}\ln \left[{Q_{q,\bar
q}(u+is,u+i\bar s)\, Q_{-q,-\bar q}(u+is,u+i\bar s)}\right]\bigg|_{u=0},
\label{Energy-I}
\ee
which is explicitly real. This expression was obtained from the first relation in
\re{T-fund}. Starting from the second relation in \re{T-fund}, one can obtain
another representation for the energy
\be
E_N(q,\bar q)=- \Im\frac{d}{du}\ln \left[u^{2N}Q_{q,\bar q}(u+i\bar s^*,u+is^*)
Q_{-q,-\bar q}(u+i\bar s^*,u+i s^*)\right]\bigg|_{u=0}.
\label{Energy-II}
\ee
We notice that, according to \re{Q-pole}, the $Q-$function entering this
expression has the $N-$th order pole at $u=0$. The factor $u^{2N}$ compensates
this pole and, as a consequence, the energy is determined by the subleading
$\CO(\epsilon)$ terms in the r.h.s.\ of \re{Q-pole}.

One can verify \cite{DKM-s} that at $N=2$ the substitution of \re{Q-2-2} into
Eqs.~\re{Energy-I} and \re{Energy-II} leads to the well-known expression for the
energy \re{E-N=2}.

\subsection{Analytical properties}

Let us examine the analytical properties of the Baxter operator,
$\mathbb{Q}_+(u,\bar u)$, and its eigenvalues, $Q_{q,\bar q}(u,\bar u)$, as
functions of the spectral parameters $u$ and $\bar u$. We remind that the
spectral parameters have to satisfy the condition $i(u-\bar u)=n$ with $n$ being
integer. It becomes convenient to parameterize their possible values as
$u=-in/2+\lambda$ and $\bar u=in/2+\lambda$ with $\lambda$ being arbitrary
complex.  The operator $\mathbb{Q}_+(\lambda-in/2,\lambda+in/2)$ becomes a function of integer $n$ and
continuous complex $\lambda$.

In this Section we shall determine analytical
properties of $\mathbb{Q}_+(\lambda-in/2,\lambda+in/2)$ on the complex $\lambda-$plane for fixed $n$.
To start with, we
consider Eq.~\re{Q-values} and substitute the $\mathbb{Q}-$operator by its
integral representation with the kernel given by the first relation in \re{kern+}
\be
\int d^2 y \,
\prod_{k=1}^N
[z_k-y_{k-1}]^{-s-iu}\,[z_{k-1}-y_{k-1}]^{-s+iu}\,
\widehat{\Psi}_{\vec p,\{q,\bar q\}}(\vec y) = Q_{\{q,\bar q\}}(u,\bar u) \,\Psi_{\vec p,\{q,\bar q\}}
(\vec z)\,,
\label{Q-int-sing}
\ee
where the notation was introduced for the function
\be
\widehat{\Psi}_{\vec p,\{q,\bar q\}}(\vec y) = \int d^2 w\,
\Psi_{\vec p,\{q,\bar q\}}(\vec w) \prod_{k=1}^N [w_{k}-y_{k}]^{2s-2}
=\left(\frac{n_s^2+4\nu_s^2}{\pi^2}\right)^{-N/2}\,\left[\mathbb{U}\Psi_{\vec
p,\{q,\bar q\}}\right](\vec y)\,.
\label{intertwin}
\ee
This relation is well-known as the $SL(2,\mathbb{C})$ intertwining transformation
with $\mathbb{U}$ being the corresponding unitary operator \cite{group}. It maps
the state $\Psi^{(s,\bar s)}(\vec z)$ belonging to the space $V\otimes ...\otimes
V$, with $V\equiv V^{(s,\bar s)}$, into the unitary equivalent state
$\Psi^{(1-s,1-\bar s)}(\vec z) =\left[\mathbb{U}\Psi^{(s,\bar s)}\right](\vec
z)$.

The singularities of the eigenvalues $Q_{\{q,\bar
q\}}(\lambda-in/2,\lambda+in/2)$ in $\lambda$ are in one-to-one correspondence
with divergences of the integral in the l.h.s.\ of \re{Q-int-sing}. The latter
originate from two different integration regions: $|\vec y_k -\vec z_{k+1}|\to 0$
and $|\vec y_k -\vec z_k| \to 0$. In the first region, one expands the integrand
in \re{Q-int-sing} in powers of $\vec w_k\equiv\vec y_k -\vec
z_{k+1}$, assuming that
$\widehat{\Psi}_{\vec p,\{q,\bar q\}}(\vec y)-\widehat{\Psi}_{\vec p,\{q,\bar
q\}}(\vec z)\sim\sum_{k}c_k\,w_k^{m_k-1}\bar w_k^{\bar m_k-1}$, and calculates
the l.h.s.\ of \re{Q-int-sing} as the product of integrals of the form
\be
\int_{|w_k|<\epsilon} d^2 w_k\, w_k^{-s-iu}\bar w_k^{-\bar s-i\bar u}
\,w_k^{m_k-1}\bar w_k^{\bar m_k-1}\sim\frac{2\pi\delta_{s+iu-m_k,\,\bar s+i\bar u-\bar m_k}}{s+iu-m_k}
\ee
with $m_k$ and $\bar m_k$ being positive integer. Thus, integration over each
$\vec y_k\sim \vec w_{k+1}$ in \re{Q-int-sing} brings a simple pole in the
spectral parameter located at $s+iu-m_k=\bar s+i\bar u-\bar m_k=0$. Taking their
product we find that the resulting $N-$fold integral over $\vec y_k$ in the
l.h.s.\ of \re{Q-int-sing} acquires the poles in the spectral parameter $\lambda$
of the order not higher than $N$. Similarly, calculating the contribution from
the second region $\vec y_k\sim\vec z_{k}$ we find another (infinite) set of
poles located at $s-iu-m_k=\bar s-i\bar u-\bar m_k=0$.

\begin{figure}[t]
\centerline{{\epsfxsize8.0cm\epsfbox{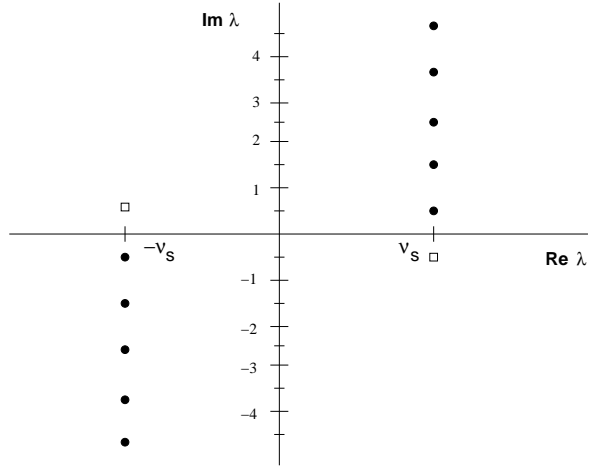}}}
\caption[]{The position of poles (fat points) of the Baxter operator
$\mathbb{Q}_+(\lambda-in/2,\lambda+in/2)$ in the complex $\lambda-$plane at
$n=0$. The $SL(2,\mathbb{C})$ spin of particles, $s=(1+n_s)/2+i\nu_s$, is chosen
in such a way that $n_s=0$ and $\nu_s>0$. Two squares correspond to the values of
the spectral parameters $(u,\bar u)=(\pm is,\pm i\bar s)$.}
\label{Fig-poles}
\end{figure}

Combining together the contribution from both regions, we conclude that the
eigenvalue of the Baxter $\mathbb{Q}-$operator, $Q_{\{q,\bar q\}}(u,\bar u)$, is
an analytical function on the complex plane except the infinite set of points
\be
\left\{u_m^{+}=i(s-m),\ \bar u_{\bar m}^{+}=i(\bar s-\bar m)\right\}\,;\qquad
\left\{u_m^{-}=-i(s-m),\
\bar u_{\bar m}^{-}=-i(\bar s-\bar m)\right\}
\label{poles-u}
\ee
with $m,\,\bar m=1,2, ...$, at which it has the poles of the order not higher
than $N$. Using the above parameterization of the spectral parameters,
$u=-in/2+\lambda$ and $\bar u=in/2+\lambda$, we obtain that the poles can be
enumerated by a pair of positive integers $(m,\,\bar m)$
\be
n_{m,\bar m}^\pm=\mp(n_s+\bar m - m)\,,\quad
\lambda_{m,\bar m}^\pm = \mp\left(\nu_s + i\frac{m+\bar m-1}2 \right)
\label{poles-l}
\ee
with $s=(1+n_s)/2+i\nu_s$ and $m,\bar m=1,2,...$~. Then, for arbitrary integer
$n=n_{m,\bar m}^\pm$ the function $Q_{\{q,\bar q\}}(\lambda-in/2,\lambda+in/2)$
has the $N-$th order poles situated along the imaginary axis in the
$\lambda-$plane at the points $\lambda=\lambda_{m,\bar m}^\pm$ as shown in
Fig.~\ref{Fig-poles}. These two infinite sets of poles lie on the different sides
from the real axis and do not overlap. Since $|\Im\lambda_{m,\bar m}^\pm|\ge
1/2$, the eigenvalues of the Baxter $\mathbb{Q}-$operator do not have poles
within the strip $-1/2 < \Im
\lambda < 1/2$.

As was shown in the previous section, different observables of the model -- the
Hamiltonian and quasimomentum, Eqs.~\re{Energy-I}, \re{Energy-II} and \re{c-Q},
are expressed in terms of the Baxter operator $\mathbb{Q}_+$ and its derivatives
defined at the special points $(u=\pm is,\bar u=\pm i\bar s)$ and $(u=\pm i\bar
s^*,\bar u=\pm is^*)$. In the former case, the points belong to the region of
analyticity of the $\mathbb{Q}_+-$operator (see Fig.~\ref{Fig-poles}) and,
therefore, the above operators are well-defined. In the latter case, the $N-$th
order pole of the Baxter $\mathbb{Q}_+-$operator  is compensated by the factor
$u^{2N}$ entering the second relation in \re{T-fund}.

Analyzing the singularities of the $\mathbb{Q}_+-$operator we have used the
integral representation of the kernel defined by the left diagram in
Fig.~\ref{Q1-f}. One could use instead another representation, corresponding to
the right diagram on the same figure. In this way, one finds that the Feynman
integrals have poles at $u=\pm i (s+m)$ and $\bar u=\pm i(\bar s+\bar m)$ with
$m$ and $\bar m$ positive integer, that are different from \re{poles-u}. In
addition, one has to take into account that the prefactor $A(u,\bar u)$ also
generates singularities. One finds that $A(u,\bar u)$ produces the poles at the
points \re{poles-u} and, at the same time, it vanishes at $u=\pm i (s+m)$ and
$\bar u=\pm i(\bar s+\bar m)$, thus compensating the singularities of the Feynman
integral.

Summarizing the analytical properties of the Baxter operator, we expand
$\mathbb{Q}_+(u,\bar u)$ into the sum over poles \re{poles-l}
\ba
\mathbb{Q}_+(\lambda-in/2,\lambda+in/2)\!\!\!&=&\!\!\!\sum_{m,\bar
m=1}^\infty\delta_{n,n^+_{m,\bar m}}
\mathbb{R}_{m,\bar m}^+\left[\frac{1}{(\lambda-\lambda_{m,\bar m}^+)^N}
+\frac{i\,\mathbb{E}_{m,\bar m}^+}{(\lambda-\lambda_{m,\bar m}^+)^{N-1}}+
...\right]
\nonumber
\\
&& \phantom{\sum\!\!\!} +\delta_{n,n^-_{m,\bar m}}\mathbb{R}^-_{m,\bar
m}\left[\frac{1}{(\lambda-\lambda_{m,\bar m}^-)^N}+\frac{i\,\mathbb{E}_{m,\bar
m}^-}{(\lambda-\lambda_{m,\bar m}^-)^{N-1}} + ...\right]\!,
\label{poles-exp}
\ea
where the operators $\mathbb{R}_{m,\bar m}^\pm$ and $\mathbb{E}_{m,\bar m}^\pm$
define the corresponding residues and ellipses denote the contribution of the
$k-$th order poles $(k\le N-2)$.

The operators $\mathbb{R}_{m,\bar m}^\pm$ and $\mathbb{E}_{m,\bar m}^\pm$ share
all properties of the operator $\mathbb{Q}_+(u,\bar u)$. In particular, they
commute with each other and are diagonalized by the eigenstate ${\Psi}_{\vec
p,\{q,\bar q\}}(\vec z)$. Replacing the operators $\mathbb{R}_{m,\bar m}^\pm$ and
$\mathbb{E}_{m,\bar m}^\pm$ by their corresponding eigenvalues, ${R}_{m,\bar
m}^\pm(q,\bar q)$ and ${E}_{m,\bar m}^\pm(q,\bar q)$, respectively, one obtains
from \re{poles-exp} the pole expansion of the eigenvalues of the
$\mathbb{Q}_+-$operator, $Q_{q,\bar q}(u,\bar u)$. The properties of the
eigenvalues, Eq.~\re{Q-eig-prop}, are translated into the following relations
\be
R_{m,\bar m}^\pm(q,\bar q)=(-1)^N \e^{i\theta_{q,\bar q}} R_{m,\bar
m}^\mp(-q,-\bar q)\,,\qquad E_{m,\bar m}^\pm(q,\bar q)= - E_{m,\bar
m}^\mp(-q,-\bar q)\,.
\label{R-quasi}
\ee
Similarly, substituting $u=u_m^++\epsilon$ and $\bar u=\bar u_m^++\epsilon$ into
\re{Q-phase} and matching $Q_{q,\bar q}(u_m^++\epsilon,\bar u_m^++\epsilon)$ into
\re{poles-exp}, one arrives at
\ba
R_{m,\bar m}^+(q,\bar q)&=&\e^{i\omega_{q,\bar q}} {C^N_{m,\bar m}}
\left[Q_{q,\bar q}(i(s+\bar m-1),i(\bar s+m-1))\right]^*
\nonumber
\\[3mm]
E_{m,\bar m}^+(q,\bar q)&=&N\varepsilon_{m,\bar m}-i\frac{d}{d\epsilon}\ln
\left[Q_{q,\bar q}(i(s+\bar m-1)+\epsilon,i(\bar
s+m-1)+\epsilon)\right]^*\bigg|_{\epsilon=0}.
\label{E-plus}
\ea
Here, $C_{m,\bar m}=-C_s/[\pi^2 B(m,1-2s)B(\bar m,1-2\bar s)]$ is expressed in
terms of the Euler $B-$function, the constant $C_s$ is defined in \re{Q-pole}
and $\varepsilon_{m,\bar m}=\psi(1-2s+m)+\psi(1-2\bar s+\bar m)-\psi(m)-\psi(\bar
m)-\psi(1-2\bar s)+\psi(2\bar s)$. We verify, using
\re{E-plus} and \re{Q-is}, that $R_{1,1}^+(q,\bar q)=\e^{iw_{q,\bar q}}C_s^N$ in
accordance with \re{Q-pole}.

Remarkably enough, the spectrum of the model can be found from the pole expansion
of the $\mathbb{Q}_+-$operator, Eq.~\re{poles-exp}. Indeed, according to
\re{R-quasi}, the quasimomentum $\theta_{q,\bar q}$ can be expressed in terms of
the residue functions $R^\pm_{1,1}(\pm q,\pm\bar q)$. Substituting \re{poles-exp}
into \re{Energy-II}, we obtain that the energy $E_N(q,\bar q)$ is related to the
residue functions $E^\pm_{1,1}(\pm q,\pm\bar q)$
\be
E_N(q,\bar q) = - \Re\left[ E_{1,1}^-(q,\bar q) + E_{1,1}^-(-q,-\bar q) \right] =
\Re\left[ E_{1,1}^+(q,\bar q) + E_{1,1}^+(-q,-\bar q) \right]\,.
\ee
Making use of \re{E-plus} it is easy to see that this expression coincides with
\re{Energy-I}.

\subsection{Asymptotic behavior}

As we have seen in the previous Section, the eigenvalues of the
$\mathbb{Q}-$operators, $Q^{_{(+)}}_{q,\bar q}(u,\bar u)$, are meromorphic
functions of the spectral parameters, $u=\lambda-in/2$ and $\bar u=\lambda+in/2$,
with a (infinite) series of poles in $\lambda$ located parallel to the imaginary axis
outside the strip $|\Im\lambda|<1/2$. Let us find their asymptotic behavior at
large $\lambda$ along the horizontal axis, $\Re\lambda\to\infty$ for $\Im
\lambda={\rm fixed}$. In this limit, we may neglect the imaginary part of
$\lambda$ and choose the spectral parameter to be real, $\lambda=\nu$, with
$\nu\to \infty$.

We substitute $u=-in/2+\nu$ and $\bar u=in/2+\nu$ into the relation \re{Q-values}
and examine its both sides in the limit of large real $\nu$. Since the
$\mathbb{Q}-$operator commutes with the $SL(2,\mathbb{C})-$generators,
Eq.~\re{Q-inv}, it can not depend on the momentum operator $\vec p$ and the same
is true for its eigenvalues, $Q^{_{(+)}}_{q,\bar q}(u,\bar u)$. Using this
property we put $\vec p=0$ in \re{Q-values}. The reason for this particular
choice of the momentum is that the state $\Psi_{\vec p=0,\{q,\bar q\}}(\vec
z)\equiv\Psi_{\{q,\bar q\}}^{_{(0)}}(\vec z)$ possesses additional symmetry
properties -- it diagonalizes the operators $S_0$ and $\bar S_0$, and at the same
time it is annihilated by the operators $p=iS_-$ and $\bar p=i\bar S_-$
\be
S_-\,\Psi_{\{q,\bar q\}}^{(0)}(\vec z)=0\,,\qquad (S_0-h)\,\Psi_{\{q,\bar
q\}}^{(0)}(\vec z)=0
\ee
and similar relations hold in the antiholomorphic sector. These properties imply
that, firstly, $\Psi_{\{q,\bar q\}}^{(0)}(\vec z)$ is translation invariant and,
as a consequence, it depends on the differences of the coordinates and, secondly,
it has a definite scaling dimension
\be
\Psi^{(0)}(z_k+\epsilon,\bar z_k +\bar\epsilon)=\Psi^{(0)}(z_k,\bar z_k
)\,,\qquad
\Psi^{(0)}(\lambda z_k,\bar\lambda \bar z_k )
=\lambda^{h-Ns}\bar\lambda^{\bar h-N\bar s}\Psi^{(0)}( z_k, \bar z_k )\,.
\label{p=0}
\ee

Let us determine the asymptotic behavior of the operator $\mathbb{Q}_+(u,\bar u)$
on the space of functions $\Psi^{(0)}(\vec z)$ satisfying the conditions
\re{p=0}. Using integral representation for the $\mathbb{Q}_+-$operator, Eq.~\re{Q-kernel},
with the kernel given by the second relation in \re{kern+}, one arrives at the
following relation in the limit $\nu\to\infty$
\ba
\left[\mathbb{Q}_+(u,\bar u)\Psi^{(0)}\right](\vec
z)&\stackrel{\nu\to\infty}{=}&\nu^{2N(1-s-\bar s)}\,c(s,\bar s)\,
 \prod_{k=1}^N[z_{k}-z_{k+1}]^{1-2s}\,
\label{Q-asym-int}
\\
&\times&\hspace*{-5mm}
\int d^2  w\,\prod_{k=1}^N
[w_k-z_k]^{s-1+\frac{n}2+i\nu}\,[w_k-z_{k+1}]^{s-1-\frac{n}2-i\nu}\,
\Psi^{(0)}(\vec w)
\nonumber
\ea
with $c(s,\bar s)=[\pi a(2-2s)(-1)^{n_s}]^N$. Examining the asymptotics of the
integrand at large $\nu$, one finds that the dominant contribution comes from two
different integration regions:
\be
{\rm (I)}:\quad \vec w_k=\CO(\nu)\,,\qqqquad {\rm (II)}:\quad\vec w_k-\vec
w_{k+1}=\CO(1/\nu)
\label{regions}
\ee
with $k=1,.., N$, in which the $\nu-$dependence of the integrand cancels out in
the product of the $z-$dependent factors.

In the first region, at large $\vec w_k$, one rescales the integration variables
as $w_k=y_k\nu$ and $\bar w_k=\bar y_k\nu$ and applies the identity
\be
[\nu y_k-z_k]^{i\nu}[\nu y_k-z_{k+1}]^{-i\nu} \sim \exp(-i (z_k-z_{k+1})/y_k-i
(\bar z_k-\bar z_{k+1})/\bar y_k)
\equiv \e^{-2i\Re((z_k-z_{k+1})/y_k)} \,.
\ee
Taking into account the scaling properties of the functions $\Psi^{(0)}$,
Eq.~\re{p=0}, we obtain from \re{Q-asym-int}
\be
\left[\mathbb{Q}_+(u,\bar u)\Psi^{(0)}\right]_{\rm I}
(\vec z)
\stackrel{\nu\to\infty}{=}
\nu^{h+\bar h-N(s+\bar s)}\times c(s,\bar s)\,
\widetilde{\Psi}_{\rm I}^{(0)}
(\vec z)\,,
\label{Q-I}
\ee
where the subscript $(\rm I)$ refers to the integration region in \re{regions}.
Here, the notation was introduced for the function $\widetilde{\Psi}_{\rm
I}^{(0)}$ depending on the differences of the coordinates $\vec z_k-\vec
z_{k+1}$. It is obtained from the function $\Psi^{(0)}\lr{\vec z}$ through the
transformation generated by the operator ${\cal S}_{\rm I}$ defined as
\ba
\widetilde{\Psi}_{\rm I}^{(0)}
(\vec z)&=& [{\cal S}_{\rm I}\,\Psi^{(0)}](\vec z)
\label{S-I}
\\
&=&\prod_{k=1}^N [z_k-z_{k+1}]^{1-2s}\int d^2 y_0\int{\cal D} y
\,\e^{-2i\Re\sum_k(z_{k}-z_{k+1})/(y_k-y_0)}
\prod_{k=1}^N \,[y_k-y_0]^{2(s-1)} \,\Psi^{(0)}\lr{\vec y}\,,
\nonumber
\ea
where ${\cal D} y =d^2 y_1 ... d^2 y_N\,\delta(\sum_{i=1}^N \vec y_i/N)$ is the
integration measure on the space of translation invariant functions. According to
\re{S-I}, the operator ${\cal S}_{\rm I}$ transforms $\Psi^{{(0)}}(\vec z)$ into
another translation invariant function. Then, it follows from \re{Q-I} that the
contribution of the region $({\rm I})$ to the asymptotic behavior of the operator
$\mathbb{Q}_+(u,\bar u)$ on the space of the functions satisfying \re{p=0} is
given by
\be
\left[\mathbb{Q}_+(\nu-in/2,\nu+in/2)\right]_{\rm I}
~\stackrel{\nu\to\infty}{=}~\nu^{h+\bar h-N(s+\bar s)}\times c(s,\bar s)\,{\cal
S}_{\rm I}
\label{Q-asym-I}
\ee
with the operator ${\cal S}_{\rm I}$ defined in \re{S-I}.

In the second region, Eq.~\re{regions}, all $\vec w_k$ approach the same point
$\vec z_0$, but its position on the plane can be arbitrary. It is convenient to
introduce   the ``center-of-mass'' coordinate $\vec z_0$ and define the relative
coordinates $\vec y_k$
\be
\vec z_0=\frac1{N}(\vec w_1 + ... +\vec w_N)\,,\qquad
\vec y_k=\nu(\vec w_k -\vec z_0) = \CO(\nu^0)\,,
\ee
such that $\sum_k \vec y_k=0$. Changing the integration variables in
\re{Q-asym-int} from $(\vec w_1,...,\vec w_N)$ to $(\vec z_0,\vec y_1,...,\vec y_N)$, one applies the identity
\be
[z_0+y_k/\nu-z_k]^{i\nu}[z_0+ y_{k-1}/\nu-z_k]^{-i\nu} \sim
\exp(-i(y_k-y_{k-1})/(z_k-z_0))
\ee
and uses the translation invariance of the function $\Psi^{(0)}$ to find after
some algebra
\be
\left[\mathbb{Q}_+(u,\bar u)\Psi^{(0)}\right]_{\rm II}(\vec
z)~\stackrel{\nu\to\infty}{=}~ \nu^{1-h+1-\bar h-N(s+\bar s)}\times c(s,\bar s)\,
\widetilde{\Psi}_{\rm II}^{(0)}(\vec z)\,.
\ee
Here, $\widetilde{\Psi}_{\rm II}^{(0)}(\vec z)$ is translation invariant function
related to $\Psi^{(0)}(\vec z)$ through the transformation generated by the
second operator ${\cal S}_{\rm II}$ defined as
\ba
\widetilde{\Psi}_{\rm II}^{(0)}(\vec z)&=&
[{\cal S}_{\rm II}\,{\Psi}^{(0)}](\vec z)
\label{S-II}
\\
&=&
\prod_{k=1}^N [z_k-z_{k+1}]^{1-2s}
\int d^2 z_0 \prod_{k=1}^N [z_0-z_k]^{2(s-1)}\,
\int {\cal D} y\, \e^{-2i\Re\sum_k (y_k-y_{k-1})/(z_k-z_0)}
\Psi^{(0)}(\vec y)
\nonumber
\ea
with the measure ${\cal D} y$ defined in \re{S-I}. Then, the contribution of the
region $({\rm II})$ to the asymptotic behavior of the $\mathbb{Q}_+-$operator can
be expressed as
\be
\left[\mathbb{Q}_+(\nu-in/2,\nu+in/2)\right]_{\rm II}~\stackrel{\nu\to\infty}{=}~\nu^{1-h+1-\bar h-N(s+\bar
s)}\times c(s,\bar s)\ {\cal S}_{\rm II}\,.
\label{Q-asym-II}
\ee
We notice that this relation looks similar to \re{Q-asym-I} and it can be
obtained from the latter by replacing the conformal spin of the state, $h\to 1-h$
and $\bar h\to 1-\bar h$, and substituting the operator ${\cal S}_{\rm I}$ by its
counterpart \re{S-II}. Moreover, comparing \re{S-I} and \re{S-II} one finds that
the kernels of the operators are related to each other as
\be
{\cal S}_{\rm I}(\vec{z}\,|\,\vec{y})=\mathbb{M}\,
\prod_{k=1}^N [z_k-z_{k+1}]^{1-2s}\, {\cal S}_{\rm II}(\vec{y}\,|\,\vec{z})
\,\prod_{k=1}^N [y_k-y_{k+1}]^{2s-1}
\ee
with the operator of mirror permutations $\mathbb{M}$ defined in \re{P-cyclic}.

Finally, combining together two contributions, Eqs.~\re{Q-asym-I} and
\re{Q-asym-II}, we obtain the following asymptotic behavior of the
operator $\mathbb{Q}_+$
\be
\mathbb{Q}_+(\nu-in/2,\nu+in/2)~\stackrel{\nu\to\infty}{=}~c(s,\bar s)\left[\nu^{h+\bar h-N(s+\bar
s)}\ {\cal S}_{\rm I}+\nu^{1-h+1-\bar h-N(s+\bar s)}\ {\cal S}_{\rm II}\right]
\,.
\label{Q-asym}
\ee
We would like to stress that this relation is valid on the space of functions
satisfying the conditions \re{p=0}.%
\footnote{This implies, in particular, that the states belonging to this space
have an infinite norm \re{SL2-norm} and therefore one can not define the
operators conjugated to ${\cal S}_{\rm I/II}$ with respect to the scalar product
\re{SL2-norm}.}
One finds the properties of the operators ${\cal S}_{\rm I/II}$ by requiring that
the asymptotic expressions for the $\mathbb{Q}-$operators have to satisfy the
relations established in Sect.~3.2. In particular, it follows from
\re{com-Q-Q} and \re{com-Q-Q-dag} that these operators commute with each other
and with the $\mathbb{Q}-$operators
\be
 [{\cal S}_{\rm I},\mathbb{Q}_\pm(u,\bar u)]
 =[{\cal S}_{\rm II},\mathbb{Q}_\pm(u,\bar u)]
 =[{\cal S}_{\rm I},{\cal S}_{\rm II}]=0\,.
\ee

Let us compare \re{Q-asym} with the general expression for the asymptotic
behavior of the eigenvalues of the operator $\mathbb{Q}_+(\nu-in/2,\nu+in/2)$
that follows from \re{Q-phase}
\be
Q^{(+)}_{\{q,\bar q\}}(\nu-in/2, \nu+in/2)~\stackrel{\nu\to\infty}{=}~
|\,Q^{(+)}_{\{q,\bar q\}}|\,\e^{iw_{q,\bar q}/2} \e^{iN(\varphi_s-\pi n_s)/2}
\,
\nu^{-2i N\nu_s}
\label{Q-asym-gen}
\ee
with $s=(1+n_s)/2+i\nu_s$ and the phase $\varphi_s$ defined in \re{Q-phase}.
Matching \re{Q-asym} into this expression we require
that the $\nu-$dependent part of the phase should not depend on the conformal
spin of the state $h=(1+n_h)/2+i\nu_h$ and $\bar h=1-h^*$. This condition is
satisfied provided that the eigenvalues of the operators ${\cal S}_{\rm I/II}$
have the same absolute value and differ only by a phase
\be
\left[{\cal S}_{\rm I/II}\,\Psi^{(0)}_{q,\bar q}\right](\vec z_1,\vec z_{2},...,\vec z_N) =
{\Upsilon_{q,\bar q}}
\e^{i\,({\Omega_{q,\bar q}}\pm{\Theta_{q,\bar q}})}\Psi^{(0)}_{q,\bar q}(\vec z_1,\vec z_{2}, ...,\vec
z_N)\,.
\ee
As a consequence, the corresponding eigenvalues of the $\mathbb{Q}_+-$operator
\re{Q-asym} have the following asymptotic behavior
\be
Q_{q,\bar q}^{(+)}(\nu-in/2,\nu+in/2)~\stackrel{\nu\to\infty}{=}
 2 \nu^{1-N(1+2i\nu_s)}\,c(s,\bar s)\, {\Upsilon_{q,\bar q}}
 \e^{i{\Omega_{q,\bar q}}}\cos\lr{{\Theta_{q,\bar q}}+2\nu_h\ln\nu}
\label{Q-asym-eig}
\ee
with $\Upsilon_{q,\bar q}$, $\Theta_{q,\bar q}$ and $\Omega_{q,\bar q}$ being
real. Its comparison with \re{Q-asym-gen} yields (up to an additive $(q,\bar q)-$independent
correction to $\Omega_{q,\bar q}$ coming from the normalization factors)
\be
\Omega_{q,\bar q}=\frac12 w_{q,\bar q}\,,\qquad
|\,Q^{(+)}_{\{q,\bar q\}}|\sim \nu^{1-N}\cos\lr{{\Theta_{q,\bar
q}}+2\nu_h\ln\nu}\,.
\ee
We recall that $w_{q,\bar q}$ was defined in \re{Q-phase} as eigenvalue of the
operator $\mathbb{W}$.

Since eigenvalues of the $\mathbb{Q}-$operator do not depend on the momentum of
the state $\vec p$, the asymptotic behavior \re{Q-asym-eig} holds for arbitrary
eigenstate $\Psi_{\vec p,\{q,\bar q\}}(\vec z)$. This is in distinction with
\re{Q-asym} that is valid only on the space of functions satisfying the conditions \re{p=0}.

In this Section, we have constructed the Baxter $\mathbb{Q}-$operator and have
shown that the main physical observables of the system, like Hamiltonian,
quasimomentum operator, transfer matrices etc, can be expressed in terms of a
single operator $\mathbb{Q}_+(u,\bar u)$. Thus, the problem of finding the energy
spectrum of the model is reduced to solving the second order finite-difference
Baxter equations, Eqs.~\re{def-Qh} and \re{def-Qa}, on the eigenvalues of this
operator. The construction of the corresponding eigenstates $\Psi_{\vec
p,\{q,\bar q\}}(\vec z_1,...,\vec z_N)$ will be the subject of the next Section.
We would like to stress that solving the Baxter equations on the eigenvalues of
the $\mathbb{Q}-$operator, we have to impose the additional conditions on their
solutions that follow from the analytical properties of the
$\mathbb{Q}-$operator, Eq.~\re{poles-exp}, and its asymptotic behavior at
infinity, Eq.~\re{Q-asym-eig}. For a general solution to the Baxter equation,
depending on the total set of the integrals of motion, $\{q,\bar q\}$, these
conditions can be satisfied provided that the values of $\{q,\bar q\}$ are
quantized. The explicit form of the quantization conditions will be discussed in
the forthcoming publication \cite{DKM-s}.

\section{Separation of Variables}

For the system of $N=2$ particles, the eigenstates can be found exactly,
Eq.~\re{WF-N=2}, due to the $SL(2,\mathbb{C})$ invariance of the Hamiltonian. For
arbitrary $N$, due to complete integrability of the model, the functions
$\Psi_{\vec p,\{q,\bar q\}}(\vec z)$ can be defined as simultaneous eigenstates
of the total set of the integrals of motion, $\vec p$ and $(q_k,\vec q_k)$ with
$k=2,...,N$. The latter are given by the $k-$th order differential operators
acting on holomorphic and antiholomorphic coordinates of the particles. Instead
of going through the solution of the resulting system of the differential
equations on $\Psi_{\vec p,\{q,\bar q\}}(\vec z)$ we shall employ the method of
the Separation of Variables (SoV) developed by Sklyanin
\cite{SoV}. It allows to find the integral representation for the eigenstates of
the model by going over to the representation of the separated coordinates
$\vec{\mybf{x}}=(\vec x_1,...,\vec x_{N-1})$
\footnote{Here, it proves convenient to define the transformation
$\Phi \to \Psi$ to be anti-linear.}
\be
\Psi_{\vec p,\{q,\bar q\}}(\vec z) = \int d\mybf{x}\,\mu(\vec x_1,...,\vec x_{N-1})
\, U_{\vec p,\vec x_1,...,\vec x_{N-1}}(\vec z_1,...,\vec z_N)\,
\lr{\Phi_{\{q,\bar q\}}(\vec x_1,...,\vec x_{N-1})}^*\,,
\label{SoV-gen}
\ee
where $U_{\vec p,\vec x}(\vec z)$ is the kernel of the unitary operator
corresponding to this transformation
\be
U_{\vec p,\vec x_1,...,\vec x_{N-1}}(\vec z_1,...,\vec z_N) =
\vev{\vec z_1,...,\vec z_N|\vec p,\vec x_1,...,\vec x_{N-1}}\,,
\label{SoV-unit}
\ee
and $\Phi_{\{q,\bar q\}}(\vec x_1,...,\vec x_{N-1})$ is the eigenstate of the
Hamiltonian in the SoV representation
\be
\Phi_{\{q,\bar q\}}(\vec x_1,...,\vec x_{N-1})\delta(\vec p-\vec p\,')
=\vev{\Psi_{\vec p,\{q,\bar q\}}|\vec p\,',\vec x_1,...,\vec x_{N-1}} =
\int d{z} \, U_{\vec p\,',\vec x_1,...,\vec x_{N-1}}(\vec z)\,
\lr{\Psi_{\vec p,\{q,\bar q\}}(\vec z)}^*
\,.
\label{Phi-def}
\ee
Here, we introduced the standard notation for the bra and ket vectors on the
quantum space of the system and defined the eigenstates in the different
representations by projecting them out onto the corresponding basis of the
states, $\vev{\Psi_{\vec p,\{q,\bar q\}}|\vec p\,',\vec x_1,...,\vec x_{N-1}}$
and $\vev{\Psi_{\vec p,\{q,\bar q\}}|\vec z_1,...,\vec z_{N}}$.

Remarkable feature of \re{SoV-gen} is that
its substitution into \re{Sch-eq} leads to the Schr\"odinger equation for
$\Phi_{\{q,\bar q\}}(\vec x)$ that takes the form of multi-dimensional Baxter
equation (see Eq.~\re{multi-B} below). Its solution is given by the product of
the eigenvalues of the Baxter $\mathbb{Q}-$operator with the spectral parameters
equal to holomorphic and antiholomorphic components of the separated coordinates
$\vec x_k=(x_k,\bar x_k)$
\be
\lr{\Phi_{\{q,\bar q\}}(\vec x_1,...,\vec x_{N-1})}^*
\sim Q(x_1,\bar x_1) \ldots
Q(x_{N-1},\bar x_{N-1})\,.
\label{prod-Q}
\ee
In this Section, we shall construct explicitly the SoV transformation, Eqs.
\re{SoV-gen}--\re{prod-Q}, establish the quantization conditions on the
possible values of the separated coordinates $\vec x_k$ and calculate the
integration measure $\mu(\vec x_1,...,\vec x_{N-1})$ entering \re{SoV-gen}.

\subsection{Representation of the Separated Variables}

Defining the representation of the Separated Variables for the noncompact
$SL(2,\mathbb{C})$ magnet, we follow closely the Sklyanin's approach \cite{SoV}.
This approach is based on the properties of the auxiliary monodromy operator
$T_N(u)$ defined in \re{monodromy} and, in particular, the off-diagonal component
$B_N(u)$ and its antiholomorphic counterpart. One finds from the Yang-Baxter
equations on the monodromy operator that the operator $B_N(u)$ satisfies the
following relations \cite{QISM,XXX,ABA,SoV}
\be
 [B_N(u), B_N(v)] = [S_-,B_N(u)] = 0\,,\qquad [S_0,B_N(u)]=-B_N(u)
\label{B-comm}
\ee
with $S_\alpha$ being the total $SL(2,\mathbb{C})$ holomorphic spin.

We recall that, according to Eq.~\re{B-asym}, $B_N(u)$ is given by a polynomial
in $u$ of degree $N-1$ with the operator valued coefficients. Commutativity
property \re{B-comm} implies that the operator coefficients commute with each
other and, therefore, can be diagonalized simultaneously. The eigenvalues of the
operator $B_N(u)$, being polynomials in $u$, can be parameterized by their zeros
$x_1,...,x_{N-1}$. Denoting the corresponding eigenfunction as $U_{\vec p,\vec
x}(\vec z_1,...,\vec z_N)$ and taking into account \re{B-asym}, one gets
\be
B_N(u)U_{\vec p,\vec x}(\vec z_1,...,\vec z_N) = p\,(u-x_1) ... (u-x_{N-1})
U_{\vec p,\vec x}(\vec z_1,...,\vec z_N)
\label{B-U}
\ee
and similarly in the antiholomorphic sector
\be
\bar B_N(\bar u)U_{\vec p,\vec x}(\vec z_1,...,\vec z_N) = \bar p(\bar u-\bar x_1) ...
(\bar u-\bar x_{N-1}) U_{\vec p,\vec x}(\vec z_1,...,\vec z_N)\,.
\label{B-bar-U}
\ee
Here, $\vec{\mybf{x}}=(\vec x_1,...,\vec x_{N-1})$, with $\vec x_k=(x_k,\bar x_k)$,
denotes the zeros of the eigenvalues of the operators $B_N(u)$ and $\bar B_N(\bar
u)$.

The common eigenfunctions of the operators $B_N(u)$ and $\bar B_N(\bar u)$ define
the basis on the quantum space of the system, $U_{\vec p,\vec x}(\vec
z_1,...,\vec z_N)$, which is parameterized by the momentum $\vec p$ and the set
of zeros $\vec{\mybf{x}}$ . Using this basis, one may expand the eigenstates of
the Hamiltonian, $\Psi_{\vec p,\{q,\bar q\}}(\vec z)$ over the states $U_{\vec
p,\vec x}(\vec z_1,...,\vec z_N)$ to arrive at the expansion similar to
\re{SoV-gen} \footnote{Taking into account \re{A-dagger}, one notices that this
basis consists of the eigenstates of two mutually commuting hermitian operators,
$B(u)+\bar B(u^*)$ and $i(B(u)-\bar B(u^*))$, and therefore it is expected to be
complete.}. As was shown by Sklyanin \cite{SoV}, the resulting expansion defines
the representation of the Separated Variables.

Since the eigenfunction in the separated coordinates, Eq.~\re{prod-Q}, is
symmetric under permutation of any pair $\vec x_k$ and $\vec x_j$, we impose the
same condition on the solutions to \re{B-U} and \re{B-bar-U}
\be
U_{\vec p, \vec x_1 ... \vec x_k ... \vec x_j ... \vec x_{N-1}}(\vec z)
=
U_{\vec p, \vec x_1 ... \vec x_j ... \vec x_k ... \vec x_{N-1}}(\vec z)\,.
\label{U-sym}
\ee

The operators $B_N(u)$ and $\bar B_N(\bar u)$ are conjugated to each other with
respect to the $SL(2,\mathbb{C})$ scalar product, Eq.~\re{A-dagger}. Together
with \re{B-U} and \re{B-bar-U} this implies that the same relation holds between
their eigenvalues leading to
\be
x_k^* = \bar x_k\,,\qquad k=1,2,...,N-1\,.
\label{x-real}
\ee
together with $p^*=\bar p$. This becomes the first constraint on the possible
values of the separated coordinates. As a consequence of \re{x-real}, the
solutions to \re{B-U} and \re{B-bar-U} corresponding to different $\vec x_k$ and
$\vec p$ are orthogonal to each other with respect to the scalar product
\re{SL2-norm}
\ba
\vev{\vec {\mybf{x}}',\vec p\,'|\vec{\mybf{x}},\vec p}
&=&
\int d^2 \vec{{z}} \, U_{\vec p,\vec{\mybf{x}}}(\vec z_1,...,\vec z_N)
\left(U_{\vec p',\vec {\mybf x}\,'}(\vec z_1,...,\vec z_N)\right)^*
\nonumber
\\
&=&(2\pi)^N\,\delta^{(2)}(p-p\,')\left\{\delta(\mybf{x}-\mybf{x'}) +
\cdots~ \right\}\,\frac{\mu^{-1}(\vec{\mybf{x}})}{(N-1)!}\,,
\label{U-unit}
\ea
where ellipses denote the sum of terms involving all permutations of the vectors
inside the set $\mybf{x}=(\vec x_1, ...,\vec x_{N-1})$ that are needed to restore
the symmetry property \re{U-sym}. To give the meaning to the
$\delta(\mybf{x}-\mybf{x'})$ one has to specify the possible values of the
separated coordinates (see Eqs.~\re{x-quan} and \re{reg-delta} below).

\subsubsection{Sklyanin's operator zeros}

Following \cite{SoV}, one can interpret parameters $x_k$ in \re{B-U} as eigenvalues of
certain operators $\widehat x_k$ and represent the operator $B_N(u)$ as
\be
B_N(u) = iS_- (u-\widehat x_1) \ldots (u-\widehat x_{N-1})\,.
\label{B-x}
\ee
One also defines similar operators $\widehat{\bar x}_k$ in the antiholomorphic
sector, so that $B_N(\widehat x_k)=\bar B_N(\widehat{\bar x}_k)=0$. According to
\re{B-comm}, the operator zeros $\widehat x_k$, defined in this way, satisfy
the commutation relations
\be
 [\widehat x_k,\widehat x_j]=[S_-,\widehat x_k]=[S_0,\widehat x_k]=0
\label{x-prop}
\ee
and, therefore, they can be diagonalized simultaneously
\be
\widehat x_k\, U_{\vec p,\vec{\mybf{x}}}(\vec z) = x_k\, U_{\vec p,\vec{\mybf{x}}}(\vec z)
\,,\qquad
iS_-\, U_{\vec p,\vec{\mybf{x}}}(\vec z) = p\, U_{\vec p,\vec{\mybf{x}}}(\vec
z)\,.
\label{x-U}
\ee
To find the Schr\"odinger equation on the expansion coefficients $\Phi_{\{q,\bar
q\}}(\vec x_1,...,\vec x_{N-1})$ in \re{SoV-gen} we introduce the operators
$X_j^\pm$
\be
X_j^+ = A_N(\widehat x_j\hookleftarrow u)\,,\qquad X_j^- = D_N(
\widehat x_j\hookleftarrow u)\,,
\label{X's}
\ee
where $\widehat x_j\hookleftarrow u$ stands for the substitution of the spectral
parameter by the operator from the {\it right\/} \cite{SoV} (see Eq.~\re{A-dec}
below).%
\footnote{In \cite{SoV}, similar operators, $\widehat X_j^\pm$, were defined by
substituting $u=\widehat x_j$ in \re{X's} from the {\it left\/}. Both sets of the
operators are related to each other as $\widehat X_j^\mp=X_j^\pm \prod_{j\neq
k}(\hat x_k-\hat x_j\pm i)/(\hat x_k-\hat x_j)$ and it is more convenient to our
purposes to use the operators $X_j^\pm$.} The operators ${X}_j^\pm$ and their
antiholomorphic counterparts ${\bar X}_j^\pm$, $j=1,...,N-1$, have the following
remarkable properties \cite{SoV}
\be
X^\pm_k\, \widehat x_j = (\widehat x_j\mp i\delta_{jk})\, X^\pm_k\,,\qquad
\bar X^\pm_k\, \widehat{\bar x}_j = (\widehat{\bar x}_j\mp i\delta_{jk})\, \bar X^\pm_k\,
\ee
and
\be
X^\pm_k
\,U_{\vec p,\vec{\mybf{x}}}(\vec z)=(x_k\pm is)^N U_{\vec p\,,\,\vec{\mybf{x}}\pm
i\,\vec{\mybf{e}}_k}(\vec{{z}})\,,\qquad \bar X^\pm_k
\,U_{\vec p,\vec{\mybf{x}}}(\vec z)=(\bar x_k\pm i\bar s)^N U_{\vec p\,,\,\vec{\mybf{x}}\pm
i\,\vec{\mybf{\bar e}}_k}(\vec{{z}})\,,
\label{shift}
\ee
where the vectors $\vec{\mybf{e}}_k$ and $\vec{\mybf{\bar e}}_k$ add a unit to
the holomorphic and antiholomorphic components of the vector $\vec x_k=(x_k,\bar
x_k)$, respectively,
\be
\vec{\mybf{x}}\pm i\,\vec{\mybf{e}}_k=(\vec x_1, ..., \vec x_k \pm i \vec e_k, ...,
\vec x_N)\,,\qquad \vec e_k=(1,0)\,,
\label{e-vect}
\ee
and similar for $\vec{\mybf{x}}\pm i\,\vec{\bar{\mybf{e}}}_k$ with $\vec {\bar
e}_k=(0,1)$. Then, the equation on $\Phi_{\{q,\bar q\}}(\vec x_1,...,\vec
x_{N-1})$ follows from the following operator identity
\ba
&&\vev{\Psi_{\vec p,\{q,\bar q\}}|t_N(\hat x_k\hookleftarrow u)|\vec
p\,',{\mybf{\vec x}}} = t_N(x_k)\vev{\Psi_{\vec p,\{q,\bar q\}}|\vec
p\,',{\mybf{\vec x}}}
\nonumber
\\[2mm]
&&\hspace*{20mm}=(x_k+is)^N\vev{\Psi_{\vec p,\{q,\bar q\}}|\vec p\,',\mybf{\vec
x}+i\,\vec{\mybf{ e}}_k} +(x_k-is)^N\vev{\Psi_{\vec p,\{q,\bar q\}}|\vec
p\,',\mybf{\vec x}-i\,\vec{\mybf{ e}}_k}\,,
\ea
where, in the first relation, we took into account that $\Psi_{\vec p,\{q,\bar
q\}}(\vec z)$ diagonalizes the auxiliary transfer matrix $t_N(u)=A_N(u)+D_N(u)$
for arbitrary $u$ and, in the second relation, we substituted $t_N(\hat
x_k\hookleftarrow u)=X^+_k+X^-_k$ and applied \re{shift}.

Substituting \re{Phi-def} into the last relation, we find that the expansion
coefficients $\Phi_{\{q,\bar
q\}}(\vec{\mybf{x}})$ have to satisfy the multi-dimensional Baxter equation%
\footnote{Since $\vec{\mybf{x}}\pm i\,\vec{\mybf{e}}_k$ does not satisfy \re{x-real},
arriving at this equation one assumes certain analyticity properties of the
function $\Phi_{\{q,\bar q\}}(\vec{\mybf{x}})$ outside the region
\re{x-real} that will be discussed below.}
\be
t_N(x_k)\Phi_{\{q,\bar q\}}(\vec{\mybf{x}}) = (x_k+is)^N  \Phi_{\{q,\bar
q\}}(\vec{\mybf{x}}+ i\,\vec{\mybf{e}}_k) +(x_k-is)^N \Phi_{\{q,\bar
q\}}(\vec{\mybf{x}}- i\,\vec{\mybf{e}}_k)
\label{multi-B}
\ee
with $k=1,2,...,N-1$ and $t_N(x_k)$ being the eigenvalue of the auxiliary
transfer matrix \re{t-aux}, depending on the quantum number $q_k$. Similar
relation holds in the antiholomorphic sector. Thus, in the representation,
defined by the functions $U_{\vec p,\vec{\mybf{x}}}(\vec z)$, the Schr\"odinger
equation on the states $\Psi_{\vec p,\{q,\bar q\}}(\vec z)$ becomes equivalent to
the system of separated, one-dimensional finite difference equations on
$\Phi_{\{q,\bar q\}}(\vec{\mybf{x}})$. It remains unclear, however, what are the
possible values of separated coordinates $\vec{\mybf{x}}$ and how does the
integration measure $\mu(\vec{\mybf{x}})$ in \re{SoV-gen} look like. Both
questions will be answered in the next Section.

One can find the recurrence relations on the integration measure
$\mu(\vec{\mybf{x}})$ by taking into account the properties of the operator
$A_N(u)$ under conjugation, Eq.~\re{A-dagger}. Since $A_N(u)$ is a polynomial in
$u$ of degree $N$, it can be uniquely reconstructed from its values at
$u=\widehat x_k$ with $k=1,...,N-1$, Eq.~\re{X's}, and the asymptotic behavior at
large $u$, Eq.~\re{B-asym},
\ba
A_N(u)=(u+iS_0+\sum_k\widehat x_k)\prod_{j=1}^{N-1}(u-\widehat x_j)
+X_k^+\sum_{k=1}^{N-1}\prod_{j\neq k}\frac{u-\widehat x_j}{\widehat x_k-\widehat
x_j}\,.
\label{A-dec}
\ea
Writing similar expansion for $\bar A_N(\bar u)$ and imposing the condition
$A_N(u) = [\bar A_N(u^*)]^\dagger$, one obtains the relation between the
operators $(\bar X_k^+)^\dagger$ and $X_k^+$
\be
\lr{\bar X_k^\pm}^\dagger = X_k^\pm \prod_{j\neq k} \frac
{\widehat x_k-\widehat x_j\pm i}{\widehat x_k-\widehat x_j}\,.
\label{X-bar}
\ee
Here, we also included the relation between $(\bar X_k^-)^\dagger$ and $X_k^-$
that comes from the analysis of the operator $D_N(u)$. Combining together
\re{X-bar}, \re{x-U} and \re{shift}, one obtains
\be
\lr{\bar X_k^\pm}^\dagger  U_{\vec p,\vec{\mybf{x}}}(\vec z)=(x_k\pm is)^N
\prod_{j\neq k}\frac{x_k-x_j\pm i}{x_k-x_j}\,
U_{\vec p\,,\,\vec{\mybf{x}}\pm i\,\vec{\mybf{e}}_k}(\vec{{z}})\,.
\ee
Then, examining the identity
\be
\int d^2 \vec{{z}} \,\left(\bar X_k^\pm U_{\vec p\,',\vec {\mybf x}\,'}(\vec
z)\right)^*U_{\vec p,\vec{\mybf{x}}} (\vec z)=
\int d^2 \vec{{z}} \, \left(U_{\vec p\,',\vec {\mybf x}\,'}(\vec z)\right)^*
\lr{\bar X_k^\pm}^\dagger U_{\vec p,\vec{\mybf{x}}}(\vec z)\,,
\ee
we calculate its both sides using \re{shift},
\re{x-U} and \re{X-bar}, and compare the coefficients in front of
$\delta(\vec{\mybf{x}}-\vec{\mybf{x}}'\pm i\vec{\mybf{e}}_k)$ in the resulting
expressions. In this way, we arrive at the following finite-difference equation
on the integration measure
\be
\frac{\mu(\vec{\mybf{x}}\pm i\vec{\mybf{e}}_k)}{\mu(\vec{\mybf{x}})}
=\prod_{j\neq k}\frac{x_k-x_j\pm i}{x_k-x_j}\,.
\label{eq-measure}
\ee
This relation constraints the dependence of the measure on the holomorphic
coordinates $x_k$. One gets similar relation in the antiholomorphic sector by
replacing $x_k \to \bar x_k$.

The solutions to the equations \re{multi-B} and \re{eq-measure} are defined up to
multiplication by an arbitrary periodic function $f(\vec{\mybf{x}})$, such that
$f(\vec{\mybf{x}}\pm i\vec{\mybf{e}}_k)=f(\vec{\mybf{x}})$. The algebraic
approach described in this Section, does not allow to fix this ambiguity and,
therefore, the construction of the SoV transformation remains uncomplete. To
avoid this obstacle, we shall construct in the next Section the explicit
expression for the basis functions $U_{\vec p,\vec{\mybf{x}}}(\vec z)$.

\subsection{Construction of the SoV transformation}

The transition functions to the SoV representation, $U_{\vec
p,\vec{\mybf{x}}}(\vec z)$, are simultaneous solutions to the system of equations
\re{B-U}--\re{U-unit}. Let us start with the conditions \re{B-U} and \re{B-bar-U}. Since $B_N(u)$
is a polynomial of degree $N-1$ in the spectral parameter $u$, its general
expression at arbitrary $u$ can be reconstructed from its special values at $N-1$
distinct points that we choose as $u=x_k$. Then, Eqs.~\re{B-U} and
\re{B-bar-U} become equivalent the system of equations
\be
B_N(x_k)\, U_{\vec p,\vec{\mybf{x}}}(\vec z_1,...,\vec z_N) = \bar B_N(\bar
x_k)\, U_{\vec p,\vec{\mybf{x}}}(\vec z_1,...,\vec z_N) =0\,,\qquad k=1,2, ...,
N-1\,.
\label{B-red}
\ee
One of the reasons for this particular choice of the spectral parameters is that
similar equations have already appeared in our analysis of the
$\mathbb{Q}-$operator, Eqs.~\re{Psi-prop} and \re{Psi}.

Indeed, using \re{Psi} and \re{Psi-prop}, we find that the function
\ba
U_{\vec p,\vec{\mybf{x}}}(\vec{{z}}) &=& \int d^2{{y}}\,
\Psi^{(s,\bar s)}_{x_1,\bar x_1}(\vec z_1,...,\vec z_N\,|\,\vec y_2,...,\vec y_N)
Z_{\vec p,\vec x_2,...,\vec x_{N-1}} (\vec{{y}})
\nonumber
\\
&=& \lim_{\vec y_1\to\infty} [y_1]^{-2s}
\int d^2 {{y}}\,
Y_{x_1,\bar x_1}^{(s,\bar s)}(\vec z_1,...,\vec z_N\,|\,\vec y_1,\vec
y_2,...,\vec y_N) Z_{\vec p,\vec x_2,...,\vec x_{N-1}} (\vec{{y}})
\label{U-repr}
\ea
satisfies the relations \re{B-red} at $k=1$ for arbitrary weight function
$Z(\vec{{y}})$ and $\vec{{y}}=(\vec y_2,...,\vec y_N)$. It remains to show that
the same relations hold for $k\ge 2$ and, in addition, to ensure the symmetry
property \re{U-sym}. It is easy to see that the former requirement is equivalent
to the latter property combined with $B_N(x_1)\, U_{\vec p,\vec{\mybf{x}}}(\vec
z)=\bar B_N(\bar x_1)\, U_{\vec p,\vec{\mybf{x}}}(\vec z)=0$. Thus, to satisfy
the equations \re{B-red} for $k\ge 2$, it is enough to require that the integral
in r.h.s.\ of \re{U-repr} should be symmetric under permutations $\vec
x_1\leftrightarrow\vec x_k$. This condition imposes constraints on the possible
form of the weight function $Z(\vec{{y}})$ in
\re{U-repr}. Supposing that such weight function exists, one applies the last two
relations in Eq.~\re{Psi-prop} to obtain
\ba
A_N(x_k)\,U_{\vec p\,,\,\vec{\mybf{x}}}(\vec{{z}})&=&(x_k+is)^N U_{\vec
p\,,\,\vec{\mybf{x}}+i\,\vec{\mybf{e}}_k}(\vec{{z}})\,,
\nonumber
\\
D_N(x_k)\,U_{\vec p\,,\,\vec{\mybf{x}}}(\vec{{z}})&=&(x_k-is)^N U_{\vec
p\,,\,\vec{\mybf{x}}-i\,\vec{\mybf{e}}_k}(\vec{{z}})\,,
\ea
where the vector $\vec{\mybf{e}}_k$ was defined in \re{e-vect}. These relations
coincide with the operator identities \re{shift}.

\begin{figure}[t]
\centerline{{\epsfxsize12.0cm \epsfbox{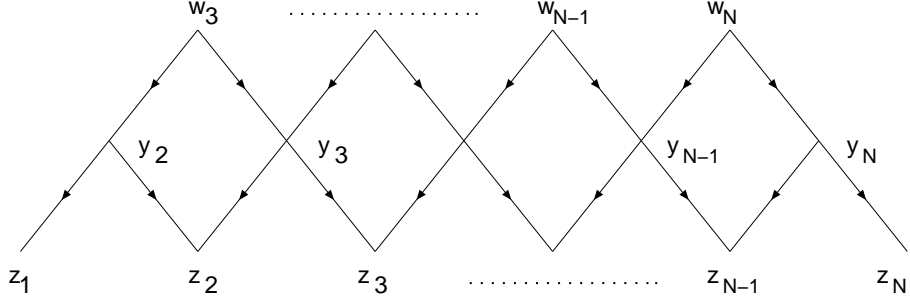}}}
\caption[]{Diagrammatical representation of the convolution of two $\Lambda-$functions
defined in Eqs.~\re{Psi-conv}.}
\label{Fig-rhomb}
\end{figure}

As a hint to finding the solution for $Z_{\vec p,\vec x_2,...,\vec x_{N-1}}$, we
notice a similarity between \re{U-repr} and the definition of the kernel
$X_{v,\bar v; u,\bar u}(\vec z \,|\,\vec w)$, Eq.~\re{X-def}. For $v=x_1$ and
$\bar v=\bar x_1$ both expressions involve the same $Y-$function but defined for
different values of $\vec y_1$. Most importantly, the $X-$function is symmetric
under permutation of the spectral parameters, Eq.~\re{perm-sym}, and it is this
property that we want to impose on \re{U-repr}. Matching the second relation in
\re{U-repr} into \re{X-def} we choose the weight function as
\be
Z_{\vec p,\vec x_2,...,\vec x_{N-1}} (\vec{{y}})= (a(s+ix_2,\bar s-i\bar
x_2))^{N-1}
\int d{w}\, \Lambda^{(1-s,1-\bar s)}_{x_2,\bar x_2}(\vec y_2,...,\vec y_N\,|\,\vec
w_3,...,\vec w_N) Z_{\vec p,\vec x_3,...,\vec x_{N-1}} (\vec{{w}})\,.
\ee
Substituting this ansatz into \re{U-repr} one finds that the dependence of
$U_{\vec p,\vec{\mybf{x}}}(\vec{{z}})$ on $\vec x_1$ and $\vec x_2$ is factorized
into the following integral
\ba
&&\left[\Lambda^{(s,\bar s)}_{N-1}(\vec x_1)\Lambda^{(1-s,1-\bar s)}_{N-2}(\vec
x_2)\right] (\vec z_1,...,\vec z_N\,|\,w_3,...,\vec w_N)
\label{Psi-conv}
\\
&&\hspace*{30mm}\equiv
\int d^2{{y}}\,
\Lambda^{(s,\bar s)}_{x_1,\bar x_1}(\vec z_1,...,\vec z_N\,|\,\vec y_2,...,\vec y_N)
\Lambda^{(1-s,1-\bar s)}_{x_2,\bar x_2}(\vec y_2,...,\vec y_N\,|\,\vec
w_3,...,\vec w_N)\,,
\nonumber
\ea
where subscript in the l.h.s.\ refers to the number of right arguments of the
$\Lambda-$function. We would like to stress that the number of left arguments of
the function $\Lambda^{(s,\bar s)}_{x_1,\bar x_1}$ is larger by 1 than the number
of its right arguments. Therefore, in the convolution of two $\Lambda-$functions
this difference increases to 2. Diagrammatical representation of \re{Psi-conv} is
shown in Fig.~\ref{Fig-rhomb}. The symmetry of \re{U-repr} under permutation of
$\vec x_1$ and $\vec x_2$ comes from the following remarkable property of
\re{Psi-conv}
\be
\left[\Lambda^{(s,\bar s)}_{N-1}(\vec
x_1)\Lambda^{(1-s,1-\bar s)}_{N-2}(\vec x_2)\right] =\lr{\frac{a(s+ix_1,\bar
s-i\bar x_1)}{a(s+ix_2,\bar s-i\bar x_2)}}^{N-1}
\left[\Lambda^{(s,\bar s)}_{N-1}(\vec x_2)\Lambda^{(1-s,1-\bar
s)}_{N-2}(\vec x_1)\right]\,,
\label{Psi-perm}
\ee
which can be considered as a generalization of the similar property of the
$X-$function, Eq.~\re{perm-sym}. Replacing $(s,\bar s)\to (1-s, 1-\bar s)$ in
\re{Psi-perm}, one obtains another useful relation
\be
\left[\Lambda^{(1-s,1-\bar s)}_{N-1}(\vec
x_1)\Lambda^{(s,\bar s)}_{N-2}(\vec x_2)\right] =
\lr{\frac{a(s+ix_1,\bar s-i\bar
x_1)}{a(s+ix_2,\bar s-i\bar x_2)}}^{1-N}\left[\Lambda^{(1-s,1-\bar s)}_{N-1}(\vec
x_2)\Lambda^{(s,\bar s)}_{N-2}(\vec x_1)\right]\,,
\label{Psi-perm-bar}
\ee
which is valid for arbitrary $N$, $\vec x_1$ and $\vec x_2$.

\begin{figure}[t]
\centerline{{\epsfxsize12.0cm \epsfbox{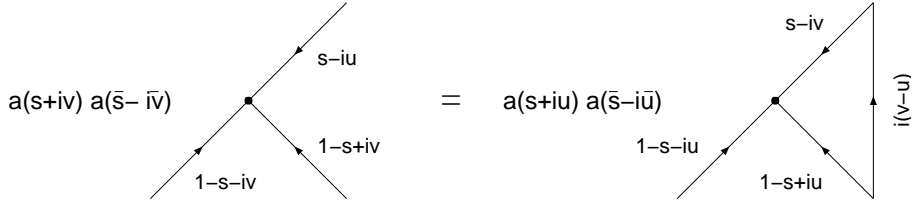}}}
\caption[]{Special case of the permutation identity shown in Fig.~\ref{perm} and
obtained by sending one of the external points to infinity.}
\label{Fig-perm-inf}
\end{figure}

The proof of \re{Psi-perm} and \re{Psi-perm-bar} goes along the same lines as
those for the $X-$function, Eq.~\re{perm-sym}, and it is based on the permutation
identities shown in Figs.~\ref{perm} and \ref{Fig-perm-inf}. Namely, one inserts
two auxiliary lines with the indices $\pm i(x_1-x_2)$ into the leftmost rhombus
in Fig.~\ref{Fig-rhomb} and moves one of the lines to the right of the diagram,
systematically applying the permutation identity, Fig.~\ref{perm}. In contrast
with the previous case, the chain of rhombuses is not periodic and two auxiliary
lines can not annihilate with each other and remain attached to the right- and
leftmost rhombuses. Nevertheless, each of these lines disappears due to the
identity shown in Fig.~\ref{Fig-perm-inf}.

Applying the identities \re{Psi-perm} and \re{Psi-perm-bar}, it becomes
straightforward to write the expression for the transition function $U_{\vec
p,\vec{\mybf{x}}}(\vec{{z}})$ which is consistent with \re{U-repr} and is
symmetric under permutation of any pair of the separated coordinates,
Eq.~\re{U-sym},
\be
U_{\vec p,\vec{\mybf{x}}}(\vec{{z}})=c_N(\vec{\mybf{x}})
\lr{{\vec p}^{\,2}}^{(N-1)/2}\int d^2 w_N
\e^{2i\vec p\cdot \vec w_N}\,U_{\vec{\mybf{x}}}(\vec{{z}}; \vec w_N)\,,
\label{U-gen}
\ee
where $2(\vec p\cdot \vec w_N)\equiv p \,w_N+\bar p \,\bar w_N$,
\be
U_{\vec{\mybf{x}}}(\vec{{z}}; \vec w_N)=
\left[\Lambda^{(s,\bar s)}_{N-1}(\vec x_1)\Lambda^{(1-s,1-\bar s)}_{N-2}(\vec
x_2)\Lambda^{(s,\bar s)}_{N-3}(\vec x_3)\ldots \Lambda_1^{(s,\bar s)}(\vec
x_{N-1})
\right](\vec z_1,...,\vec z_N|\vec w_N)
\label{U-even}
\ee
for {\it even\/} $N$, and
\be
U_{\vec{\mybf{x}}}(\vec{{z}}; \vec w_N)=
\left[\Lambda^{(s,\bar s)}_{N-1}(\vec x_1)\Lambda^{(1-s,1-\bar s)}_{N-2}(\vec
x_2)\Lambda^{(s,\bar s)}_{N-3}(\vec x_3)\ldots \Lambda_1^{(1-s,1-\bar s)}(\vec
x_{N-1})
\right](\vec z_1,...,\vec z_N|\vec w_N)
\label{U-odd}
\ee
for {\it odd\/} $N$. Here, the convolution involves the product of $(N-1)$
functions $\Lambda_{N-k}(\vec x_k)$ with alternating spins $(s,\bar s)$ and
$(1-s,1-\bar s)$. The following comments are in order.

Since the $\Lambda-$functions are translation invariant, their convolution in
\re{U-even} and \re{U-odd} depends on the differences $\vec z_k-\vec w_N$ and, as
a consequence, $(iS_--p) U_{\vec p,\vec{\mybf{x}}}(\vec{{z}})=0$ in agreement
with \re{x-U}. The additional factor $\lr{{\vec p}^{\,2}}^{_{(N-1)/2}}$
in \re{U-gen} restores the scaling dimension of $U_{\vec p,\vec{\mybf{x}}}$.%
\footnote{Since the separated variables are dimensionless (see Eq.~\re{x-prop}),
it follows from \re{U-unit} that the scaling dimension of $U_{\vec
p,\vec{\mybf{x}}}$ is equal to $(N-1)$.}

Throughout this section we have tacitly assumed that $\Lambda^{(s,\bar
s)}_{x_k,\bar x_k}(\vec z_1,...,\vec z_N\,|\,\vec y_2,...,\vec y_N)$ is a
well-defined function of the $\vec z-$ and $\vec y-$vectors on the plane. As we
have seen in Sect.~2.2.1, this condition constraints the possible values of the
spectral parameters, Eq.~\re{u-bar u}. Repeating the same analysis for the
function $\Lambda^{(s,\bar s)}_{x_k,\bar x_k}$, defined in \re{Psi}, we find that
the separated coordinates have to satisfy the relation \re{u-bar u} with $u$ and
$\bar u$ replaced by $x_k$ and $\bar x_k$, respectively. Together with
\re{x-real} this leads to the following quantization conditions on the separated
coordinates
\be
x_k=\nu_k-\frac{in_k}2\,,\qquad \bar x_k=\nu_k+\frac{in_k}2
\label{x-quan}
\ee
with $\nu_k$ real and $n_k$ integer. We would like to remind that it is for these
values of the spectral parameters that the $R-$matrix is a unitary operator,
Eq.~\re{R-unit} and \re{u-unit}.

The normalization factor $c_N(\vec{\mbox{\boldmath${x}$}})$ in \re{U-gen} ensures the symmetry property \re{U-sym}
of the expression \re{U-gen}. It compensates the additional factors in the
r.h.s.\ of \re{Psi-perm} and \re{Psi-perm-bar} and is given for $N\ge 3$ by
\be
c_N(\vec{\mybf{x}})= \prod_{k=1}^{[(N-1)/2]} \left[a(s+ix_{2k},\bar s-i\bar
x_{2k})\right]^{N-k}\prod_{k=1}^{[N/2-1]} \left[a(s+ix_{2k+1},\bar s-i\bar
x_{2k+1})\right]^k
\label{c-N}
\ee
while $c_2(\vec x_1)=1$. Using the definition of the $a-$function, Eq.~\re{a-a},
together with the relations, $\bar s=1-s^*$ and $\bar x_k=x_k^*$, one finds that
the r.h.s.\ of \re{c-N} is a phase factor, $|c_N(\vec{\mybf{x}})|=1$. Its
explicit expression at $N=3,\,4$ looks as follows
\ba
c_3(\vec x_1,\vec x_2)&=&\left[a(s+ix_{2},\bar s-i\bar x_{2})\right]^{2}\,,
\nonumber
\\[3mm]
c_4(\vec x_1,\vec x_2,\vec x_3)&=&\left[a(s+ix_{2},\bar s-i\bar
x_{2})\right]^{3}\left[a(s+ix_{3},\bar s-i\bar x_{3})\right]\,.
\ea

\begin{figure}[t]
\centerline{{\epsfxsize10.0cm\epsfbox{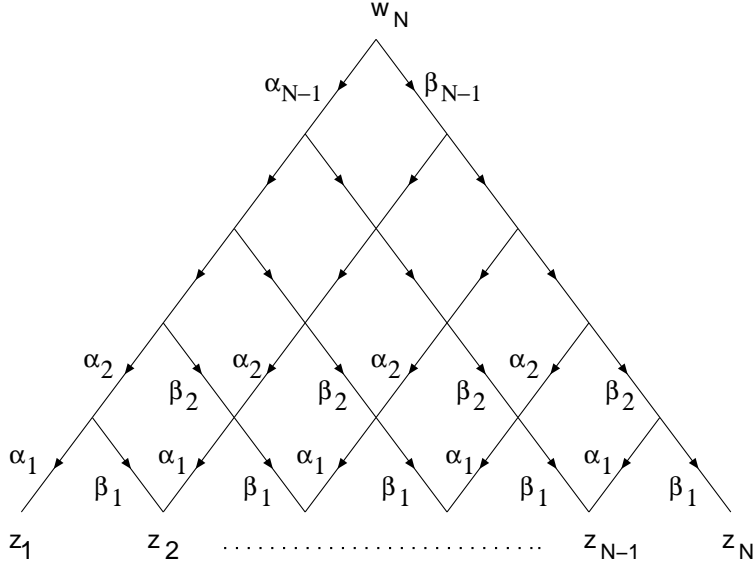}}}
\caption[]{Diagrammatical representation of the transition function,
$U_{\vec p,\vec{\mybf{x}}}(\vec z_1,...,\vec z_N)$. The momentum $\vec p$ flows
into the diagram through the top vertex $\vec w_N$. The indices $\alpha_k$ and
$\beta_k$ parameterize the corresponding two-dimensional propagators (see
Eq.~\re{prop}) and are defined differently for even and odd $k$:
$\alpha_{2k}=1-s-ix_{2k}$, $\beta_{2k}=1-s+ix_{2k}$ and
$\alpha_{2k+1}=s-ix_{2k+1}$, $\beta_{2k+1}=s+ix_{2k+1}$. Integration over the
positions of $(N+1)(N-2)/2$ intermediate vertices is tacitly assumed.}
\label{Fig-SoV}
\end{figure}

The expressions \re{U-even} and \re{U-odd} can be represented as the ``Pyramide
du Louvre'' \cite{Louvre} diagram shown in Fig.~\ref{Fig-SoV}. This diagram
consists of $(N-1)-$rows, in accordance with the total number of
$\Lambda-$functions in Eqs.~\re{U-even} and \re{U-odd}. The $k-$th row of the
pyramid is built from the lines carrying the indices $\alpha_k$ and $\beta_k$
that depend on the separated coordinate $\vec x_k$ and are defined in
Fig.~\ref{Fig-SoV}. It is attached to the $(k+1)-$th row through
$(N-k)-$vertices. At $N=2$ the expression for the pyramid looks like
\ba
&&U_{\vec p,\vec x_1}(\vec{{z}})=|\vec p\,| \int d^2 w_2
\e^{2i\vec p\cdot \vec w_2}\,U_{\vec{x}_1}(\vec{{z}}; \vec w_2)\,,
\nonumber
\\
\quad
&&U_{\vec{x}_1}(\vec{{z}}; \vec w_2)=[z_1-w_2]^{s-ix_1}[z_2-w_2]^{s+ix_1}\,.
\label{U-N=2}
\ea
At $N=3$ the corresponding expression for $U_{\vec{x}_1,\vec x_2}(\vec{{z}}; \vec
w_3)$ is equal to the product of two (nonunique) star diagrams, which can be
expressed in terms of $\,{}_2F_1-$hypergeometric series. We do not present here its
explicit form, since it is more convenient to our purposes to use the integral
representation \re{U-gen}.

\subsection{Integration measure}

The transition functions $U_{\vec p,\vec{\mybf{x}}}(\vec{{z}})$ defined in
\re{U-gen} are the eigenstates of two mutually commuting hermitian operators,
$B(u)+\bar B(u^*)$ and $i(B(u)-\bar B(u^*))$, Eqs.~\re{B-U} and \re{B-bar-U}. As such,
they should be orthogonal to each other with respect to the
$SL(2,\mathbb{C})$ scalar product \re{SL2-norm}. The general form of their
orthogonality condition, consistent with the symmetry properties \re{U-sym}, is
given by \re{U-unit} and it involves the integration measure
$\mu(\vec{\mybf{x}})$. To find $\mu(\vec{\mybf{x}})$ we substitute the transition
function \re{U-gen} into the orthogonality condition \re{U-unit} and match the
result of integration into the r.h.s.\ of \re{U-unit}.

Let us evaluate the scalar product \re{U-unit} using the diagrammatical
representation of $U_{\vec p,\vec{\mybf{x}}}(\vec{{z}})$ shown in
Fig.~\ref{Fig-SoV}. The conjugated function $(U_{\vec
p,\vec{\mybf{x}}}(\vec{{z}}))^*$ is represented by the same pyramid diagram, in
which the exponents $\alpha_k$ and $\beta_k$ are replaced by the corresponding
conjugated exponents
\be
\alpha_k \to \bar \alpha_k^*=1- \alpha_k\,,\qquad \beta_k \to \bar {\beta_k}^*=1-
\beta_k\,.
\label{regul}
\ee
It becomes convenient to flip horizontally the conjugated pyramid diagram, so
that the point $\vec w_N$ will be located at the bottom of the diagram and the
points $\vec z_k$ at the top. The scalar product
\be
\vev{\vec{\mybf{x}},\vec
w_N|\vec{\mybf{x}}',\vec w_N'}=
\int d^2z\,
U_{\vec{\mybf{x}}}(\vec{{z}}; \vec w_N)\lr{U_{\vec{\mybf{x}}'}(\vec{{z}};
\vec w_N')}^*
\label{scal-prod-w}
\ee
is obtained by sewing the pyramid and its conjugated counterpart through the
points $\vec z_1,...,\vec z_N$. The resulting Feynman diagram has the form of a
big rhombus built out of $(N-1)^2-$elementary rhombuses (see
Fig.~\ref{Fig-measure} below) with the top and bottom vertices located at the
points $\vec w_N$ and $\vec w_N'$, respectively. Notice that the indices
$\alpha_k$ and $\beta_k$ in the upper and lower part of this rhombus depend on
the separated coordinates $\vec{\mybf{x}}$ and $\vec{\mybf{x}}'$, respectively.

\begin{figure}[t]
\centerline{{\epsfxsize5.0cm\epsfbox{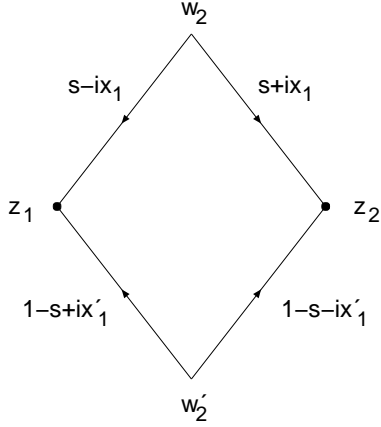}}}
\caption[]{Diagrammatical representation of the scalar product of two pyramids
at $N=2$.}
\label{Fig-sp-N=2}
\end{figure}

Let us first calculate \re{scal-prod-w} at $N=2$. Substituting \re{U-N=2} into
\re{scal-prod-w} one obtains the rhombus diagram shown in Fig.~\ref{Fig-sp-N=2}.
Integration over $\vec z_1$ and $\vec z_2$ can be easily performed using the
chain relation \re{chain-rule}, leading to the expression independent on $\vec
w_2$ and $\vec w_2'$, $\vev{\vec{x}_1,\vec w_2|\vec{x}_1',\vec w_2'}\sim |\vec
w_2-\vec w_2'|^0$. Its Fourier transformation with respect to $\vec w_2$ and
$\vec w_2'$ leads to a singular expression.\footnote{This has to do with the fact
that \re{scal-prod-w} should be understood as a distribution, that is one has to
integrate the r.h.s.\ of \re{scal-prod-w} with a smooth function of
$\vec{\mybf{x}}$ and perform the integration over $\vec z$ afterwards.} To
identify the corresponding generalized function, one has to regularize the
underlying two-dimensional Feynman integrals. The singularities appear due to the
fact that the sum of four indices, corresponding to the sides of the rhombus,
coincides with the space-dimension, $d_{\vec z} =2$. This suggests to regularize
the Feynman integrals by shifting the indices as
\be
(\alpha_k,\,\bar \alpha_k)\to  (\alpha_k -\varepsilon,\,
\bar\alpha_k -\varepsilon)\,,\qquad
(\beta_k,\,\bar \beta_k)\to  (\beta_k -\varepsilon,\,
\bar\beta_k -\varepsilon)
\label{shift-reg}
\ee
with $\varepsilon$ real positive. The sum of four indices inside the regularized
rhombus becomes $2-4\varepsilon$ and the Feynman integrals are well-defined.
Their calculation is based on the chain relation \re{chain-rule} and it leads to
\ba
\vev{\vec{x}_1,\vec w_2|\vec{x}_1',\vec w_2'}_{\varepsilon}=
\pi^2 [w_2-w_2']^{4\varepsilon}\hspace*{-6mm}&&\hspace*{-3mm}
a(s-ix_1-\varepsilon,1-s+ix_1'-\varepsilon,1+i(\bar x_1-\bar x_1')+2\varepsilon)
\nonumber
\\
\hspace*{-6mm}&\times&\hspace*{-3mm} a(s+ix_1-\varepsilon,1-s-ix_1'-\varepsilon,1-i(\bar
x_1-\bar x_1')+2\varepsilon)\,,
\label{reg-scal-prod}
\ea
where the subscript ${\varepsilon}$ in the l.h.s.\ indicates the regularization.
Going over to the momentum representation and using \re{Fourier}, we calculate
the Fourier transform of \re{reg-scal-prod} and obtain the expression involving
the factor $a(-4\varepsilon)=\Gamma(1+4\varepsilon)/\Gamma(-4\varepsilon)$ that
vanishes as $\varepsilon\to 0$. To get a nonzero result, the smallness of
$a(-4\varepsilon)$ should be compensated by one of the factors in
\re{reg-scal-prod}. Examining \re{reg-scal-prod} one finds that
this happens at $\vec x_1=\vec x_1'$ since
$a(1+2\varepsilon)=\Gamma(-2\varepsilon)/\Gamma(1+2\varepsilon)\sim
1/\varepsilon$.\footnote{There exist other possibilities to get the factor $\sim
1/\varepsilon$ by putting, for instance, $\bar x_1=i(1-\bar s)$ for $x_1\neq
-is$, but the corresponding values of the separated coordinates do not satisfy
the quantization condition \re{x-quan}.} Carefully examining the limit
$\varepsilon\to 0$ one obtains
\be
\lim_{\varepsilon\to 0}a(-4\varepsilon,1+i(\bar x_1-\bar x_1')+2\varepsilon,1-i(\bar x_1-\bar
x_1')+2\varepsilon) = 2\pi
\delta_{n_1,n_1'}\delta(\nu_1-\nu_1')\equiv 2\pi\delta(\mybf{x}_1-\mybf{x}_1')\,,
\label{reg-delta}
\ee
where $x_1=\nu_1-in_1/2$ and $x_1'=\nu_1'-in_1'/2$. Finally, combining together
Eqs.~\re{reg-scal-prod} and \re{reg-delta}, we obtain the orthogonality condition
at $N=2$
\be
\vev{\vec {\mybf{x}}_1',\vec p'\,|\vec{\mybf{x}}_1,\vec p\,}=\lim_{\varepsilon\to 0}{\vec p\,}^2
\int d^2w_2\,d^2w_2' \e^{2i\vec
w_2\vec p-2i\vec w_2'\vec p'}\vev{\vec{x}_1,\vec w_2|\vec{x}_1',\vec
w_2'}_{\varepsilon} = (2\pi)^2 \delta^{(2)}(p-p')
\delta(\mybf{x}_1-\mybf{x}_1')\frac{\pi^4}2
\,.
\ee
Its comparison with \re{U-unit} yields the expression for the measure at $N=2$
\be
\mu(\mybf{x}_1)={2/\pi^{4}}\,.
\ee

\begin{figure}[t]
\centerline{{\epsfxsize17.0cm\epsfbox{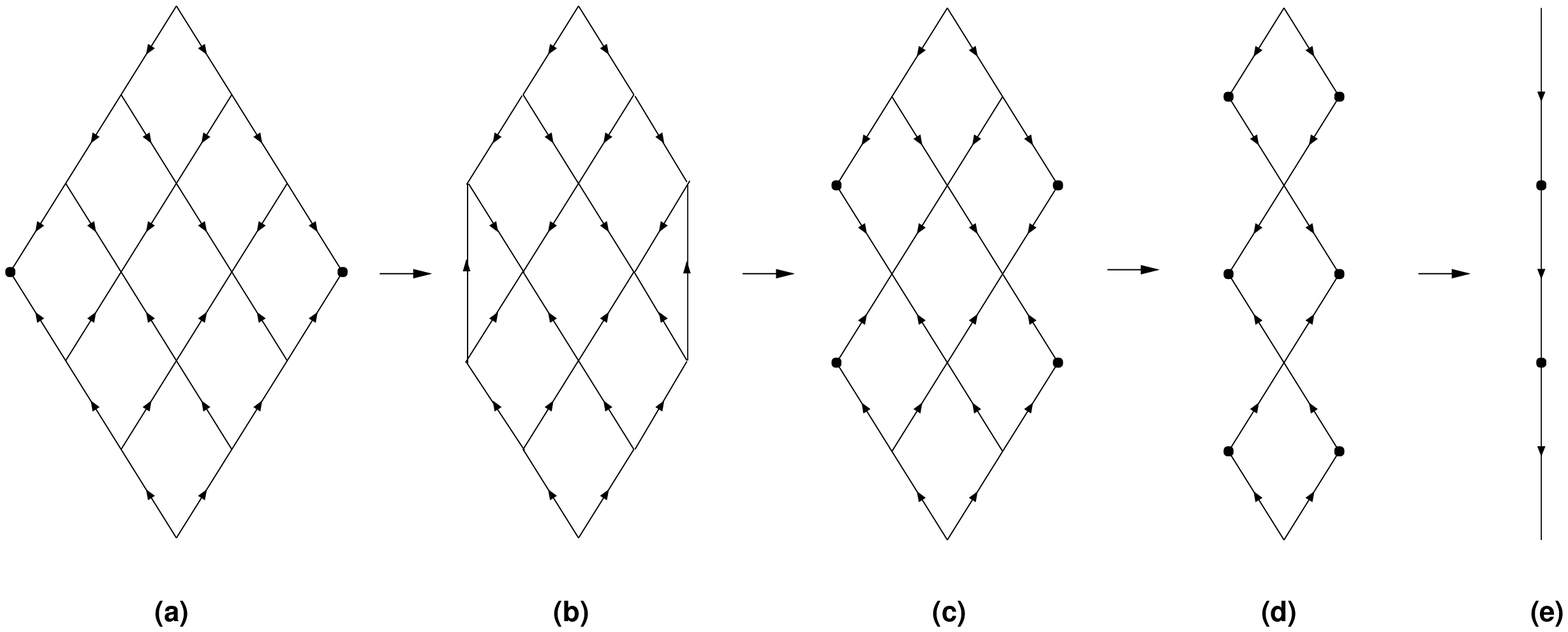}}}
\caption[]{Diagrammatical calculation of the scalar product of two pyramids. The fat
points indicate the vertices that can be integrated out using the chain relation
shown in Fig.~\ref{chain}.}
\label{Fig-measure}
\end{figure}

The calculation of the scalar product at higher $N$ goes along the same lines and
is shown schematically in Fig.~\ref{Fig-measure}. Examining the rhombus diagram
shown in Fig.~\ref{Fig-measure}a, one notices that all sewing vertices $z_2, ...,
z_{N-1}$, except the left- and rightmost ones, $z_1$ and $z_N$, respectively,
have four lines attached to them. At the same time, only two lines are joined at
the points $z_1$ and $z_N$. The integration over these two points can be
performed using the chain relation, similar to the previous case, producing two
vertical lines with the indices $\pm i(x_1-x_1')$ (see Fig.~\ref{Fig-measure}b),
which can be moved horizontally through the diagram in the direction of each
other, thanks to the permutation identity shown in Fig.~\ref{perm}, until they
meet and annihilate each other. In this way, one arrives at the diagram shown in
Fig.~\ref{Fig-measure}c. It differs from Fig.~\ref{Fig-measure}a in that two
sewing vertices $\vec z_1$ and $\vec z_N$ disappeared and the spectral parameters
$\vec x_1$ and $\vec x_1'$ were interchanged. We notice that in this diagram
there are already four vertices which can be integrated out using the chain
relation \re{chain-rule}. Further simplification amounts to repetition of the
steps that we just described. Each next step reduces the number of vertices in
the diagram and interchanges the spectral parameters $\vec x_k
\leftrightarrow {\vec x_j}^\prime$. Continuing this procedure, one obtains
the diagram shown in Fig.~\ref{Fig-measure}d, in which $(N-1)$ elementary
rhombuses are aligned along the vertical axis and are attached to their neighbors
through the common vertices. Each of these rhombuses corresponds to the Feynman
diagram in Fig.~\ref{Fig-sp-N=2} defining the scalar product at $N=2$,
Eq.~\re{reg-scal-prod}. The only difference with the previous case is that the
separated variables parameterizing the $k-$th rhombus from the top are given by
$\vec x_{N-k}$ and $\vec x_k'$ instead of $\vec x_1$ and $\vec x_1'$. We expect
that the chain of rhombuses should produce the contribution $\sim
\prod_{k=1}^N\delta(\vec x_{N-k}-\vec x_k')$. To calculate it carefully, one has
to regularize the Feynman integrals according to \re{regul}. Applying
\re{reg-scal-prod}, we replace each rhombus by a single line with the index
$-4\varepsilon$ (see Fig.~\ref{Fig-measure}e) and integrate over their
$(N-2)-$common vertices using the chain relation \re{chain-rule}. Going over to
the limit $\varepsilon\to 0$ and taking into account the identity \re{reg-delta},
we combine together different factors coming from \re{c-N} and
\re{reg-scal-prod}, perform the Fourier transformation
\re{U-gen} and finally obtain the following relation
\ba
\lim_{\varepsilon\to 0}\,\vev{\vec{\mybf{x}},\vec
p\,|\vec{\mybf{x}}',\vec p'}_\varepsilon &=& (2\pi)^N\,\delta^{(2)}( p- p')\,
\delta^{(2)}( x_1- x_{N-1}')\delta^{(2)}( x_2- x_{N-2}') \ldots\delta^{(2)}( x_{N-1}- x_1')
\nonumber
\\
&\times& \frac12{\pi^{N^2}}\prod_{j,k=1\atop j>k}^{N-1}a(1+i(x_k-x_j),1-i(\bar
x_k-\bar x_j))\,.
\label{normP}
\ea
In arriving at this relation we integrated out, using the chain relation
\re{chain-rule}, all vertices of the pyramid except the top and bottom vertices.
Each integrated vertex brought the factor $\sim\pi a(1+i(x_k-x_{N-j}'))$ and
their product appeared in the second line in \re{normP}.

Comparing \re{normP} with the general expression for the scalar product,
Eq.~\re{U-unit}, we notice that it is not symmetric under permutations of the
separated variables $\vec x_k\leftrightarrow \vec x_j$. The reason for this is
that performing the calculation of \re{normP} we have tacitly assumed that $\vec
x_1$ is different from $\vec x_1^{\,\prime}$, $\vec x_2^{\,\prime}$,..., $\vec
x_{N-2}^{\,\prime}$, thus eliminating all terms in \re{U-unit} except the one
entering \re{normP}. Indeed, as we have seen in the case of the $N=2$ scalar
product, the limit $x_1\to x_1'$ requires the regularization of the exponents
\re{shift-reg}. Although this regularization makes all Feynman integrals well defined, it
transforms the unique Feynman diagrams (stars and triangles) into nonunique ones
and, therefore, it does not allow to apply the permutation identity (see
Fig.~\ref{perm}). Since in our calculation we applied the permutation identity to
all pairs of the separated coordinates $(\vec x_1,\vec x_{N-k}')$, $k=2,...,N-1$,
except $(\vec x_1,\vec x_{N-1}')$ we have to assume that $|\vec x_1-\vec
x_{N-k}'|\neq 0$.

Matching \re{normP} into \re{U-unit} we obtain the following expression for the
integration measure
\be
\mu(\vec{\mybf{x}})=\frac{2\pi^{-N^2}}{(N-1)!}\prod_{j,k=1\atop j>k}^{N-1}|a(1+i(x_k-x_j)|^{-2}
=\frac{2\pi^{-N^2}}{(N-1)!}\prod_{j,k=1\atop j>k}^{N-1}{|\vec x_k-\vec x_j|^2}\,,
\label{measure-x}
\ee
which is valid for $N\ge 3$. Here, $|\vec x_k-\vec x_j|^2=(x_k-x_j)(\bar x_k-\bar
x_j)$ and the relations \re{a-a} and \re{x-real} have been taking into
account. Replacing $\vec x_k$ by their quantized values \re{x-quan} one gets
\be
\mu(\vec{\mybf{x}})=\frac{2\pi^{-N^2}}{(N-1)!}\prod_{j,k=1\atop j>k}^{N-1}
\left[(\nu_k-\nu_j)^2+\frac14(n_k-n_j)^2\right]\,.
\label{measure-exp}
\ee
We verify that the obtained expression for the integration measure satisfies the
relation \re{eq-measure}.

Having determined the integration measure in the SoV representation, we can use
\re{U-unit} to write the completeness condition
\be
\int d^2 p \int d \vec{\mybf{x}}\,\mu(\vec{\mybf{x}}) \,
\left(U_{\vec p,\vec {\mybf{x}}}(\vec z_1',...,\vec z_N')\right)^*\,
U_{\vec p,\vec{\mybf{x}}}(\vec z_1,...,\vec z_N) =(2\pi)^N\,
\delta^{(2)}({z}-{z'})\,,
\label{U-comp}
\ee
where the integration over quantized values of the separated variables, $\int d
\vec{\mybf{x}}$, implies the summation over integer $n_k$ and integration
over continuous $\nu_k$ defined in \re{x-quan}
\be
\int d \vec{\mybf{x}}=\prod_{k=1}^{N-1}
\lr{\sum_{n_k=-\infty}^\infty\int_{-\infty}^\infty d\nu_k}\,.
\label{int-dx}
\ee

Since the eigenstates $\Psi_{\vec p,\{q,\bar q\}}(\vec z)$ are orthogonal to each
other with respect to the $SL(2,\mathbb{C})$ scalar product \re{SL2-norm}, the
same should be true for the functions $\Phi_{\{q,\bar q\}}(\vec x_1,...,\vec
x_{N-1})$ in the SoV representation. To obtain the corresponding orthogonality
condition one substitutes \re{SoV-gen} into \re{SL2-norm} and uses
Eq.~\re{U-unit}. In this way one arrives at the Plancherel formula
\be
\vev{\Psi_{\vec p,\{q,\bar q\}}|\Psi_{\vec p\,',\{q',\bar q'\}}}
=(2\pi)^N\delta(\vec p-\vec p\,') \vev{\Phi_{\{q,\bar q\}}|\Phi_{\{q',\bar q'\}}}
=(2\pi)^N\delta(\vec p-\vec p\,')\delta(\mybf{q}-\mybf{q'})\,,
\ee
where the notation was introduced for the norm of states in the SoV
representation
\be
\vev{\Phi_{\{q,\bar q\}}|\Phi_{\{q',\bar q'\}}}
\equiv\int d\mybf{x}\,\mu(\mybf{x})\,
\lr{\Phi_{\{q,\bar q\}}(\vec x_1,...,\vec
x_{N-1})}^*\Phi_{\{q',\bar q'\}}(\vec x_1,...,\vec
x_{N-1})=\delta(\mybf{q}-\mybf{q'})\,.
\label{Phi-norm}
\ee
Here, $\Phi_{\{q,\bar q\}}(\vec x_1,...,\vec x_{N-1})$ is a completely symmetric
function of the separated variables $\vec x_k$, satisfying the multi-dimensional
Baxter equations, Eq.~\re{multi-B}, in the holomorphic and antiholomorphic sectors. As we will show
in Sect.~4.4, the solution of these equations is factorized into the product of
the eigenvalues of the $\mathbb{Q}-$operators that have been studied in Sect.~3.
Therefore, substitution of \re{prod-Q} into \re{Phi-norm} yields the
orthogonality condition for the solutions of the Baxter equation. Its explicit
form depends on the normalization factor present in the r.h.s.\ of \re{prod-Q}
and it will be established in the next Section.

\subsection{Relation to the Baxter $\mathbb{Q}-$operator}

The construction of the SoV representation, performed in Sect.~4.2, is similar in
many respects to that of the Baxter $\mathbb{Q}-$operators in Sect.~3. In
addition, the eigenfunctions in the SoV representation, $\Phi_{\{q,\bar q\}}(\vec
x_1,...,\vec x_{N-1})$, and the eigenvalues of the Baxter $\mathbb{Q}-$operators
satisfy similar finite-difference Baxter equations. This suggests \cite{KS} that there
should exist the relation between two different objects -- the transition
function to the SoV representation, $U_{\vec p,\vec{\mybf{x}}}(\vec{{z}})$,
defined in \re{U-gen}, and the Baxter operators $\mathbb{Q}_\pm(u,\bar u)$ given
by Eqs.~\re{kern+} and \re{kern-}. In this Section, we shall establish such
relation (see Eq.~\re{SoV-Bax} below) and use it to prove \re{prod-Q}.

\begin{figure}[t]
\centerline{{\epsfysize4.9cm\epsfbox{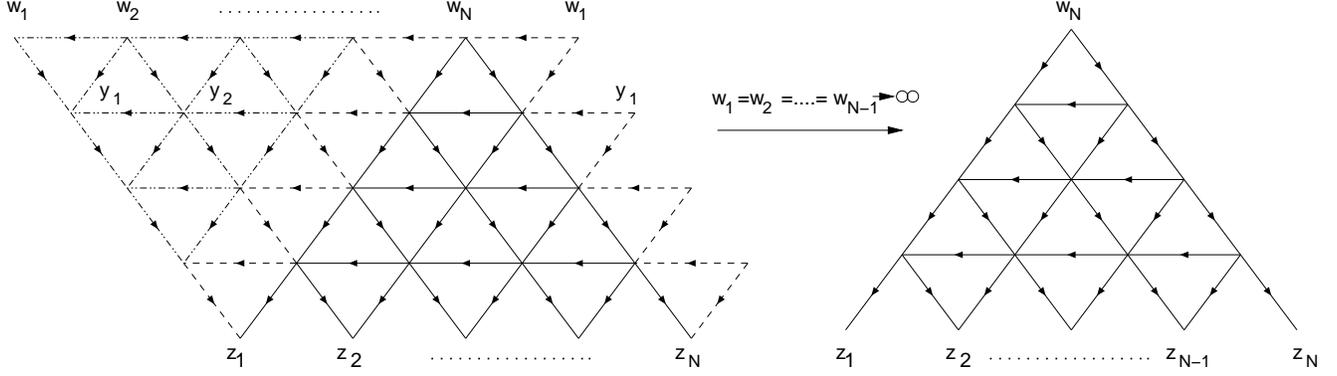}}}
\caption[]{Reduction of the product of $(N-1)$ $\mathbb{Q}-$operators, Eq.~\re{U-Q},
into the pyramid diagram. The $k-$th row of triangles in the left diagram is
given by Fig.~\ref{Q2-f} for $u=x_k$. The triangles shown by the dash-dotted
lines collapse into the $\delta-$functions at $\vec w_1=\vec w_2=...=\vec
w_{N-1}$ due to the reduction formulae. The triangles shown by the dashed lines
are replaced by a power of $[w_1]$ as $\vec w_1\to\infty$. }
\label{Fig-reduction}
\end{figure}

To start with, let us consider the product of $N-1$ Baxter operators
$\mathbb{Q}_-(u,\bar u)$
\ba
Q_{\vec x_1,...,\vec x_{N-1}}(\vec z\,|\,\vec w)&=&
\left[\mathbb{Q}_{-} (x_1,\bar x_1)\ldots
\mathbb{Q}_{-}(x_{N-1},\bar x_{N-1})\right](\vec z\,|\,\vec w)
\nonumber
\\
&=&\int d^2 \vec y ...\, d^2 \vec y\,'\, Q^{(-)}_{x_1,\bar x_1}(\vec z\,|\,\vec
y\,')\ldots Q^{(-)}_{x_{N-1},\bar x_{N-1}}(\vec y\,|\,\vec w)\,,
\label{U-Q}
\ea
where $\vec y=(\vec y_1,...,\vec y_N)$ and the kernel $Q^{_{(-)}}_{u,\bar u}(\vec
z_1,...,\vec z_N\,|\,\vec w_1,...,\vec w_N)$ was defined in \re{kern-}. Using the
diagrammatical representation of the $Q^{_{(-)}}_{x_1,\bar x_1}(\vec z\,|\,\vec
y\,')$, the left diagram in Fig.~\ref{Q2-f}, we associate \re{U-Q} with the
Feynman diagram, in which $(N-1)-$periodic chains of the unique triangles (one
for each $\mathbb{Q}_-$) are glued together through their common vertices as
shown in the left diagram in Fig.~\ref{Fig-reduction}. Comparing this diagram
with the one shown in Fig.~\ref{Fig-SoV}, we notice their similarity -- the
lowest row defining the $\vec z-$dependence look alike, as well as their
difference -- the number of vertices is higher in the former diagram; the latter
diagram does not have periodic boundary conditions along each row. In addition,
we notice that $Q_{\vec{\mybf{x}}}(\vec{{z}}\,|\,\vec w_1,...,\vec w_{N-1},\vec
w_N)$ depends on the $(N-1)-$additional arbitrary vectors $\vec w_1,...,\vec
w_{N-1}$ as compared to $U_{\vec{\mybf{x}}}(\vec{{z}}; \vec w_N)$, defined in
\re{U-even} and \re{U-odd}. As we will see in a moment, in order to obtain the
transition function $U_{\vec{\mybf{x}}}(\vec{{z}}; \vec w_N)$ from \re{U-Q},
these vectors have to be chosen in a particular way -- one has to put $\vec
w_1=\vec w_2=...=\vec w_{N-1}$ and take the limit $w_1\to\infty$ afterwards
\be
U_{\vec p,\vec{\mybf{x}}}(\vec{{z}}) = c_{\vec p} \lim_{\vec
w_1\to\infty}\,[w_1]^{2(N-1)(1-s)} \int d^2 w_N \, \e^{i p w_N + i\bar p \bar
w_N} Q_{\vec{\mybf{x}}}(\vec{{z}}\,|\,\vec w_1,...,\vec w_{1},\vec w_N)\,.
\label{SoV-Bax}
\ee
Here, $c_{\vec p}$ is the normalization factor depending only on the momentum
$\vec p$.

The proof of \re{SoV-Bax} is based on the reduction formulae obtained in
Sect.~3.3.4 and it goes as follows. The dependence of $Q_{\vec x_1,...,\vec
x_{N-1}}(\vec z\,|\,\vec w)$ on $\vec w_1,...,\vec w_{N-1}$ is carried by the
rightmost $Q-$kernel in \re{U-Q}, or equivalently by the $N$ unique triangles in
the upper row in Fig.~\ref{Fig-reduction}. At $\vec w_1=\vec w_2=...=\vec
w_{N-1}$, $(N-2)$ of these triangles shrink into $\delta-$functions (see
Eq.~\re{uniq-deg}), leading to significant simplification of the corresponding
kernel. Indeed, systematically applying the reduction formula \re{Q-red-minus}
one finds that $Q^{_{(-)}}_{x_{N-1},\bar x_{N-1}}(\vec y\,|\,\vec w)\sim
\delta(\vec w_1-\vec y_1)\delta(\vec w_1-\vec y_2)...\delta(\vec w_1-\vec
y_{N-2})$. Substituting this expression into \re{U-Q}, one finds that the
adjacent kernel $Q^{_{(-)}}_{x_{N-2},\bar x_{N-2}}(\vec y\,''\,|\,\vec y)$ is
evaluated at $\vec y_1=\vec y_2=...=\vec y_{N-2}$ and the same reduction formulae
can be applied again reducing the number of unique triangles in the row next to
the upper one by $N-3$. Repeating this procedure one finds that the substitution
$\vec w_1=\vec w_2=...=\vec w_{N-1}$ in the upper, $(N-1)-$th row of the diagram,
creates an avalanche of simplifications throughout the whole diagram, in which
$(k-1)$ unique triangles disappear in the $k-$th row $(k=2,3,...,N-1)$ as shown
in Fig.~\ref{Fig-reduction}. The resulting diagram still depends on the arbitrary
vector $\vec w_1$, which defines the position of the vertex of the right- and
leftmost unique triangles in each row. In the limit $\vec w_1\to\infty$, these
triangles scale as a power of $[w_1]$, so that each row of the diagram
contributes the factor $[w_1]^{-2(1-s)}$. Combining together the factors coming
from $(N-1)$ rows of the diagram, one finds that the r.h.s.\ of \re{SoV-Bax} has
a finite limit as $\vec w_1\to\infty$. The corresponding Feynman diagram has the
form shown in the r.h.s.\ of Fig.~\ref{Fig-reduction}. Still, it differs from the
pyramid diagram, Fig.~\ref{Fig-SoV}, by additional horizontal lines with the
indices $(2s-1)$.

To demonstrate the equivalence between two different diagrams,
Figs.~\ref{Fig-reduction} and \ref{Fig-SoV}, we apply the same trick as was used
in Sect.~3.2 in establishing the commutativity condition for the Baxter
$\mathbb{Q}-$operators (see Fig.~\ref{QQ}). Namely, we start with the diagram in
Fig.~\ref{Fig-SoV} and insert two {\it horizontal\/} lines with the opposite
indices, $(2s-1)$ and $(1-2s)$, into all rhombuses with the indices
$(\alpha_1,\beta_1,\alpha_2,\beta_2)$, in the lower part of the diagram. Applying
the permutation identity, Fig.~\ref{perm}, we move $(N-2)$ auxiliary lines with
the indices $(2s-1)$ vertically to the upper part of the diagram until they
either reach the boundary of the pyramid and disappear in virtue of the
permutation identity, Fig.~\ref{Fig-perm-inf}, or end up in the rhombus with the
indices $(\alpha_{N-2},\beta_{N-2},\alpha_{N-1},\beta_{N-1})$ just below the top
vertex $\vec w_N$. In the latter case, one gets rid of the auxiliary line by
applying the star-triangle relation to the unique triangle with the indices
$(\alpha_{N-1},\beta_{N-1},2s-1)$. Up to the additional line with the index
$2(1-s)$ attached to the top of the pyramid, the resulting diagram looks like the
original pyramid with one row less. The latter has been replaced by the lowest
row of the diagram in Fig.~\ref{Fig-reduction}. Applying the same transformations
to the reduced pyramid with $(N-2)-$rows, we reconstruct one more layer of the
diagram Fig.~\ref{Fig-reduction}. We continue this procedure iteratively $N-3$
times until we arrive at the reduced pyramid containing a single rhombus.
Inserting, as before, two lines with the indices $(1-2s)$ and $(2s-1)$ into the
rhombus, we get rid of the latter line by applying the uniqueness relation to the
triangle containing the top of the pyramid. The resulting diagram coincides with
the right diagram in Fig.~\ref{Fig-reduction}, in which the chain of the
additional lines with the index $2(1-s)$ is attached to its top vertex $\vec
w_N$. The total number of lines in this chain, $[(N-1)/2]$, is equal to the
number of rhombuses in the original pyramid, in which two opposite vertices are
located on the vertical axis going through the top vertex $w_N$. Going over to
the momentum representation, the whole chain is reduced to a power of the momenta
$((-1)^{n_s}p^{1-2s}\bar p^{1-2\bar s})^{[(N-1)/2]}$ and one recover the
relation \re{SoV-Bax}. Carefully collecting all factors and comparing the
expression for the r.h.s.\ of \re{SoV-Bax} with Eq.~\re{U-gen}, we calculate the
normalization constant in \re{SoV-Bax} as
\be
c_{\vec p}= \lr{(-1)^{n_s}p^{1-2s}\bar p^{1-2\bar s}}^{[(N-1)/2]}\left[\pi
a(2(1-s))\right]^{-(N-1)(N-2)/2}\,.
\label{c-p}
\ee
It comes about from applying the reduction formulae \re{Q-red-minus} to
$(N-1)(N-2)/2$ collapsing unique triangles in Fig.~\ref{Fig-reduction} and
compensates the difference of the scaling dimensions in the both sides of the
relation \re{SoV-Bax}.

The relation \re{SoV-Bax} between $U_{\vec p,\vec{\mybf{x}}}(\vec{z})$ and the
Baxter $\mathbb{Q}-$operator can be rewritten in the operator form by introducing
the special state $\ket{\omega_{\vec z_0,\vec p}\,}$, which depends on two
arbitrary vectors, $\vec z_0$ and $\vec p$, and is defined in the coordinate
representation as follows \footnote{Notice that this state is not normalizable
with respect to the $SL(2,\mathbb{C})$ scalar product \re{SL2-norm} and,
therefore, it does {\it not\/} belong to the quantum space of the system. Keeping
this in mind, we shall use the same scalar product notation for the convolution
of the wave function $\omega_{\vec z_0,\vec p}(\vec z)$ with an arbitrary test
function, $\vev{\Psi|\omega_{\vec z_0,\vec p}\,}=\int d^2{z}
(\Psi(\vec{{z}}))^*\omega_{\vec z_0,\vec p}(\vec z)$, assuming that the integral
is convergent.}
\be
\vev{{\vec{z}}|\omega_{\vec z_0,\vec p}\,}=c_{\vec p}\,[z_0]^{2(N-1)(1-s)}
\e^{ipz_N +i \bar p\bar z_N}\,\prod_{k=1}^{N-1}{\delta(\vec z_k-\vec z_0)}\,.
\label{omega}
\ee
Using this state, one finds from \re{SoV-Bax}
\be
U_{\vec p,\vec{\mybf{x}}}(\vec{{z}} )=\lim_{\vec z_0\to\infty}
\vev{{\vec{z}}\,|\,\mathbb{Q}_{-} (x_1,\bar x_1)\ldots
\mathbb{Q}_{-}(x_{N-1},\bar x_{N-1})\,|\,\omega_{\vec z_0,\vec p}\,}\,.
\label{U-Q-ops}
\ee

Many properties of the SoV representation follow from this relation. In
particular, the symmetry property \re{U-sym} comes from the commutativity of the
$\mathbb{Q}-$operators. Substituting \re{U-Q-ops} into \re{Phi-def}, we take into
account that the eigenstates $\Psi_{\vec p,\{q,\bar q\}}(\vec z)$ diagonalize the
operator $\mathbb{Q}_-(u,\bar u)$
\ba
\Phi_{\{q,\bar q\}}(\vec{\mybf{x}})\delta(\vec p-\vec p\,')
&=&\vev{\Psi_{\vec p,\{q,\bar q\}}|\,\mathbb{Q}_{-} (x_1,\bar x_1)\ldots
\mathbb{Q}_{-}(x_{N-1},\bar x_{N-1})\,|\,\omega_{\vec z_0=\infty,\vec p\,'}\,}
\nonumber
\\
&=&c_{\Phi}\,\delta^{(2)}(p-p\,')\prod_{k=1}^{N-1} Q^{(-)}_{q,\bar q}(x_k,\bar
x_k)\,, \label{Phi-Q-min}
\ea
where the normalization factor is defined as $c_{\Phi}=\int d^2 p
\,\vev{\Psi_{\vec p,\{q,\bar q\}}|\,\omega_{\vec z_0=\infty,\vec p\,'}\,}$ and
$Q^{_{(-)}}_{q,\bar q}(u,\bar u)$ stands for the eigenvalue of the operator
$\mathbb{Q}_-(u,\bar u)$. This relation replaces the ansatz \re{prod-Q} and it
provides the expression for $\Phi_{\{q,\bar q\}}(\vec{\mybf{x}})$ that
automatically satisfies the multi-dimensional Baxter equation \re{multi-B}.
Applying the conjugation relations \re{Q+dagger}, one expresses the wave function
in the separated coordinates, $\Phi_{\{q,\bar q\}}(\vec{\mybf{x}})$ in terms of
the eigenvalues $Q_{q,\bar q}\equiv Q^{_{(+)}}_{q,\bar q}(u,\bar u)$ (up to
irrelevant normalization factor $c_{\Phi} [a(1-2s)/\pi]^{N(N-1)}$)  as
\be
\Phi_{\{q,\bar q\}}(\vec{\mybf{x}})= \prod_{k=1}^{N-1} [a(s+ix_k,\bar s-i\bar
x_k)]^N \,\lr{Q_{q,\bar q}(x_k,\bar x_k)}^*\,.
\label{Phi-Q-ops}
\ee
We recall that the possible values of the separated coordinates,
$x_k=\nu_k-in_k/2$ and $\bar x_k=x_k^*$, are parameterized by integer $n_k$ and
real $\nu_k$, Eq.~\re{x-quan}. The explicit form of $\Phi_{\{q,\bar
q\}}(\vec{\mybf{x}})$ at $N=2$ can be found from \re{Q-2-2}.

The properties of  $\Phi_{\{q,\bar q\}}(\vec{\mybf{x}})$ can be established using
the results for the functions $Q_{q,\bar q}(u,\bar u)$ obtained in Sect.~3. The
analytical properties of $\Phi_{\{q,\bar q\}}(\vec{\mybf{x}})$ on the complex
$\nu_k-$plane follow from Eq.~\re{poles-exp} and Fig.~\ref{Fig-poles}. In
particular, $\Phi_{\{q,\bar q\}}(\vec{\mybf{x}})$ does have any singularities on
the real $\nu_k-$axis and, applying \re{Q-asym-eig}, one finds its the asymptotic
behavior at large $\nu_k$ and fixed $n_k$ as $\Phi_{\{q,\bar
q\}}(\vec{\mybf{x}})=\CO(\nu_k^{_{1-N(1+2i\nu_s)}})$, leading to
$\mu(\mybf{x})|\Phi_{\{q,\bar q\}}(\vec{\mybf{x}})|^2\sim 1/\nu_k^2$. These
properties are in accord with the fact that $\Phi_{\{q,\bar q\}}(\vec{\mybf{x}})$
should be normalizable with respect to the scalar product \re{Phi-norm}.

Substituting \re{Phi-Q-ops} into \re{SoV-gen}, we obtain the integral
representation of the eigenstates of the system, $\Psi_{\vec p,\{q,\bar q\}}(\vec
z)$. As shown in the Appendix C,  this representation can be rewritten, thanks to
remarkable properties of the transition function $U_{\vec
p,\vec{\mybf{x}}}(\vec{{z}})$, in a concise form \cite{FK}
\be
{\Psi_{\vec p,\{q,\bar q\}}}(\vec z) = \prod_{k=1}^{N-1}Q_{q,\bar q}(\widehat
x_k,\widehat{\bar x}_k)\ {\Omega_{\vec p}}(\vec z)\,,
\label{Psi-ABA}
\ee
where
\be
{\Omega_{\vec p}}(\vec z)=\int d^2 z_0\,\e^{2i\vec p\cdot\vec z_0}\,
\prod_{k=1}^N {(z_k-z_0)^{-2s}(\bar z_k-\bar z_0)^{-2\bar s}}
\ee
is the so-called pseudovacuum state and $Q_{q,\bar q}(\hat x_k,\hat{\bar x}_k)$
stands for the eigenvalue of the Baxter operator $\mathbb{Q}_+(u,\bar u)$ with
the spectral parameters substituted by the operator zeros. The representation
\re{Psi-ABA} is well-known in the theory of the $SL(2,\mathbb{R})$ spin magnets.
There, the relevant solutions to the Baxter equation are given by polynomials in
the spectral parameters and \re{Psi-ABA} becomes equivalent to the highest-weight
representation for the eigenstates in the ABA approach. Eq.~\re{Psi-ABA} provides
a generalization of the above representation to the noncompact $SL(2,\mathbb{C})$
magnets, for which the ABA is not applicable and the Baxter equations have
nonpolynomial solutions for $Q_{q,\bar q}(u,\bar u)$.

As a byproduct of \re{Phi-Q-ops}, we substitute the relation \re{Phi-Q-ops} into
the scalar product \re{Phi-norm} and obtain the orthogonality condition on the
space of solutions to the Baxter equation
\be
\int d\mybf{x}\,\mu(\mybf{x})\,\prod_{k=1}^{N-1}Q_{q',\bar q'}(x_k,\bar x_k)\,
\lr{Q_{q,\bar q}(x_k,\bar x_k)}^*=\delta(\mybf{q}-\mybf{q'})\,,
\ee
where the integration measure $\int d\mybf{x}\,\mu(\mybf{x})$ was defined in
\re{measure-x} and \re{int-dx}.

\section{Conclusions}

In this paper, we have studied completely integrable quantum
mechanical model of $N$ interacting spinning particles in two-dimensional space
that has previously
appeared in high-energy QCD as describing the properties of the $N-$gluon
compound states in the multi-color limit. Applying the quantum inverse scattering
method, we identified this model as a generalization of one-dimensional
homogenous ${\rm XXX}$ Heisenberg spin magnet to infinite-dimensional principal
series representation of the $SL(2,\mathbb{C})$ group.

The model has a number of properties in common with two-dimensional conformal
field theories \cite{BPZ}. Introducing holomorphic and antiholomorphic
coordinates on a two-dimensional plane one finds that the the kernels of the
$R-$matrix and the Baxter $Q-$operator are factorized into the product of two
functions depending separately on the holomorphic and antiholomorphic
coordinates, while the Hamiltonian is given by the sum of two commuting operators
acting in the different sectors. In spite of this, the Hamiltonian dynamics of
the model can not be reduced to independent dynamics in the holomorphic and
antiholomorphic sectors. The interaction between two sectors is ensured by the
condition for the kernels of the operators and their eigenstates to be
well-defined functions on a two-dimensional plane. This condition leads to the
additional constraints on the properties of the model. In particular, it leads to
the quantization of the integrals of motion of the system, the separated
coordinates as well as the spectral parameters of the $R-$matrix and the Baxter
$Q-$operator. Using the global $SL(2,\mathbb{C})$ invariance of the model, one
can represent the eigenstates of the $N-$particle Hamiltonian, $\Psi_{q,\bar
q}(\vec z_1-\vec z_0,\vec z_2-\vec z_0, ...,\vec z_N-\vec z_0)$, as correlators
of $N+1$ quasi-primary fields in a two-dimensional conformal field theory. The
latter are given by the sum of products of the holomorphic and antiholomorphic
conformal blocks. This structure is general enough and, as was shown in
Sect.~3.3.5, it also holds for the eigenvalues of the Baxter $Q-$operator. This,
in turn, suggests that the Baxter $Q-$operators constructed in this paper should
be related to similar operators defined in the conformal field theory \cite{BLZ}.

We have argued in this paper that all integrability properties of the model are
encoded in the properties of the $Q-$operator, satisfying the Baxter relations.
Namely, the Hamiltonian as well as transfer matrices are expressed in terms of a
single operator $\mathbb{Q}(u,\bar u)$ and, therefore, their spectrum is
determined by the eigenvalues of the $Q-$operator. In addition, going over
through Sklyanin's construction of the Separated Variables we have demonstrated
that the unitary transformation to these variables is also determined by the same
operator. This leads to the integral representation for the eigenstates of the
model with the wave function in the separated coordinates given by the product of
the eigenvalues of the $Q-$operator.

Thus, the spectral problem for the noncompact $SL(2,\mathbb{C})$ spin magnet
turns out to be equivalent to the problem of finding the eigenvalues of the
Baxter $Q-$operator. The latter are defined as solutions to the Baxter equations,
\re{def-Qh} and \re{def-Qa}, supplemented by additional conditions on their analytical
properties and asymptotic behavior at infinity that have been established in
Sects.~3.5 and 3.6, respectively. The solution satisfying these conditions was
found in the simplest case of the system of $N=2$ particles. Constructing
solutions to the Baxter equation for $N\ge 3$, we find that in order to satisfy
the above conditions the values of the integrals of motion have to be
constrained. The explicit form of the corresponding quantization conditions is
known only for the polynomial solutions of the Baxter equation \cite{K1,K2,DKM}.
Their generalization to arbitrary, nonpolynomial solutions will be presented
elsewhere \cite{DKM-s}.

\section*{Acknowledgements}

This work was supported by the grant of Spanish Ministry of Science (A.M.)
and by the grant 00-01-005-00 of the Russian Foundation for Fundamental Research
(A.M. and S.D.).

\section*{Note added}

When this paper was written, we learned about the recent work by H.~J.~de~Vega
and L.N.~Lipatov, in which similar results were obtained -- the unitary
transformation to the separated variables was constructed, using different
approach, for the $SL(2,\mathbb{C})$ magnet of spin $(s=1,\bar s=0)$ with $N=2$
and $N=3$ sites; the explicit solution to the Baxter equation was found at $N=2$
and its generalization to higher $N$ was discussed.

\appendix
\renewcommand{\theequation}{\Alph{section}.\arabic{equation}}
\setcounter{table}{0}
\renewcommand{\thetable}{\Alph{table}}

\section{Appendix: Uniqueness relations}

Analyzing the noncompact spin magnet we encounter various two-dimensional
integrals over the positions of particles which we represent in terms of the
Feynman diagrams. Such interpretation turns out to be very useful as it allows to
apply powerful methods of calculation the Feynman integrals well-known in
perturbation QCD. More precisely, our calculations are heavily based on the
``method of uniqueness'' \cite{uniq,uniq1}. This method has been originally
developed for the calculation of $d-$dimensional Feynman integral and we shall
use its simplified version at $d=2$. In this Appendix we collect all necessary
formulae - the chain and the star-triangle relations, which form the basis of the
method of uniqueness. Remarkable feature of the method is that subsequently
applying these two relations one is able to obtain analytical expressions for
complicated Feynman integrals without doing any integration.

Throughout the paper we use the following compact notation for the ``propagator''
\be
\label{prop}
\frac{1}{[z-w]^\alpha}\equiv
\frac{1}{(z-w)^\alpha (\bar z-\bar w)^{\bar \alpha}}=
\frac{(\bar z-\bar w)^{\alpha -\bar \alpha}}{|z-w|^{2\alpha}}=
\frac{( z- w)^{\bar \alpha -\alpha}}{|z-w|^{2\bar \alpha}}
=\frac{(-1)^{\alpha-\bar\alpha}}{[w-z]^\alpha}\,,
\ee
with $\alpha-\bar \alpha=n$, and represent it graphically by the arrow directed
from $w$ to $z$ and with the index $\alpha$ attached to it as shown in
Fig.~\ref{R-figure}. Going over to the Fourier transformation we define the
propagator in the momentum representation
\be
\label{Fourier}
\int d^2 w \frac{\e^{i(w\bar q+\bar w q)}}{[w]^\alpha}
=
\pi\,i^{\alpha-\bar\alpha}\,a(\alpha) \,
\frac{1}{q^{1-\bar\alpha}{\bar q}^{1-\alpha}}
\,.
\ee
Here, the notation was introduced for the function
\be
\label{a-a}
a(\alpha)=\frac{\Gamma(1-\bar \alpha)}{\Gamma(\alpha)}\,,
\qquad
a(\bar\alpha)=\frac{\Gamma(1-\alpha)}{\Gamma(\bar\alpha)}\,,
\qquad
a(\alpha,\beta,...) = a(\alpha)a(\beta) ... \,.
\ee
It has the following properties
\be
a(\alpha)a(1-\bar \alpha)=1\,,\qquad
\frac{a(1+\alpha)}{a(\alpha)}=-\frac1{\alpha\bar\alpha}\,,\qquad
a(\alpha)=(-1)^{\alpha-\bar\alpha} a(\bar\alpha) = \frac{(-1)^{\alpha-\bar\alpha}
}{a(1-\alpha)}\,,
\label{a-prop}
\ee
where in the last relation $\alpha-\bar\alpha$ has to be integer. Taking the
limit of \re{Fourier} for $\alpha=\bar\alpha=i\varepsilon$ as $\varepsilon\to 0$
we obtain the following convenient representation for the delta function
\be
\label{delta-f}
\delta^{(2)}(w)=\lim_{\epsilon\to 0}\frac{a(i\epsilon)}{\pi}
\frac1{[w]^{1-i\varepsilon}}\,.
\ee

\begin{figure}[ht]
\centerline{{\epsfxsize10.0cm\epsfbox{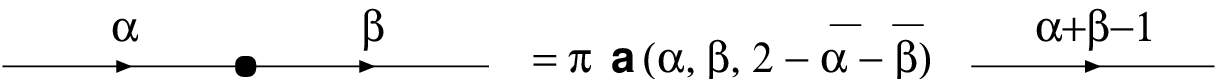}}}
\vspace*{0.5cm}
\caption[]{The chain relation.}
\label{chain}
\vskip 1cm
\end{figure}

\begin{figure}[ht]
\centerline{{\epsfxsize12.0cm\epsfbox{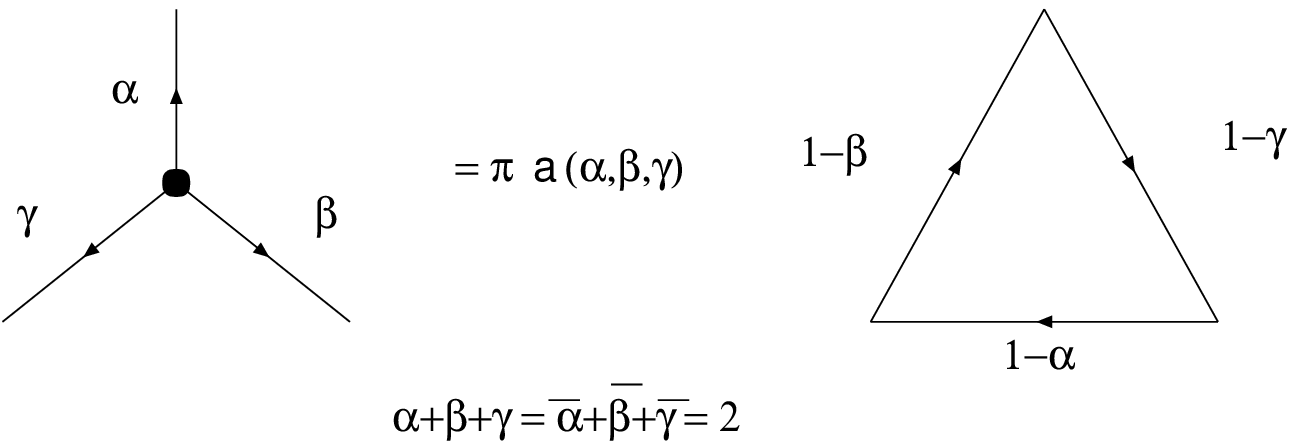}}}
\vspace*{0.5cm}
\caption[]{The uniqueness star-triangle relation.}
\label{uniq-f}
\end{figure}

The method of uniqueness is based on the following two relations shown
diagrammatically in Fig.~\ref{chain}~and~\ref{uniq-f}
\begin{itemize}
\item{{\bf Chain relation}:}
\be\label{chain-rule}
\int d^2 w \frac{1}{[x-w]^\beta\,[w-z]^\alpha}=
\pi \,(-1)^{\gamma-\bar\gamma}\, a(\alpha,\beta,\gamma)\frac{1\,\,\,\,}{
[x-z]^{\alpha+\beta-1}}\,,
\ee
\end{itemize}
where $\gamma=2-\alpha-\beta$ and $\bar\gamma=2-\bar\alpha-\bar\beta$. At
$\alpha+\beta=\bar\alpha+\bar\beta=2$ this relation can be further simplified
using \re{delta-f} as
\be\label{chain-rule-delta}
\int d^2 w \frac{1}{[x-w]^{2-\alpha}\,[w-z]^\alpha}=
\pi^2 \, a(\alpha,2-\alpha)\,\delta^{(2)}(x-z)\,.
\ee
\begin{itemize}
\item{{\bf Star-triangle relation}:}
\be
\label{uniq}
\int d^2 w \frac{1}{[z-w]^\alpha\,[x-w]^\beta\, [y-w]^\gamma}=
\frac{\pi\,  a( \alpha,\beta, \gamma)}
{[x-z]^{1-\gamma}\,[z-y]^{1-\beta}\,[y-x]^{1-\alpha}}
\ee
\end{itemize}
with $\alpha+\beta+\gamma=\bar \alpha+\bar\beta+\bar\gamma=2$.

%

The l.h.s.\ of \re{uniq} is described by the ``star'' diagram in which three
propagators are glued together at the same point $\vec w$. The r.h.s.\ of
\re{uniq} correspond to the ``triangle'' diagram. We shall call the star and
triangle diagrams {\it unique\/} if, firstly, the sums of the holomorphic indices
attached to three propagators in these diagrams coincides with the sum of the
antiholomorphic indices and, secondly, it is equal to 2 and 1, respectively.
Then, the star-triangle relation allows to replace the unique star diagram by
unique triangle and vice versa as shown in Fig.~\ref{uniq-f}.

In the limit $\vec y\to \vec x$, when the end-points of the unique star diagram
approach each other, or equivalently the line in the unique triangle diagram with
the index $1-\alpha$ shrinks into a point, one finds from \re{chain-rule-delta}
\be
\int d^2 w \frac{1}{[z-w]^\alpha\,[x-w]^\beta\, [y-w]^\gamma}\stackrel{\vec y\to \vec
x}{=}\pi^2 \, a(\alpha,2-\bar \alpha)\,\delta^{(2)}(z-x)\,.
\label{uniq-deg}
\ee

To prove \re{chain-rule} it is sufficient to take the Fourier transform of the
both sides of the relation w.r.t.\ $\vec x$ and apply \re{Fourier}. The relation
\re{uniq} follows from \re{chain-rule}. To see this one replaces the coordinates
in \re{chain-rule} as $w\to 1/(w-y)$, $x\to 1/(x-y)$, $z\to 1/(z-y)$ and
similarly for the antiholomorphic coordinates. After a simple calculation one
reproduces Eq.~\re{uniq}.

\section{Appendix: Relation between  $Q-$operator and transfer matrices}

It is well-known in the theory of compact spin magnets that the transfer matrices
of high spin are related to those of the lowest spin through the fusion relations
\cite{KRS,KR}. Moreover, solving the hierarchy of the fusion relations one can
express the transfer matrices of arbitrary spin in terms of the Baxter
$\mathbb{Q}-$operator \cite{BLZ,P,DKM}. In this Appendix, we shall generalize
these relations to the noncompact $SL(2,\mathbb{C})$ spin chains.

Our starting point is the relation between the transfer matrices and the product
of $\mathbb{Q}-$operator, Eq.~\re{T-Q+Q-}. Using this relation we can establish
some general properties of the transfer matrices.

Since the $SL(2,\mathbb{C})$ representations of the spins $(s,\bar s)$ and
$(1-s,1-\bar s)$ are unitary equivalent, we expect that the corresponding
transfer matrices should be related to each other. Indeed, one finds using
\re{T-Q+Q-} together with \re{rho-T} that the transfer matrices
$\mathbb{T}^{(s_0,\bar s_0)}(u,\bar u)$ and ${\mathbb{T}^{(1-s_0,1-\bar
s_0)}(u,\bar u)}$ differ by a c-valued coefficient function
\be
\frac{\mathbb{T}^{(s_0,\bar s_0)}(u,\bar u)}{\mathbb{T}^{(1-s_0,1-\bar s_0)}(u,\bar u)}
=\frac{\rho_{_T}^{(s_0,\bar s_0)}(u,\bar u)}{\rho_{_T}^{(1-s_0,1-\bar
s_0)}(u,\bar u)} =\left[\frac{a(s-1+s_0+iu,\bar s-\bar s_0 - i\bar
u)}{a(s-s_0+iu,\bar s-1+\bar s_0 - i\bar u)}
\right]^N\,.
\label{T-inter}
\ee
Here, we should not worry about ordering of the transfer matrices due their
commutativity.

Making use of \re{Q+dagger}, we obtain the expression for the transfer matrix
\re{T-Q+Q-} entirely in terms of the $\mathbb{Q}_+-$operator in two different
forms
\ba
\mathbb{T}^{(s_0,\bar s_0)}(u,\bar u ) \hspace*{-10mm}&&
\nonumber
\\
&=&
\rho_{_Q}^{(s_0,\bar s_0)}(u,\bar u)\times
\left[\mathbb{Q}_+\lr{\bar u^*-is_0,u^*-i\bar s_0}\right]^\dagger
\mathbb{Q}_+\lr{u+ i s_0, \bar u+ i\bar s_0}
\label{T-QQ}
\\
&=&
\widetilde\rho_{_Q}^{(s_0,\bar s_0)}(u,\bar u)\times
\left[\mathbb{Q}_+\lr{\bar u^*-i(1-s_0),u^*-i(1-\bar s_0)}\right]^\dagger
\mathbb{Q}_+\lr{u+ i (1-s_0), \bar u+ i(1-\bar s_0)}
\nonumber
\ea
Here, the normalization factors are given by
\ba
\rho_{_Q}^{(s_0,\bar s_0)}(u,\bar u)&=&\left[(-1)^{2i(u-\bar u)}\frac{n_s^2+4\nu_s^2}{\pi^4}
\frac{a(s-1+s_0+iu,\bar s-\bar s_0-i\bar u)}{a(s-s_0+iu,\bar s+\bar s_0-i\bar u)}
\right]^N\,,
\label{rho-Q}
\\
\widetilde\rho_{_Q}^{(s_0,\bar s_0)}(u,\bar u)&=&\left[(-1)^{2i(u-\bar u)}\frac{n_s^2+4\nu_s^2}{\pi^4}
{a(s_0-s+iu,\bar s-\bar s_0-i\bar u)}
\right]^N\,.
\nonumber
\ea
It is straightforward to verify that \re{T-QQ} satisfies the intertwining
relation \re{T-inter} and the normalization factors $\rho_{_T}^{(s_0,\bar s_0)}$,
$\rho_{_Q}^{(s_0,\bar s_0)}$ and $\widetilde\rho_{_Q}^{(s_0,\bar s_0)}$ are
related to each other as
\be
\widetilde\rho_{_Q}^{(s_0,\bar s_0)}(u,\bar u)=\rho_{_Q}^{(1-s_0,1-\bar s_0)}(u,\bar u)\,\rho_{_T}^{(s_0,\bar
s_0)}(u,\bar u)/\rho_{_T}^{(1-s_0,1-\bar s_0)}(u,\bar u)\,.
\ee

Using \re{T-QQ} we can find the relation between the transfer matrix and its
conjugate counterpart. Conjugating the both sides of the first relation in
\re{T-QQ} we can either match it into itself using the relations \re{Q-mirror}, or
into the second relation in \re{T-QQ}. In this way, we obtain two equivalent
representations
\ba
 [\mathbb{T}^{(s_0,\bar s_0)}(u,\bar u )]^\dagger
 &=&\mathbb{M}\, \mathbb{T}^{(s_0,\bar s_0)}(-\bar u^*, -u^*)\,\mathbb{M}\times
 \frac{(\rho_{_Q}^{(s_0,\bar s_0)}(u,\bar u))^*}{\rho_{_Q}^{(s_0,\bar s_0)}(-\bar u^*,-u^*)}
\\
 &=& \mathbb{T}^{(s_0,\bar s_0)}(\bar u^*+i, u^*+i)\times
 \frac{(\rho_{_Q}^{(s_0,\bar s_0)}(u,\bar u))^*}{\widetilde\rho_{_Q}^{(s_0,\bar s_0)}(\bar u^*+i,
 u^*+i)}\,.
\label{fusion}
\ea

Finally, let us show that the transfer matrices of different spin satisfy
nonlinear fusion relations. To this end, one considers the product of two
transfer matrices of different spins as well as different values of the spectral
parameters and expresses it in terms of the $\mathbb{Q}-$operators using
\re{T-QQ}. The resulting expressions involve four $\mathbb{Q}-$operators including
two conjugated operators. Using their commutativity, we may regroup them into two
pairs of $\mathbb{Q}^\dagger\mathbb{Q}-$operators and apply the relations
\re{T-QQ} to convert each pair into the transfer matrix. In this way, we arrive at
the following fusion relation
\ba
&&\widehat{\mathbb{T}}^{(s_0-i\varrho,\bar s_0-i\bar
\varrho)}(u+\sigma,\bar u+\bar \sigma)\
\widehat{\mathbb{T}}^{(s_0+i\varrho,\bar s_0+i\bar \varrho)}(u-\sigma,\bar u-\bar \sigma)
\nonumber
\\[2mm]
=&&
\widehat{\mathbb{T}}^{(s_0-i\sigma,\bar s_0-i\bar \sigma)}(u+\varrho,\bar u+\bar \varrho)\
\widehat{\mathbb{T}}^{(s_0+i\sigma,\bar s_0+i\bar \sigma)}(u-\varrho,\bar u-\bar \varrho)\,,
\label{T-T}
\ea
where $\widehat{\mathbb{T}}^{(s_0)}(u) = \mathbb{T}^{(s_0,\bar s_0)}(u,\bar u
)/\rho_{_Q}^{(s_0,\bar s_0)}(u,\bar u)$. Another fusion relation follows from the
Baxter equation \re{def-Qh}. Multiplying its both sides by conjugated
$\mathbb{Q}-$operator and applying \re{T-QQ} we find
\be
t_N(u)~\widehat{\mathbb{T}}^{(s_0,\bar s_0)}(u,\bar u)
=(u+is)^N\,\widehat{\mathbb{T}}^{(s_0+\frac12,\bar s_0)}\lr{u-\frac{i}2,\bar u}
+(u-is)^N\,\widehat{\mathbb{T}}^{(s_0-\frac12,\bar s_0)}\lr{u+\frac{i}2,\bar u}
\label{t-T}
\ee
and similar relation in the antiholomorphic sector
\be
\bar t_N(\bar u)~\widehat{\mathbb{T}}^{(s_0,\bar s_0)}(u,\bar u)
=(\bar u+i\bar s)^N\,\widehat{\mathbb{T}}^{(s_0,\bar s_0+\frac12)}\lr{u,\bar
u-\frac{i}2} +(\bar u-i\bar s)^N\,\widehat{\mathbb{T}}^{(s_0,\bar
s_0-\frac12)}\lr{u,\bar u+\frac{i}2}\,.
\label{bar t-T}
\ee
There is the following important difference between these relations and \re{T-T}.
According to the definition of the transfer matrix $\mathbb{T}^{(s_0,\bar
s_0)}(u,\bar u)$, Eq.~\re{t-N}, the spins $(s_0,\bar s_0)$ parameterize the
auxiliary space $V^{(s_0,\bar s_0)}$ and for the $SL(2,\mathbb{C})$
representation of the principal series they satisfy the conditions \re{s-cond}.
We notice that for the transfer martices in the r.h.s.\ of \re{t-T} and
\re{bar t-T} these conditions are not satisfied and, therefore, they do not belong
to the same family of the transfer matrices of the $SL(2,\mathbb{C})$ principal
series. At the same time, appropriately choosing the parameters $(\varrho,\bar
\varrho)$ and $(\sigma,\bar\sigma)$, one finds from \re{T-T} the fusion
relations between the transfer matrices within the same family.

\section{Appendix: Bethe Ansatz representation of the eigenstates}

In this appendix we prove the relation \re{Psi-ABA} which establishes the
correspondence between \re{SoV-gen} and the representation of the eigenstates in
the Algebraic Bethe Anstatz \cite{QISM,XXX,ABA}. The ABA representation is based
on the existence of the special pseudovacuum state $\Omega(\vec z)$, which is
annihilated by the spin operators of all particles \re{SL2-spins}
\be
\Omega(\vec z)=\prod_{k=1}^N {z_k^{-2s}\bar z_k^{-2\bar s}}\,,\qquad
S_+^{(k)}\,\Omega(\vec z)=\bar S_+^{(k)}\,\Omega(\vec z)=0\,.
\ee
The remarkable feature of this state is \cite{QISM,XXX,ABA} that it brings the
Lax operators \re{Lax} to a upper-triangle form and, as a consequence,
diagonalizes the auxiliary transfer matrices \re{t-aux}
\be
t_N(u)\,\Omega(\vec z-\vec z_0) = \left[(u+is)^N+(u-is)^N\right]\Omega(\vec
z-\vec z_0)
\label{t-pseudo}
\ee
with $\vec z_0$ being arbitrary vector reflecting the $SL(2,\mathbb{C})$
invariance of $t_N(u)$. The operator $\bar t_N(\bar u)$ satisfies similar
relation. Then, the complete integrability of the model implies that $\Omega(\vec
z-\vec z_0)$ diagonalizes the Baxter $\mathbb{Q}-$operators and, as a
consequence, the Hamiltonian of the model. This does not mean, however, that the
pseudovacuum state can be identified as one of the eigenstates of the model.
Examining its $SL(2,\mathbb{C})$ transformation properties, Eqs.~\re{SL2-trans}
and \re{SL2-trans-z}, one verifies that $\Omega(\vec z-\vec z_0)$ does not belong
to the quantum space of the system (see footnote to Eq.~\re{omega}).
Nevertheless, one can use its properties to establish different relations
including Eq.~\re{Psi-ABA}.

Going over to the momentum representation, we define the state
\be
\Omega_{\vec p}(\vec z_1,...,\vec z_N) = c_{\vec p}^\Omega\int d^2
z_0\,\e^{2i\vec p\cdot\vec z_0}\Omega(\vec z-\vec z_0)
\label{Omega}
\ee
with the normalization factor $c_{\vec p}^\Omega=(-1)^{n_s}p^{2s-1}\bar p^{2\bar
s-1}/(\pi^3 a(2(1-s)))$ chosen for the later convenience. Let us examine its
convolution with the transition function defined in \re{U-Q-ops}
\be
\int d^2{z}\, (\Omega_{\vec p}(\vec{{z}}))^*U_{\vec p\,',\vec{\mybf{x}}}(\vec{{z}})
=\lim_{\vec z_0=\infty}\vev{\Omega_{\vec p}\,|\,\mathbb{Q}_{-} (x_1,\bar
x_1)\ldots \mathbb{Q}_{-}(x_{N-1},\bar x_{N-1})\,|\,\omega_{\vec z_0,\vec
p\,'}\,}
\label{Omega-Q-omega}
\ee
with the state $\omega_{\vec z_0=\infty,\vec p\,'}(\vec z)$ defined in
\re{omega}. Then, substituting \re{Omega} into \re{Omega-Q-omega} and calculating
the convolution $\int d^2z\, (\Omega(\vec z-\vec z_0))^*\,Q^{_{(-)}}_{u,\bar
u}(\vec z|\vec w)$, we represent the kernel of the $\mathbb{Q}-$operator by the
right diagram in Fig.~\ref{Q2-f}. The integral over the intermediate points $\vec
z_k$ takes the form \re{chain-rule-delta} and it gives rise to the
$\delta-$functions which put the centers of the star diagrams at the same point
$z_0$. The resulting diagram is given by the product of propagators connecting
$\vec z_0$ with the points $\vec z_k$ and it coincides, up to prefactor, with
$\Omega_{\vec p}$. In this way, one gets
\be
\bra{\Omega_{\vec p}\,}\,\mathbb{Q}_-(u,\bar u) = \left[\pi\,a(2-2s,s+iu,\bar s-i\bar u)\right]^N
\bra{\Omega_{\vec p}\,}
\equiv A(u,\bar u)\bra{\Omega_{\vec p}\,}
\,.
\label{Q-Omega}
\ee
As it was expected, the pseudovacuum state diagonalizes the Baxter operator. One
can verify that its eigenvalue satisfies the Baxter equation \re{def-Qh} with
$t_N(u)$ replaced by \re{t-pseudo}.

Substituting \re{Q-Omega} into \re{Omega-Q-omega} one replaces the
$\mathbb{Q}-$operators by their corresponding eigenvalues and reduces the r.h.s.\
of \re{Omega-Q-omega} to the convolution of the pseudovacuum state with the
function $\omega_{\vec z_0,\vec p\,'}$ defined in \re{omega}. A simple
calculation leads to
\be
\lim_{\vec z_0=\infty}\vev{\Omega_{\vec p}\,|\omega_{\vec z_0,\vec p\,'}}=
c_{\vec p}\, \delta^{(2)}(p-p')\,.
\ee
Here, the normalization factor $c_{\vec p}$ is given by \re{c-p}. Finally, one
calculates \re{Omega-Q-omega} as
\be
\int d^2{z}\, (\Omega_{\vec p}(\vec{{z}}))^*U_{\vec p\,',\vec{\mybf{x}}}(\vec{{z}})
={c_{\vec p}\,\delta^{(2)}(p-p')}{\prod_{k=1}^{N-1} A(x_k,\bar x_k)}\,.
\label{Omega-U}
\ee
One can arrive at the same relation using the diagrammatical representation of
the l.h.s.\ as the pyramid diagram, Fig.~\ref{Fig-SoV}, with the additional lines
carrying the indices $2(1-s)$ attached to its end-points $\vec z_k$ and joined at
the same point $\vec z_0$.

Using the completeness condition, Eq.~\re{U-comp}, we find from \re{Omega-U}
\be
{ c_{\vec p}^*}\int{ d\mybf{x}\,\mu(\vec{\mybf{x}})\,U_{\vec
p,\vec{\mybf{x}}}(\vec{{z}})}{\prod_{k=1}^{N-1} \left(A(x_k,\bar
x_k)\right)^*}=(2\pi)^{-N}\,\Omega_{\vec p}(\vec{{z}})\,.
\ee
Here, the normalization factor in the l.h.s.\ compensates the difference in the
scaling dimensions of $U_{\vec p,\vec{\mybf{x}}}$ and $\Omega_{\vec p}$. Let us
consider an arbitrary operator $f(\hat x_k,\hat{\bar x}_k)$, depending on the
Sklyanin's operator zeros defined in Sect.~4.1.1, and apply it to the both sides
of the last relation. Since $U_{\vec p,\vec{\mybf{x}}}$ diagonalizes the
operators zeros, Eq.~\re{x-U}, this operator can be replaced in the l.h.s.\ by
its eigenvalue $f(x_k,\bar x_k)$. Then, choosing $f(x_k,\bar x_k)$ as the wave
function in the separated coordinates $\Phi_{\{q,\bar q\}}(\vec{\mybf{x}})$,
Eq.~\re{Phi-Q-ops}, we obtain from \re{SoV-gen} (up to an overall normalization
factor depending on the momentum $\vec p$)
\be
{\Psi_{\vec p,\{q,\bar q\}}}(\vec z) =(2\pi)^N { c_{\vec p}^*}\int{
d\mybf{x}\,\mu(\vec{\mybf{x}})\,U_{\vec
p,\vec{\mybf{x}}}(\vec{{z}})}{\prod_{k=1}^{N-1}\left(A(x_k,\bar
x_k)\right)^*}Q_{q,\bar q}(x_k,{\bar x}_k)= \prod_{k=1}^{N-1}Q_{q,\bar q}(\hat
x_k,\hat{\bar x}_k) {\Omega_{\vec p}}(\vec z)\,. \nonumber
\label{Psi-Omega}
\ee
Here, $Q_{q,\bar q}(\hat x_k,\hat{\bar x}_k)$ stands for the eigenvalue of the
Baxter operator with the spectral parameters substituted by the operator zeros.

The relation \re{Psi-Omega} generalizes the well-known highest-weight
representation of the eigenstates of compact spin chain magnets in the Algebraic
Bethe Ansatz
\be
\ket{\Psi_{l}^{\rm ABA}} = B(\lambda_1) ... B(\lambda_l)\ket{\Omega_{\vec p}}
\label{ABA}
\ee
with $B(u)$ being the off-diagonal element of the monodromy matrix,
Eq.~\re{monodromy}, and $\lambda_1,...,\lambda_l$ satisfying the Bethe equations.
Using \re{B-x}, one can show that two representations, Eqs.~\re{Psi-Omega} and
\re{ABA}, are equivalent if the eigenvalues of the Baxter $\mathbb{Q}-$operators
are given by polynomials of degree $l$ in the spectral parameter $u$. Going
beyond the class of polynomial solutions to the Baxter equation, one finds
\cite{FK,K1} that \re{ABA} is not applicable while the representation
\re{Psi-Omega} remains valid.

\end{document}